\documentclass[aps,prd,10pt,notitlepage,superscriptaddress,nofootinbib,numbers]{revtex4-2}

\usepackage[utf8]{inputenc}
\usepackage[T1]{fontenc}
\usepackage{lmodern}
\usepackage{microtype}
\usepackage{amsmath,amssymb,amsfonts,mathrsfs}
\usepackage{bm}
\usepackage{graphicx}
\usepackage{float}
\usepackage[dvipsnames]{xcolor}
\usepackage{hyperref}
\hypersetup{colorlinks=true,citecolor=Purple,linkcolor=Purple,urlcolor=Purple,breaklinks=true}
\usepackage[capitalize]{cleveref}
\usepackage{tikz}

\DeclareMathOperator{\sech}{sech}

\newcommand{\dd}{\mathrm{d}}
\newcommand{\FS}{F_{\mathcal S}}
\newcommand{\sig}{\sigma}
\newcommand{\Veff}{V_{\rm eff}}

\definecolor{lime}{HTML}{A6CE39}
\DeclareRobustCommand{\orcidicon}{%
	\begin{tikzpicture}
		\draw[lime, fill=lime] (0,0) 
		circle [radius=0.16] 
		node[white] {{\fontfamily{qag}\selectfont \tiny ID}};
		\draw[white, fill=white] (-0.0625,0.095) 
		circle [radius=0.007];
	\end{tikzpicture}
	\hspace{-2mm}
}

\foreach \x in {A, ..., Z}{%
	\expandafter\xdef\csname orcid\x\endcsname{\noexpand\href{https://orcid.org/\csname orcidauthor\x\endcsname}{\noexpand\orcidicon}}
}

\newcommand\orcidJonathan{{\href{https://orcid.org/0000-0001-9291-0893}{\orcidicon}}}
\newcommand\orcidEdson{{\href{https://orcid.org/0000-0001-9929-5977}{\orcidicon}}}

\begin{document}

\title{Thin-Shell Wormholes from Entropy-Induced Black-Hole Geometries}

\author{Jonathan A. Rebou\c{c}as\orcidJonathan\!\!}
\email{jalvesreboucas@ifce.edu.br}
\affiliation{Instituto Federal de Educa\c{c}\~ao, Ci\^encia e Tecnologia do Cear\'a (IFCE), Iguatu, Brazil}

\author{Edson Otoniel\orcidEdson\!\!}
\email{edson.otoniel@ufca.edu.br}
\affiliation{Universidade Federal do Cariri (UFCA), Instituto de Forma\c{c}\~ao de Educadores - IFE,  R. Oleg\'ario Emidio de Araujo S/N, Brejo Santo - CE, 63.260-000 - Brazil}

\date{\today}

\begin{abstract}
Modified black-hole entropies can induce effective spacetime geometries and thereby provide a thermodynamic route for investigating thin-shell wormholes. In this work, we construct symmetric cut-and-paste wormholes from the generic entropic lapse function $F_{\mathcal S}(r)=1-4\pi M/\mathcal S'(r)$ and formulate the Darmois--Israel junction conditions directly in terms of the lapse and of the entropy derivatives. We derive the surface stresses, shell energy-condition combinations, conservation equation, and radial effective potential, and then apply the formalism to the Bekenstein--Hawking, Barrow, Tsallis--Cirto, R\'enyi, Kaniadakis, logarithmic, loop-quantum-gravity-inspired, and exponential entropy prescriptions. The analysis shows that the symmetric construction requires negative surface energy density throughout every admissible positive-lapse domain, although entropy deformations can significantly modify the horizon structure, the allowed throat region, and the localization of the surface stresses. Within the parameter domains considered here, all examined constant-barotropic branches are linearly radially unstable, despite quantitative changes in their near-horizon scales. In contrast, a variable Chaplygin shell can support stable configurations, with the stability domains determined jointly by the entropic geometry, the throat radius, and the radial exponent of the shell equation of state. These results establish a unified framework for comparing entropy-induced black-hole geometries as thin-shell wormhole seeds and show that stability is governed not by the entropy deformation alone, but by its interplay with the dynamical response of the matter localized at the throat.
\end{abstract}

\maketitle

\tableofcontents

\section{Introduction}\label{sec:introduction}

Traversable wormholes provide a useful arena for examining the relation between spacetime topology, gravitational dynamics, and the matter content required to sustain nontrivial geometries. They represent horizonless configurations with a minimum-area surface connecting two regions of spacetime and therefore offer a controlled setting in which the interplay among causal structure, energy conditions, and possible extensions of the gravitational sector can be investigated. The modern traversable-wormhole program was established in the Morris--Thorne framework \cite{MorrisThorne1988}, while the Ellis and Bronnikov scalar-field configurations provided early explicit realizations of regular geometries with nontrivial topology \cite{Ellis1973,Bronnikov1973}. These constructions and their subsequent developments have made wormholes useful theoretical laboratories for testing the limits of classical general relativity and for assessing whether effective or quantum-inspired sectors can sustain geometries that ordinary classical matter does not readily support \cite{Visser1989,VisserBook,Radhakrishnan2024}. Possible astrophysical signatures and sensitive observational searches have also been investigated \cite{DaiStojkovic2019Observing,SimonettiEtAl2021SensitiveSearch,BambiStojkovic2021Astrophysical}.

A central difficulty is that the flare-out condition at a traversable throat is in general relativity closely tied to the violation of the null energy condition. This feature motivates the search for mechanisms that either confine the exoticity to a small region, reduce its integrated amount, or reinterpret it as an effective contribution generated by additional gravitational degrees of freedom. Examples include wormholes supported by phantom sectors \cite{Lobo2005,Sushkov2005}, modified-gravity configurations in which the geometric corrections act as an effective source \cite{Harko2013}, and constructions based on quantum-vacuum effects, brane tension, or cosmological repulsion in de Sitter space \cite{DaiMinicStojkovic2020Formation,DaiMinicStojkovic2018DeSitter}. In this sense, the study of wormholes is not only a search for exotic solutions: it is also a way of probing how the microscopic, effective, or emergent content of gravity may influence the admissible macroscopic topology of spacetime.

Recent progress has broadened the range of source sectors and gravitational frameworks in which traversable wormholes can be explored. Loop-quantum-gravity-inspired effective sources were used in Ref.~\cite{Cruz2024}, while the role of asymptotic-safety corrections in Ellis--Bronnikov geometries was examined in Ref.~\cite{Alencar2021}. Quantum-vacuum contributions have been considered through Yang--Mills Casimir sources and through thermally corrected Casimir sectors in Einstein--Gauss--Bonnet gravity \cite{Santos2024,Muniz2025}. Ref.~\cite{Crispim2026} further showed that generalized Ellis--Bronnikov geometries can arise in general relativity from combined phantom-scalar and electromagnetic fields, while Ref.~\cite{NojiriOdintsovFolomeev2024Wormholes} investigated wormholes inside stars and black holes. These examples illustrate that the support of a wormhole throat can be investigated using a variety of mechanisms, ranging from effective quantum corrections and higher-curvature terms to nonstandard but regular matter sectors.

A complementary line of research starts from phenomenologically motivated density profiles and reconstructs the corresponding Morris--Thorne geometry. Dark-matter-inspired distributions, for example, have been used in modified teleparallel gravity, loop quantum cosmology, and Dekel--Zhao halo constructions \cite{Mustafa2023,Silva2025,Khatri2025}. Matter-first strategies based on rational approximations have also been developed to preserve a prescribed source profile while maintaining the geometric requirements of asymptotic flatness and flare-out \cite{ReboucasPade2026}. Within this broader class of inspired constructions, entropy-induced density profiles have recently been employed as phenomenological inputs for smooth Morris--Thorne wormholes, together with pressures reconstructed from the field equations and throat-regularity conditions \cite{ReboucasEntropyMT2026}. These approaches are valuable because they allow one to isolate the physical implications of a chosen effective source before committing to a complete microscopic theory.

Thin-shell wormholes provide a distinct and complementary route. In the cut-and-paste construction, two exterior spacetime regions are joined across a timelike hypersurface, so that the matter required to maintain the throat is confined to an infinitesimally thin layer instead of being distributed throughout the bulk \cite{Visser1989Surgical,PoissonVisser1995}. The surface stress-energy tensor is determined by the discontinuity of the extrinsic curvature through the Darmois--Israel junction conditions \cite{Darmois1927,Lanczos1924,Israel1966}. This construction separates the bulk geometry from the matter localized on the shell and makes it possible to investigate directly how a given seed spacetime influences the surface energy conditions and the dynamics of the throat.

The existence of an admissible timelike junction does not, however, guarantee dynamical viability. The linearized stability analysis introduced in Ref.~\cite{PoissonVisser1995} showed that radial perturbations are controlled jointly by the seed geometry and the local equation of state of the shell. Subsequent analyses extended this treatment to generic spherically symmetric thin shells \cite{Eiroa2008Stability,GarciaLoboVisser2012}, as well as to cosmological, charged, dilatonic, and generalized-Chaplygin-gas backgrounds \cite{LoboCrawford2004,EiroaRomero2004,EiroaSimeone2005,Eiroa2007Chaplygin,Eiroa2009Chaplygin,Varela2015}. More recent studies have explored complementary thermodynamic criteria for Schwarzschild thin-shell wormholes \cite{ForghaniMazharimousaviHalilsoy2019}, quantum-corrected polymer black-hole seeds \cite{Javed2024PolymerTSW}, and the simultaneous dynamical and thermodynamical stability of charged shells \cite{EiroaFigueroaAguirre2024}. Collectively, these results show that the physical viability of a thin-shell wormhole depends not only on the shell equation of state, but also on the horizon structure and local derivatives of the metric function inherited from the bulk geometry.

At the same time, black-hole thermodynamics has provided a major conceptual bridge between gravitation, quantum theory, and statistical physics. The Bekenstein--Hawking entropy, Hawking radiation, Euclidean gravitational thermodynamics, and the Noether-charge formulation of horizon entropy indicate that spacetime geometry carries thermodynamic information associated with horizon degrees of freedom \cite{Bekenstein1973,Hawking1975,GibbonsHawking1977,Wald1993}. This perspective was strengthened by the derivation of Einstein's equations from local horizon thermodynamics, by the interpretation of gravitational dynamics as an equation of state, and by entropic-force proposals in which gravity emerges from microscopic holographic degrees of freedom \cite{Jacobson1995,Padmanabhan2010,Verlinde2011}. From this viewpoint, gravity can be regarded as an effective macroscopic manifestation of an underlying thermodynamic or statistical structure.

Modified black-hole entropies provide a natural setting in which this possibility can be examined. Deviations from the Bekenstein--Hawking area law have been proposed in the context of rough or fractal horizon deformations, nonextensive statistical mechanics, R\'enyi-type generalizations, relativistic generalized statistics, logarithmic quantum corrections, and exponential contributions of semiclassical or nonperturbative origin \cite{Barrow2020,Tsallis1988,TsallisCirto2013,Renyi1961,CzinnerIguchi2016,Kaniadakis2002,KaulMajumdar2000,DasMajumdarBhaduri2002,ChatterjeeGhosh2020}; broader generalized-entropy frameworks and consistency analyses have also been developed \cite{NojiriOdintsovFaraoni2022GeneralizedEntropy,NojiriOdintsovPaul2022GeneralizedEntropy,NojiriOdintsovFaraoni2021AreaLaw,ElizaldeNojiriOdintsov2025GeneralisedEntropy}. Complementary approaches interpret generalized entropy as modifying the effective gravitational coupling or the gravitational dynamics \cite{LuDiGennaroOng2025VaryingG,FiglioliaJizbaLambiase2026ThermodynamicGravity}. Ref.~\cite{Anand2026} recently established an entropy--geometry correspondence in which a chosen entropy functional determines a static and spherically symmetric black-hole metric and an associated effective anisotropic matter sector. The entropy families considered here are precisely the representative models analyzed in that correspondence: the Bekenstein--Hawking, Barrow, Tsallis--Cirto, R\'enyi, Kaniadakis, logarithmic, loop-quantum-gravity-inspired, and exponential prescriptions. Taken together, they provide a broad and physically distinct sample of departures from the area law, including power-law and fractal deformations, nonextensive and logarithmic generalizations, hyperbolic statistical corrections, and localized quantum-inspired modifications. They are therefore well suited for assessing which qualitative features of a thin-shell wormhole are sensitive to the particular manner in which the Bekenstein--Hawking entropy is deformed.

The aim of this work is to investigate the viability and linear radial stability of symmetric thin-shell wormholes constructed from entropy-induced black-hole geometries in an emergent-gravity context. Starting from a generic entropy function, we obtain the associated lapse function, as showed in Ref.\cite{Anand2026}, and construct the wormhole by gluing two exterior copies of the corresponding spacetime. We then derive the surface stresses, the shell energy-condition combinations, the conservation equation, and the radial effective potential. The general framework is applied to the entropy families listed above, with the Schwarzschild thin-shell wormhole recovered from the Bekenstein--Hawking entropy as the consistency limit. This construction makes it possible to assess whether entropy-induced deformations change the horizon structure, modify the admissible static region of the shell, reduce the magnitude of the negative surface energy, or generate stability behavior distinct from the standard Schwarzschild branch.

This paper is organized as follows. In Sec.~\ref{sec:entropic_bulk}, we introduce the entropy-induced black-hole geometry and its horizon condition. In Sec.~\ref{sec:tsw_general}, we construct the symmetric thin-shell wormhole and derive the induced metric, proper-time normalization, unit normal, and extrinsic-curvature components. Section~\ref{sec:surface_conditions} presents the surface stress tensor and the relevant energy-condition combinations. In Sec.~\ref{sec:dynamics}, we derive the conservation equation and the radial effective potential, and introduce the constant barotropic and variable Chaplygin shell models used in the stability analysis. In Sec.~\ref{sec:entropy_models}, the general framework is specialized to the different entropy prescriptions. Finally, Sec.~\ref{sec:conclusions} summarizes the main results and discusses their implications. Throughout this work, we use geometrized units $G=c=\hbar=k_B=1$. All entropy functions and their parameters are understood in Planck units, with the combinations entering logarithms, exponentials, and noninteger powers taken to be dimensionless.

\section{Entropy-induced black-hole geometry}
\label{sec:entropic_bulk}

This section defines the seed geometry used in the cut-and-paste construction. The only input is an entropy function $\mathcal S(r)$ written as a function of the areal radius.

We consider a static and spherically symmetric metric in areal coordinates,
\begin{equation}
\dd s^2=-\FS(r)\dd t^2+\frac{\dd r^2}{\FS(r)}+r^2\dd\Omega^2,
\qquad
\FS(r)=1-\frac{4\pi M}{\mathcal S'(r)}.
\label{eq:entropic_metric}
\end{equation}
Here a prime denotes differentiation with respect to $r$, and $M$ is the mass scale of the entropy--geometry prescription. It coincides
with the ADM mass only in the asymptotically Schwarzschild sectors. We adopt the entropy--geometry prescription of Ref.~\cite{Anand2026}. In that construction, the first law determines the relation between the entropy derivative and the metric function locally at the horizon. The extension of this relation to the exterior radial domain, leading to Eq.~\eqref{eq:entropic_metric}, is therefore a prescribed global continuation rather than a direct consequence of the first law away from the horizon. Throughout this work, the resulting geometries are interpreted within general relativity as being sourced by the effective anisotropic matter sector associated with the entropy deformation. Under this effective-GR interpretation, the standard Darmois--Israel junction conditions apply. If one writes $f(r)=1-Mg(r)$ and imposes $\dd M=T_H\dd \mathcal S$ with $T_H=f'(r_h)/(4\pi)$ at a simple horizon, the horizon relation is $g(r_h)=4\pi/\mathcal S'(r_h)$. Promoting this relation to the exterior radial domain gives Eq.~\eqref{eq:entropic_metric}.

The event horizon is obtained from the zero of the lapse function,
\begin{equation}
\FS(r_h)=0,
\qquad
\mathcal S'(r_h)=4\pi M.
\label{eq:horizon_general}
\end{equation}
In the thin-shell construction, the throat must be placed in a static region of the seed geometry. Thus, for a single-horizon black hole one requires $a>r_h$ and $\FS(a)>0$. If a given entropy model produces more than one positive zero of $\FS(r)$, the shell must be placed inside a positive-lapse interval. In the latter case, the present construction is restricted to the local static patch containing the shell, and no claim is made about a global causal extension connecting two asymptotically accessible exterior regions.

The derivatives of the entropic lapse will enter the pressure and the stability analysis. We use
\begin{equation}
\FS'(r)=\frac{4\pi M\mathcal S''(r)}{[\mathcal S'(r)]^2},
\qquad
\FS''(r)=4\pi M\left[\frac{\mathcal S'''(r)}{[\mathcal S'(r)]^2}-\frac{2[\mathcal S''(r)]^2}{[\mathcal S'(r)]^3}\right].
\label{eq:lapse_derivatives_entropy}
\end{equation}
These expressions allow all junction quantities to be written either in terms of the metric function or directly in terms of the entropy derivatives.

\section{Thin-shell wormhole from a generic entropy}
\label{sec:tsw_general}

This section constructs the shell and derives the junction quantities. The calculation is kept generic so that each entropy model can be inserted later through its lapse function.

We take two identical copies of the positive-lapse static region of Eq.~\eqref{eq:entropic_metric} containing the shell, denoted by $\mathcal M^+$ and $\mathcal M^-$. The two copies are identified at the timelike hypersurface
\begin{equation}
\Sigma:\quad r_+=r_-=a(\tau),
\qquad
x^\mu_\pm(\tau,\theta,\phi)=\left(t_\pm(\tau),a(\tau),\theta,\phi\right).
\label{eq:shell_embedding}
\end{equation}
The intrinsic coordinates on the shell are $\xi^i=(\tau,\theta,\phi)$, where $\tau$ is the proper time measured by an observer comoving with the throat. The induced metric is
\begin{equation}
\dd s_\Sigma^2=-\dd\tau^2+a^2(\tau)\dd\Omega^2.
\label{eq:induced_metric}
\end{equation}
For a static configuration, $a(\tau)=a_0$, the induced geometry becomes
\begin{equation}
\dd s_\Sigma^2=-\dd\tau^2+a_0^2\dd\Omega^2.
\label{eq:induced_metric_static}
\end{equation}

The proper-time normalization of the shell four-velocity $u^\mu_\pm=(\dot t_\pm,\dot a,0,0)$ is
\begin{equation}
-\FS(a)\dot t_\pm^2+\frac{\dot a^2}{\FS(a)}=-1,
\qquad
\dot t_\pm=\frac{\sqrt{\FS(a)+\dot a^2}}{\FS(a)}=\frac{\sqrt{1-4\pi M/\mathcal S'(a)+\dot a^2}}{1-4\pi M/\mathcal S'(a)}.
\label{eq:proper_time_normalization}
\end{equation}
The covariant unit normal, oriented outward from each copy, is
\begin{equation}
\eta_\mu^\pm=\pm\left(-\dot a,\frac{\sqrt{\FS(a)+\dot a^2}}{\FS(a)},0,0\right)=\pm\left(-\dot a,\frac{\sqrt{1-4\pi M/\mathcal S'(a)+\dot a^2}}{1-4\pi M/\mathcal S'(a)},0,0\right).
\label{eq:normal_entropy}
\end{equation}
It satisfies $\eta_\mu u^\mu=0$ and $\eta_\mu\eta^\mu=1$. In the static case,
\begin{equation}
\eta_\mu^\pm=\pm\left(0,\frac{1}{\sqrt{\FS(a_0)}},0,0\right)=\pm\left(0,\frac{1}{\sqrt{1-4\pi M/\mathcal S'(a_0)}},0,0\right).
\label{eq:normal_static_entropy}
\end{equation}

The extrinsic curvature is defined by
\begin{equation}
K_{ij}^{\pm}=-\eta_\mu^{\pm}\left(\frac{\partial^2x^\mu}{\partial\xi^i\partial\xi^j}+\Gamma^\mu_{\alpha\beta}\frac{\partial x^\alpha}{\partial\xi^i}\frac{\partial x^\beta}{\partial\xi^j}\right).
\label{eq:extrinsic_definition}
\end{equation}
In an orthonormal basis on the shell, the independent components are
\begin{equation}
K_{\hat\tau\hat\tau}^{\pm}=\mp\frac{\FS'(a)+2\ddot a}{2\sqrt{\FS(a)+\dot a^2}}=\mp\frac{2\ddot a+4\pi M\mathcal S''(a)/[\mathcal S'(a)]^2}{2\sqrt{1-4\pi M/\mathcal S'(a)+\dot a^2}},
\label{eq:K_tautau_entropy}
\end{equation}
\begin{equation}
K_{\hat\theta\hat\theta}^{\pm}=K_{\hat\phi\hat\phi}^{\pm}=\pm\frac{\sqrt{\FS(a)+\dot a^2}}{a}=\pm\frac{\sqrt{1-4\pi M/\mathcal S'(a)+\dot a^2}}{a}.
\label{eq:K_angular_entropy}
\end{equation}
For a static throat one obtains
\begin{equation}
K_{\hat\tau\hat\tau}^{\pm}\big|_0=\mp\frac{F'_{\mathcal S,0}}{2\sqrt{F_{\mathcal S,0}}}=\mp\frac{4\pi M\mathcal S''_0/[\mathcal S'_0]^2}{2\sqrt{1-4\pi M/\mathcal S'_0}},
\label{eq:K_tautau_static_entropy}
\end{equation}
\begin{equation}
K_{\hat\theta\hat\theta}^{\pm}\big|_0=K_{\hat\phi\hat\phi}^{\pm}\big|_0=\pm\frac{\sqrt{F_{\mathcal S,0}}}{a_0}=\pm\frac{\sqrt{1-4\pi M/\mathcal S'_0}}{a_0}.
\label{eq:K_angular_static_entropy}
\end{equation}
Here and below, a subscript zero indicates evaluation at $r=a_0$.

\section{Surface stress tensor and energy conditions}
\label{sec:surface_conditions}

This section gives the surface stresses and the energy-condition combinations. The static limits are displayed explicitly because they provide the first physical diagnostic of the shell.

The surface stress tensor is written as
\begin{equation}
S_{\hat i\hat j}=\mathrm{diag}(\sig,p,p),
\label{eq:surface_tensor}
\end{equation}
where $\sig$ is the surface energy density and $p$ is the tangential surface pressure. The Lanczos equation is
\begin{equation}
-[K_{\hat i\hat j}]+[K]\eta_{\hat i\hat j}=8\pi S_{\hat i\hat j},
\qquad
[K_{\hat i\hat j}]=K^+_{\hat i\hat j}-K^-_{\hat i\hat j}.
\label{eq:lanczos_equation}
\end{equation}
For the symmetric construction, the dynamic density and pressure are
\begin{equation}
\sig(a)=-\frac{1}{2\pi a}\sqrt{\FS(a)+\dot a^2}=-\frac{1}{2\pi a}\sqrt{1-\frac{4\pi M}{\mathcal S'(a)}+\dot a^2},
\label{eq:sigma_dynamic_entropy}
\end{equation}
\begin{equation}
p(a)=\frac{1}{8\pi}\left[\frac{\FS'(a)+2\ddot a}{\sqrt{\FS(a)+\dot a^2}}+\frac{2\sqrt{\FS(a)+\dot a^2}}{a}\right]
=\frac{1}{8\pi}\left[\frac{2\ddot a+4\pi M\mathcal S''(a)/[\mathcal S'(a)]^2}{\sqrt{1-4\pi M/\mathcal S'(a)+\dot a^2}}+\frac{2\sqrt{1-4\pi M/\mathcal S'(a)+\dot a^2}}{a}\right].
\label{eq:pressure_dynamic_entropy}
\end{equation}
For a static throat, these expressions reduce to
\begin{equation}
\sig_0=-\frac{\sqrt{F_{\mathcal S,0}}}{2\pi a_0}=-\frac{1}{2\pi a_0}\sqrt{1-\frac{4\pi M}{\mathcal S'_0}},
\label{eq:sigma_static_entropy}
\end{equation}
\begin{equation}
p_0=\frac{F'_{\mathcal S,0}}{8\pi\sqrt{F_{\mathcal S,0}}}+\frac{\sqrt{F_{\mathcal S,0}}}{4\pi a_0}
=\frac{4\pi M\mathcal S''_0/[\mathcal S'_0]^2}{8\pi\sqrt{1-4\pi M/\mathcal S'_0}}+\frac{\sqrt{1-4\pi M/\mathcal S'_0}}{4\pi a_0}.
\label{eq:pressure_static_entropy}
\end{equation}
Since the shell is located in a positive-lapse region, Eq.~\eqref{eq:sigma_static_entropy} shows that the static density is negative for any finite $a_0>0$. Therefore, the weak energy condition (WEC) on the shell is violated in the symmetric thin-shell construction.

The relevant static combinations are
\begin{equation}
\sig_0+p_0=\frac{a_0F'_{\mathcal S,0}-2F_{\mathcal S,0}}{8\pi a_0\sqrt{F_{\mathcal S,0}}}
=\frac{a_0\,4\pi M\mathcal S''_0/[\mathcal S'_0]^2-2\left(1-4\pi M/\mathcal S'_0\right)}{8\pi a_0\sqrt{1-4\pi M/\mathcal S'_0}},
\label{eq:nec_static_entropy}
\end{equation}
\begin{equation}
\sig_0+2p_0=\frac{F'_{\mathcal S,0}}{4\pi\sqrt{F_{\mathcal S,0}}}
=\frac{4\pi M\mathcal S''_0/[\mathcal S'_0]^2}{4\pi\sqrt{1-4\pi M/\mathcal S'_0}}.
\label{eq:sec_combination_static_entropy}
\end{equation}
The intrinsic shell null energy condition (NEC) is controlled by the sign of $\sig_0+p_0$, which is also one of the two inequalities required by the shell strong energy condition (SEC). The remaining trace SEC inequality is $\sig_0+2p_0\geq0$; hence, the full shell SEC holds only when both combinations are nonnegative.

\section{Conservation equation and radial dynamics}
\label{sec:dynamics}

This section derives the conservation equation and the radial equation of motion. No stability criterion is imposed before specifying the shell equation of state.

For the metric in Eq.~\eqref{eq:entropic_metric}, the conservation equation has no additional flux term. Indeed, within the effective-GR interpretation, $G^t{}_t=G^r{}_r$, and hence $T^t{}_t=T^r{}_r$, or equivalently $p_r=-\rho$; therefore, the normal bulk energy flux through the shell vanishes. Thus,
\begin{equation}
\frac{\dd}{\dd\tau}\left(4\pi a^2\sig\right)+p\frac{\dd}{\dd\tau}\left(4\pi a^2\right)=0.
\label{eq:conservation_tau_simple}
\end{equation}
Equivalently,
\begin{equation}
\dot\sig+2\frac{\dot a}{a}(\sig+p)=0,
\qquad
\sig'(a)=-\frac{2}{a}\left[\sig(a)+p(a)\right].
\label{eq:conservation_radial_simple}
\end{equation}
This relation can also be written as
\begin{equation}
p(a)=-\sig(a)-\frac{a}{2}\sig'(a).
\label{eq:pressure_from_conservation}
\end{equation}
The equation becomes a closed differential equation for $\sig(a)$ only after an equation of state is chosen.

From Eq.~\eqref{eq:sigma_dynamic_entropy}, the radial velocity is
\begin{equation}
\dot a^2=4\pi^2a^2\sig^2(a)-\FS(a)=4\pi^2a^2\sig^2(a)-1+\frac{4\pi M}{\mathcal S'(a)}.
\label{eq:adot_general_entropy}
\end{equation}
Thus the shell motion can be written as
\begin{equation}
\dot a^2+\Veff(a)=0,
\qquad
\Veff(a)=\FS(a)-4\pi^2a^2\sig^2(a)=1-\frac{4\pi M}{\mathcal S'(a)}-4\pi^2a^2\sig^2(a).
\label{eq:potential_general_entropy}
\end{equation}
This is the basic dynamical equation. The function $\sig(a)$ must be obtained from the conservation equation and from the chosen surface equation of state.

\subsection{Barotropic shell model}
\label{sec:barotropic_shell}

We first consider a constant barotropic equation of state,
\begin{equation}
p(a)=w\sig(a),
\label{eq:barotropic_eos}
\end{equation}
where $w$ is constant along a shell trajectory. Substitution into the radial conservation equation, Eq.~\eqref{eq:conservation_radial_simple}, gives
\begin{equation}
\sig(a)=\sig_0\left(\frac{a_0}{a}\right)^{2(1+w)}.
\label{eq:sigma_barotropic_entropy}
\end{equation}
Using the static junction value $\sig_0^2=F_{\mathcal S,0}/(4\pi^2a_0^2)$, the effective potential becomes
\begin{equation}
\Veff(a)=\FS(a)-\FS(a_0)\left(\frac{a_0}{a}\right)^{2+4w}
=1-\frac{4\pi M}{\mathcal S'(a)}
-\left(1-\frac{4\pi M}{\mathcal S'_0}\right)
\left(\frac{a_0}{a}\right)^{2+4w}.
\label{eq:potential_barotropic_entropy}
\end{equation}
The condition $\Veff'(a_0)=0$, or equivalently $w_0=p_0/\sig_0$, fixes the constant barotropic parameter as
\begin{equation}
w_0=-\frac{1}{2}-\frac{a_0F'_{\mathcal S,0}}{4F_{\mathcal S,0}}
=-\frac{1}{2}
-\frac{a_0\pi M\mathcal S''_0}
{[\mathcal S'_0]^2\left(1-4\pi M/\mathcal S'_0\right)}.
\label{eq:w_static_entropy}
\end{equation}
After imposing this equilibrium condition, the potential curvature is
\begin{equation}
\Veff''(a_0)
=
F''_{\mathcal S,0}
+\frac{F'_{\mathcal S,0}}{a_0}
-\frac{(F'_{\mathcal S,0})^2}{F_{\mathcal S,0}},
\label{eq:barotropic_stability_metric}
\end{equation}
or, directly in terms of the entropy function,
\begin{equation}
\begin{aligned}
\Veff''(a_0)
={}&
4\pi M\left[
\frac{\mathcal S'''_0}{[\mathcal S'_0]^2}
-\frac{2[\mathcal S''_0]^2}{[\mathcal S'_0]^3}
+\frac{\mathcal S''_0}{a_0[\mathcal S'_0]^2}
\right]-
\frac{16\pi^2M^2[\mathcal S''_0]^2}
{[\mathcal S'_0]^4\left(1-\dfrac{4\pi M}{\mathcal S'_0}\right)}.
\end{aligned}
\label{eq:barotropic_stability_entropy}
\end{equation}
The static configuration is linearly stable whenever $\Veff''(a_0)>0$.

\subsection{Variable Chaplygin shell model}
\label{sec:chaplygin_shell}

As a second matter model, we consider the variable Chaplygin equation of state adopted in Ref.~\cite{Javed2024PolymerTSW},
\begin{equation}
p(a)=\frac{\Omega}{a^n\sig(a)},
\label{eq:chaplygin_eos}
\end{equation}
where $\Omega$ is a constant parameter and $n$ controls the radial dependence of the equation of state.

Substituting Eq.~\eqref{eq:chaplygin_eos} into radial form of the conservation equation \eqref{eq:conservation_radial_simple}, one obtains
\begin{equation}
\frac{\dd}{\dd a}\left[\sig^2(a)\right]+\frac{4}{a}\sig^2(a)
=-\frac{4\Omega}{a^{n+1}}.
\label{eq:chaplygin_sigma_differential}
\end{equation}
For $n\neq4$, the solution satisfying $\sig(a_0)=\sig_0$ is
\begin{equation}
\sig^2(a)
=
\left(\frac{a_0}{a}\right)^4
\left[
\sig_0^2+\frac{4\Omega}{(4-n)a_0^n}
\right]
-\frac{4\Omega}{(4-n)a^n}.
\label{eq:chaplygin_sigma_solution}
\end{equation}
Since the symmetric thin-shell construction has $\sig_0<0$, the physical branch is selected by taking the negative square root of Eq.~\eqref{eq:chaplygin_sigma_solution}. For the special case $n=4$, the solution becomes
\begin{equation}
\sig^2(a)
=
\left(\frac{a_0}{a}\right)^4\sig_0^2
-\frac{4\Omega}{a^4}
\ln\left(\frac{a}{a_0}\right).
\label{eq:chaplygin_sigma_n4}
\end{equation}

Using Eq.~\eqref{eq:potential_general_entropy}, the effective potential for $n\neq4$ can be written as
\begin{equation}
\begin{aligned}
\Veff(a)
&=
\FS(a)
-\FS(a_0)\left(\frac{a_0}{a}\right)^2
+\frac{16\pi^2\Omega}{4-n}
\left[
a^{2-n}
-\frac{a_0^{4-n}}{a^2}
\right]
\\
&=
1-\frac{4\pi M}{\mathcal S'(a)}
-\left(1-\frac{4\pi M}{\mathcal S'_0}\right)
\left(\frac{a_0}{a}\right)^2
+\frac{16\pi^2\Omega}{4-n}
\left[
a^{2-n}
-\frac{a_0^{4-n}}{a^2}
\right].
\end{aligned}
\label{eq:chaplygin_potential}
\end{equation}
For $n=4$, the corresponding potential is
\begin{equation}
\Veff^{(n=4)}(a)
=
1-\frac{4\pi M}{\mathcal S'(a)}
-\left(1-\frac{4\pi M}{\mathcal S'_0}\right)
\left(\frac{a_0}{a}\right)^2
+\frac{16\pi^2\Omega}{a^2}
\ln\left(\frac{a}{a_0}\right).
\label{eq:chaplygin_potential_n4}
\end{equation}

The condition $\Veff(a_0)=0$ is automatically satisfied by construction. The remaining equilibrium condition, $\Veff'(a_0)=0$, gives
\begin{equation}
\Omega_0
=
a_0^n\sig_0p_0
=
-\frac{a_0^{n-2}}{16\pi^2}
\left[
a_0F'_{\mathcal S,0}
+2F_{\mathcal S,0}
\right].
\label{eq:chaplygin_omega_static}
\end{equation}
Equivalently, in terms of the entropy derivatives,
\begin{equation}
\Omega_0
=
-\frac{a_0^{n-2}}{16\pi^2}
\left[
\frac{4\pi M a_0\mathcal S''_0}{[\mathcal S'_0]^2}
+
2\left(1-\frac{4\pi M}{\mathcal S'_0}\right)
\right].
\label{eq:chaplygin_omega_entropy}
\end{equation}
Thus, for each static radius and entropy model, the Chaplygin parameter $\Omega$ is fixed by the static junction stresses, whereas $n$ remains as the parameter controlling the radial variation of the equation of state.

After imposing Eq.~\eqref{eq:chaplygin_omega_static}, the second derivative of the effective potential takes the compact form
\begin{equation}
\Veff''(a_0)
=
F''_{\mathcal S,0}
+\frac{n+1}{a_0}F'_{\mathcal S,0}
+\frac{2(n-2)}{a_0^2}F_{\mathcal S,0}.
\label{eq:chaplygin_stability_generic}
\end{equation}
Using the entropy representation of the lapse derivatives, this becomes
\begin{equation}
\begin{aligned}
\Veff''(a_0)
={}&
4\pi M
\left[
\frac{\mathcal S'''_0}{[\mathcal S'_0]^2}
-\frac{2[\mathcal S''_0]^2}{[\mathcal S'_0]^3}
+\frac{(n+1)\mathcal S''_0}{a_0[\mathcal S'_0]^2}
\right]+
\frac{2(n-2)}{a_0^2}
\left(
1-\frac{4\pi M}{\mathcal S'_0}
\right).
\end{aligned}
\label{eq:chaplygin_stability_entropy}
\end{equation}
Equation~\eqref{eq:chaplygin_stability_entropy} is valid for all values of $n$, including $n=4$, despite the logarithmic form of the corresponding potential in Eq.~\eqref{eq:chaplygin_potential_n4}. The static thin-shell configuration is linearly stable whenever $\Veff''(a_0)>0$.

\section{Entropy models: surface conditions and stability}
\label{sec:entropy_models}

This section applies each entropy function to the generic thin-shell equations and presents the corresponding static surface quantities, together with the barotropic and variable Chaplygin effective potentials and their associated stability conditions, explicitly in terms of the entropy parameters.

\subsection{Schwarzschild thin-shell wormhole}
\label{sec:schwarzschild_tswh}

The Bekenstein--Hawking entropy is given by
\begin{equation}
\mathcal S_{\rm BH}(r)=\pi r^2.
\label{eq:schwarzschild_entropy}
\end{equation}
Substituting this entropy into Eq.~\eqref{eq:entropic_metric}, we recover the Schwarzschild lapse function,
\begin{equation}
F_{\rm Schw}(r)=1-\frac{2M}{r}.
\label{eq:schwarzschild_lapse_from_entropy}
\end{equation}
The horizon is located at $r_h=2M$, and the static throat must satisfy $a_0>2M$. The static surface density and pressure are
\begin{equation}\label{eq:schwarzschild_static_stresses}
\sigma^{\rm Schw}_0=-\frac{1}{2\pi a_0}\sqrt{1-\frac{2M}{a_0}}
\end{equation}
and
\begin{equation}
p^{\rm Schw}_0
=\frac{a_0-M}{4\pi a_0^2\sqrt{1-2M/a_0}}.
\end{equation}
Thus, the density is negative for every admissible radius, while the tangential pressure is positive. Close to the horizon, the density tends to zero from below, but the pressure diverges. Therefore, moving the shell toward $2M$ reduces the magnitude of the negative surface energy at the price of an increasingly large tangential pressure. Far from the source, the shell approaches the familiar regime in which a negative surface density is balanced by a positive pressure with approximately half its magnitude.

The static intrinsic NEC combination and the remaining trace SEC combination are, respectively,
\begin{equation}\label{eq:schwarzschild_energy_combinations}
\left(\sig_0+p_0\right)_{\rm Schw}=\frac{3M-a_0}{4\pi a_0^2\sqrt{1-2M/a_0}}
\end{equation}
and
\begin{equation}
\left(\sig_0+2p_0\right)_{\rm Schw}=\frac{M}{2\pi a_0^2\sqrt{1-2M/a_0}}.
\end{equation}
The null combination changes sign at $a_0=3M$, exactly at the photon-sphere radius of the Schwarzschild geometry. Since $\sig_0<0$ for every admissible throat radius, the shell is exotic throughout the Schwarzschild branch, independently of the sign of $\sig_0+p_0$. The change of sign of $\sig_0+p_0$ at $a_0=3M$ concerns only the null energy condition evaluated along null directions tangent to the shell. Hence, in the interval $2M<a_0<3M$, the intrinsic shell NEC is satisfied, whereas the weak energy condition remains violated and the shell still carries negative surface energy. For $a_0>3M$, the shell remains exotic because $\sig_0<0$, while the additional negativity of $\sig_0+p_0$ signals a violation of the intrinsic null energy condition on the shell. Thus, the region $a_0>3M$ exhibits both negative surface energy density and intrinsic NEC violation. Although the remaining trace SEC combination $\sig_0+2p_0$ is positive throughout the Schwarzschild branch, the full shell SEC is satisfied only for $2M<a_0\leq3M$, where the intrinsic NEC combination is also nonnegative.

\begin{figure}[htp!]
    \centering
    \includegraphics[width=0.49\linewidth]{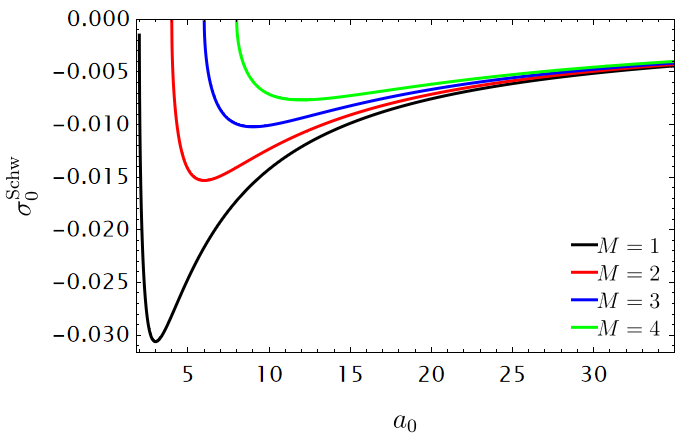}
    \includegraphics[width=0.49\linewidth]{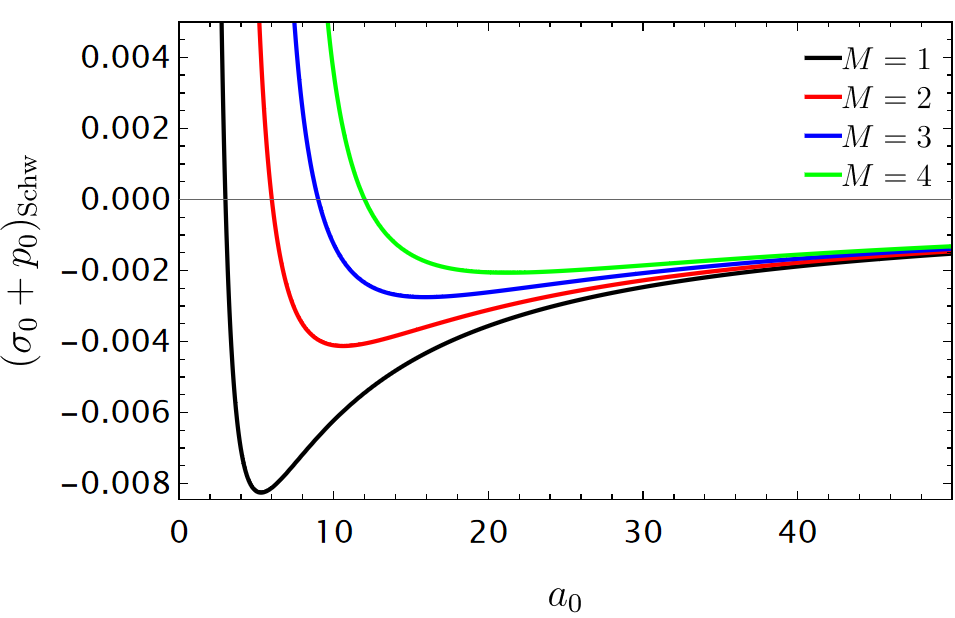}
    \includegraphics[width=0.5\linewidth]{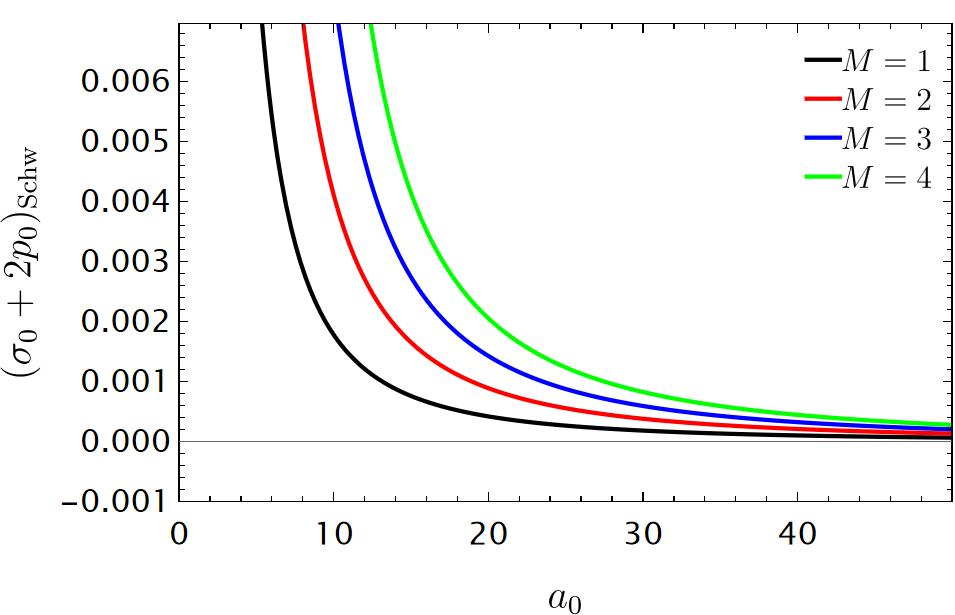}
    \caption{Static surface energy conditions for the Schwarzschild thin-shell wormhole as functions of the throat radius $a_0$. The upper-left panel shows the surface energy density $\sigma_0$, the upper-right panel displays the intrinsic shell NEC combination $\sigma_0+p_0$, and the lower panel shows the remaining trace SEC combination $\sigma_0+2p_0$. The curves correspond to $M=1$, $2$, $3$, and $4$, and are shown only in their respective admissible domains, $a_0>2M$.}
    \label{fig:ecsch}
\end{figure}

Figure~\ref{fig:ecsch} makes explicit that the Schwarzschild mass changes the radial scale of the shell observables without altering their qualitative pattern. For each value of $M$, the admissible branch begins at $a_0=2M$, the negative surface density reaches a finite minimum outside the horizon and then approaches zero at large radius, while the remaining trace SEC combination remains positive and monotonically decreases away from the near-horizon region. The plot also shows that the zero of the intrinsic NEC combination is displaced according to $a_0=3M$, so that the transition between the two shell-NEC regimes scales linearly with the black-hole mass. Thus, the curves differ mainly by their characteristic length and amplitude scales, whereas their sign structure is universal when expressed in terms of the dimensionless ratio $a_0/M$.

For the barotropic shell model, substituting the Schwarzschild lapse function into Eqs.~\eqref{eq:potential_barotropic_entropy} and \eqref{eq:w_static_entropy} gives
\begin{equation}
w_{{\rm Schw},0}
=
-\frac{a_0-M}{2(a_0-2M)},
\label{eq:schwarzschild_w_barotropic}
\end{equation}
and
\begin{equation}
V_{{\rm eff},\rm Schw}(a)
=
1-\frac{2M}{a}
-\left(1-\frac{2M}{a_0}\right)
\left(\frac{a_0}{a}\right)^{2+4w}.
\label{eq:schwarzschild_potential_barotropic}
\end{equation}
The equilibrium value $w=w_{{\rm Schw},0}$ ensures that $V_{{\rm eff},\rm Schw}(a_0)=V_{{\rm eff},\rm Schw}'(a_0)=0$. The corresponding potential curvature is
\begin{equation}
V_{{\rm eff},\rm Schw}''(a_0)
=
-\frac{2M}{a_0^2(a_0-2M)}.
\label{eq:schwarzschild_barotropic_stability}
\end{equation}
Since $a_0>2M$, one has $V_{{\rm eff},\rm Schw}''(a_0)<0$ throughout the admissible domain. Therefore, the Schwarzschild thin-shell wormhole supported by a constant barotropic shell is linearly unstable under radial perturbations. This branch provides the reference stability limit against which the entropy-induced deformations are compared.

\begin{figure}[htp!]
    \centering
    \includegraphics[width=0.6\linewidth]{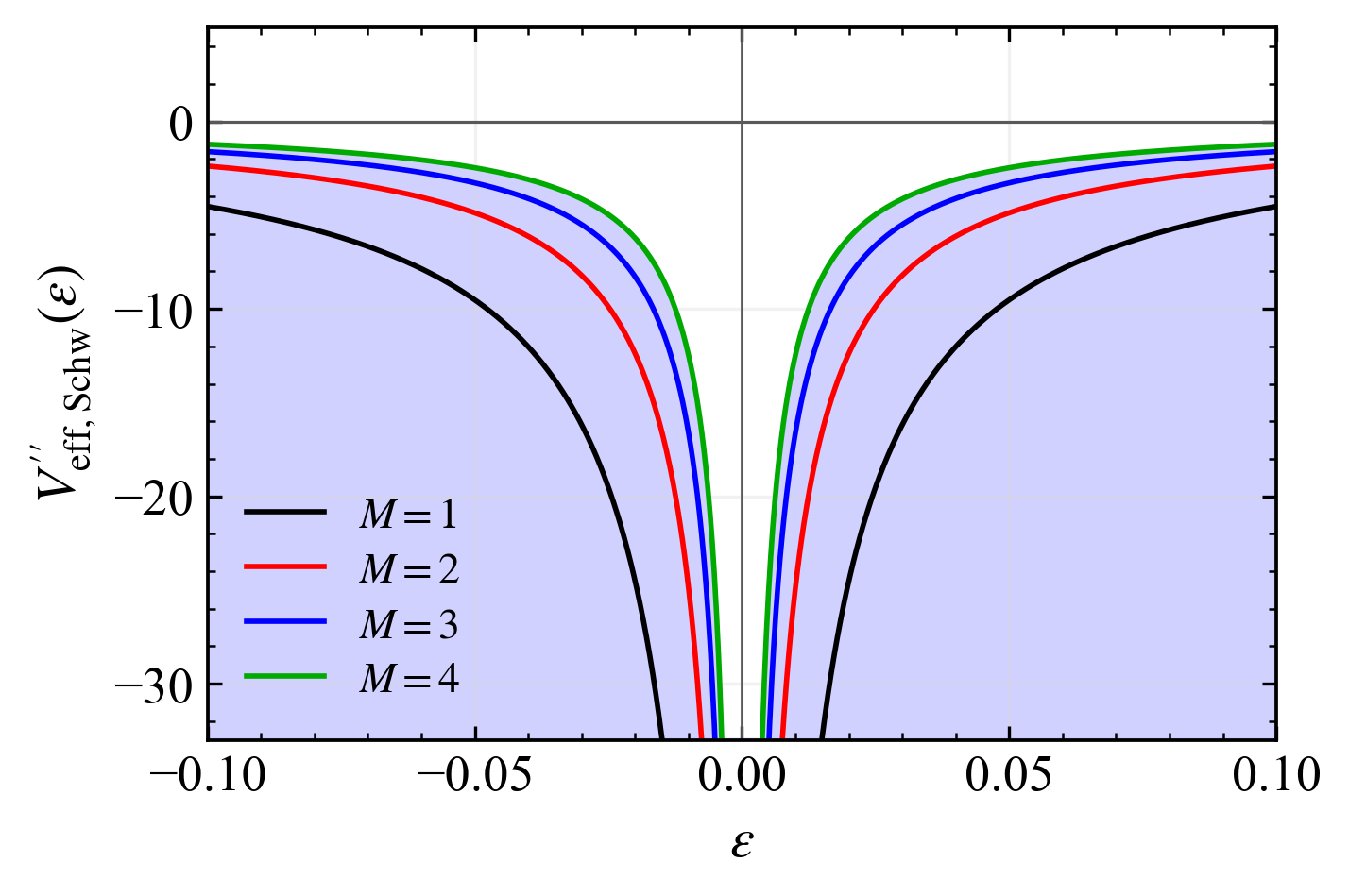}
    \caption{Effective-potential curvature $V_{{\rm eff},\rm Schw}''(\mathcal E)$ for the barotropic Schwarzschild thin-shell wormhole, with $a_0=2M+|\mathcal{E}|$. The curves correspond to $M=1$, $2$, $3$, and $4$, and remain negative throughout the physical domain, indicating linear instability.}
    \label{fig:stbsch}
\end{figure}

Figure~\ref{fig:stbsch} displays the mass dependence of the barotropic stability curvature in terms of the displacement parameter $\mathcal{E}$. Because the throat is parametrized as $a_0=2M+|\mathcal{E}|$, the two sides of the plot are not distinct physical branches: they represent the same exterior configurations at equal distances from the horizon. The divergence to negative values as $\mathcal{E}\to0$ reflects the increasing sensitivity of the shell dynamics when the throat approaches the Schwarzschild horizon. For larger $|\mathcal{E}|$, the curvature approaches zero from below, showing that the instability becomes weaker far from the horizon but is never removed. Increasing $M$ changes the magnitude and radial scale of the curves, without modifying this qualitative behavior.

For the variable Chaplygin shell model, the Schwarzschild lapse function gives the equilibrium parameter
\begin{equation}
\Omega_{{\rm Schw},0}
=
-\frac{a_0^{n-3}(a_0-M)}{8\pi^2}.
\label{eq:schwarzschild_omega_chaplygin}
\end{equation}
For $n\neq4$, substituting this result into Eq.~\eqref{eq:chaplygin_potential} yields
\begin{equation}
\begin{aligned}
V_{{\rm eff},\rm Schw}^{(C)}(a)
={}&
1-\frac{2M}{a}
-\left(1-\frac{2M}{a_0}\right)
\left(\frac{a_0}{a}\right)^2-
\frac{2a_0^{n-3}(a_0-M)}{4-n}
\left[
a^{2-n}-\frac{a_0^{4-n}}{a^2}
\right].
\end{aligned}
\label{eq:schwarzschild_potential_chaplygin}
\end{equation}
For the special case $n=4$, which must be treated separately, the potential becomes
\begin{equation}
V_{{\rm eff},\rm Schw}^{(C)}(a)
=
1-\frac{2M}{a}
-\left(1-\frac{2M}{a_0}\right)
\left(\frac{a_0}{a}\right)^2
-\frac{2a_0(a_0-M)}{a^2}
\ln\left(\frac{a}{a_0}\right).
\label{eq:schwarzschild_potential_chaplygin_n4}
\end{equation}
The value in Eq.~\eqref{eq:schwarzschild_omega_chaplygin} ensures that $V_{{\rm eff},\rm Schw}^{(C)}(a_0)=V_{{\rm eff},\rm Schw}^{{'}\,\rm (C)}(a_0)=0$. The potential curvature is
\begin{equation}
V_{{\rm eff},\rm Schw}^{{''}\,\rm (C)}(a_0)
=
\frac{2}{a_0^3}
\left[
(n-2)a_0-(n-3)M
\right].
\label{eq:schwarzschild_chaplygin_stability}
\end{equation}
Therefore, the Schwarzschild thin-shell wormhole is stable whenever
\begin{equation}
(n-2)a_0-(n-3)M>0.
\label{eq:schwarzschild_chaplygin_stability_condition}
\end{equation}
In particular, for $n=4$,
\begin{equation}
V_{{\rm eff},\rm Schw}^{{''}\,\rm (C)}(a_0)
=
\frac{2(2a_0-M)}{a_0^3}>0,
\qquad
a_0>2M.
\label{eq:schwarzschild_chaplygin_stability_n4}
\end{equation}
Hence, the variable Chaplygin shell with $n=4$ provides a linearly stable Schwarzschild thin-shell branch throughout the admissible domain.

\begin{figure}[htp!]
    \centering
    \includegraphics[width=1\linewidth]{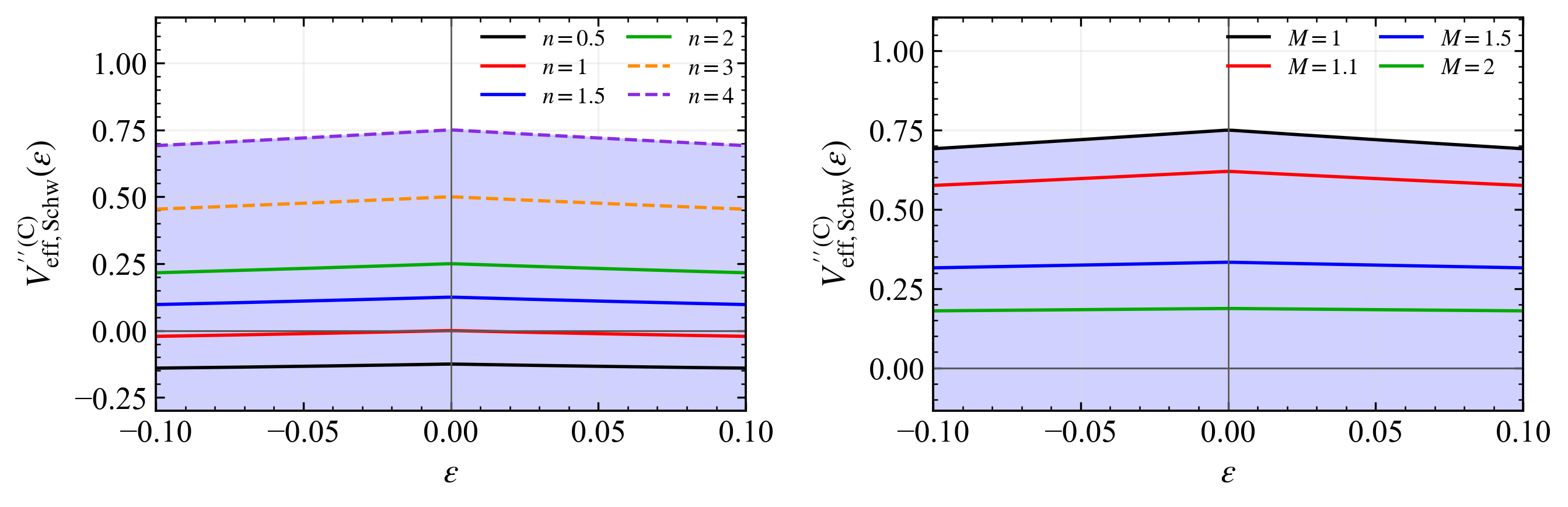}
    \caption{Effective-potential curvature $V_{{\rm eff},\rm Schw}^{{''}\,\rm (C)}(\mathcal E)$ for the variable Chaplygin Schwarzschild thin-shell wormhole, with $a_0=2M+|\mathcal{E}|$. In the left panel, $M=1$ and the curves correspond to $n=0.5$, $1$, $1.5$, $2$, $3$, and $4$. In the right panel, $n=4$ and the curves correspond to $M=1$, $1.1$, $1.5$, and $2$. The Chaplygin parameter $\Omega$ is fixed by the static equilibrium condition in each case. Positive regions of the curvature correspond to linear radial stability.}
    \label{fig:stbchsch}
\end{figure}

Figure~\ref{fig:stbchsch} highlights the qualitative change introduced by the variable Chaplygin shell relative to the barotropic case. Whereas the latter is unstable throughout the Schwarzschild exterior, the Chaplygin curvature can become positive, with the stability character controlled primarily by the exponent $n$. For fixed $M=1$, the curves with $n\leq1$ remain negative in the physical domain, while the cases $n\geq2$ are positive. The intermediate interval $1<n<2$ is especially informative: it corresponds to a near-horizon stable region followed by a transition to instability at larger throat radii. Indeed, from Eq.~\eqref{eq:schwarzschild_chaplygin_stability}, the transition occurs when
\begin{equation}
n=\frac{2a_0-3M}{a_0-M},
\end{equation}
whose value increases from $1$ close to the horizon to $2$ in the large-radius limit. Thus, for example, the positive $n=1.5$ curve in the left panel represents a stable near-horizon branch; its eventual transition occurs at $a_0=3M$, outside the narrow range of $\mathcal E$ displayed in the figure. The right panel shows that, once $n=4$ is fixed, changing the mass rescales the magnitude of the curvature without altering its positive sign. The decreasing amplitude for increasing $M$ reflects the weakening of the local potential curvature with the mass scale, while the symmetry with respect to $\mathcal E=0$ follows from the parametrization $a_0=2M+|\mathcal E|$ rather than from the existence of two distinct physical branches.

\subsection{Barrow entropy}
\label{sec:barrow_entropy}

The Barrow entropy is defined by
\begin{equation}
\mathcal S_B(r)=\left(\pi r^2\right)^{1+\Delta/2},
\qquad
0\leq\Delta\leq1.
\label{eq:barrow_entropy}
\end{equation}
For convenience, we define
\begin{equation}
\mathcal A_B\equiv\frac{4M}{(\Delta+2)\pi^{\Delta/2}}.
\label{eq:barrow_A}
\end{equation}
Substituting Eq.~\eqref{eq:barrow_entropy} into Eq.~\eqref{eq:entropic_metric}, the corresponding lapse function is
\begin{equation}
F_B(r)=1-\frac{\mathcal A_B}{r^{\Delta+1}}.
\label{eq:barrow_entropy_lapse}
\end{equation}
Its derivatives are
\begin{equation}
F'_B(r)=\frac{(\Delta+1)\mathcal A_B}{r^{\Delta+2}},
\qquad
F''_B(r)=-\frac{(\Delta+1)(\Delta+2)\mathcal A_B}{r^{\Delta+3}}.
\label{eq:barrow_lapse_derivatives}
\end{equation}
The horizon is located at
\begin{equation}
r_h^{(B)}=\mathcal A_B^{1/(\Delta+1)}
=
\left[\frac{4M}{(\Delta+2)\pi^{\Delta/2}}\right]^{\frac{1}{\Delta+1}},
\label{eq:barrow_horizon}
\end{equation}
and the static throat must satisfy $a_0>r_h^{(B)}$.

The static surface density and pressure are
\begin{equation}
\sigma_{B0}
=
-\frac{1}{2\pi a_0}
\sqrt{1-\frac{\mathcal A_B}{a_0^{\Delta+1}}},
\label{eq:barrow_static_density}
\end{equation}
and
\begin{equation}
p_{B0}
=
\frac{(\Delta+1)\mathcal A_B}
{8\pi a_0^{\Delta+2}
\sqrt{1-\dfrac{\mathcal A_B}{a_0^{\Delta+1}}}}
+
\frac{1}{4\pi a_0}
\sqrt{1-\frac{\mathcal A_B}{a_0^{\Delta+1}}}.
\label{eq:barrow_static_pressure}
\end{equation}
Thus, as in the Schwarzschild case, the shell density is negative throughout the positive-lapse region, whereas the tangential pressure is positive. At the horizon limit, $\sigma_{B0}$ tends to zero from below, while $p_{B0}$ diverges because $F_B(a_0)$ vanishes. Far from the throat, the Barrow deformation becomes subdominant and the surface stresses approach their usual asymptotic thin-shell behavior.

The static intrinsic NEC combination and the remaining trace SEC combination are
\begin{equation}
\left(\sigma_0+p_0\right)_B
=
\frac{\dfrac{(\Delta+3)\mathcal A_B}{a_0^{\Delta+1}}-2}
{8\pi a_0\sqrt{1-\dfrac{\mathcal A_B}{a_0^{\Delta+1}}}
}
\label{eq:barrow_nec_explicit}
\end{equation}
and
\begin{equation}
\left(\sigma_0+2p_0\right)_B
=
\frac{(\Delta+1)\mathcal A_B}
{4\pi a_0^{\Delta+2}
\sqrt{1-\dfrac{\mathcal A_B}{a_0^{\Delta+1}}}}.
\label{eq:barrow_sec_explicit}
\end{equation}
The WEC is violated because $\sigma_{B0}<0$ for every admissible throat radius. The intrinsic shell NEC is positive sufficiently close to the horizon and changes sign when
\begin{equation}
\frac{\mathcal A_B}{a_0^{\Delta+1}}
=
\frac{2}{\Delta+3}.
\label{eq:barrow_nec_transition}
\end{equation}
The remaining trace SEC combination remains positive throughout the physical domain because $F'_B(a_0)>0$.

\begin{figure}[htp!]
    \centering
    \includegraphics[width=0.49\linewidth]{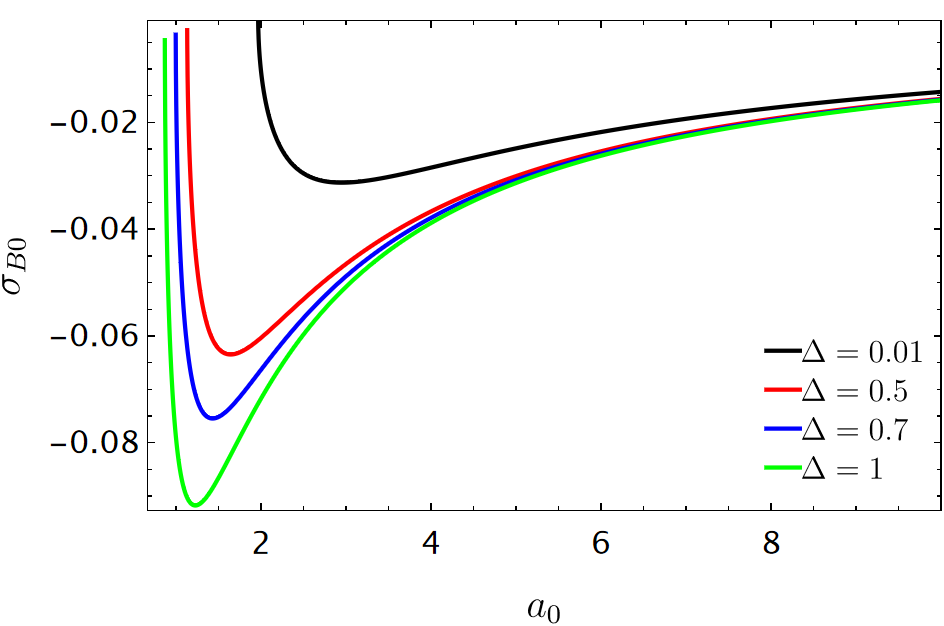}
    \includegraphics[width=0.49\linewidth]{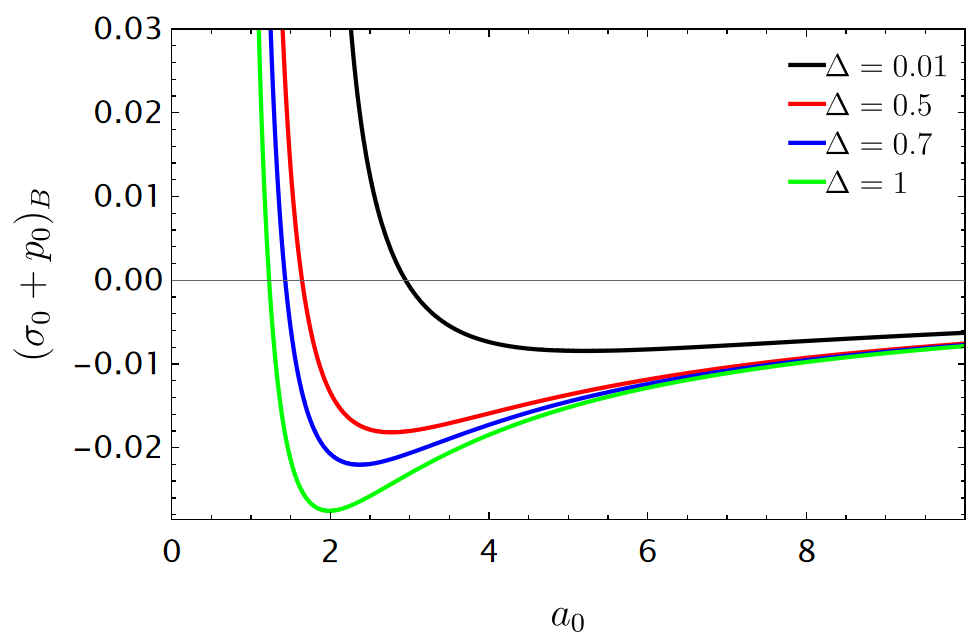}
    \includegraphics[width=0.5\linewidth]{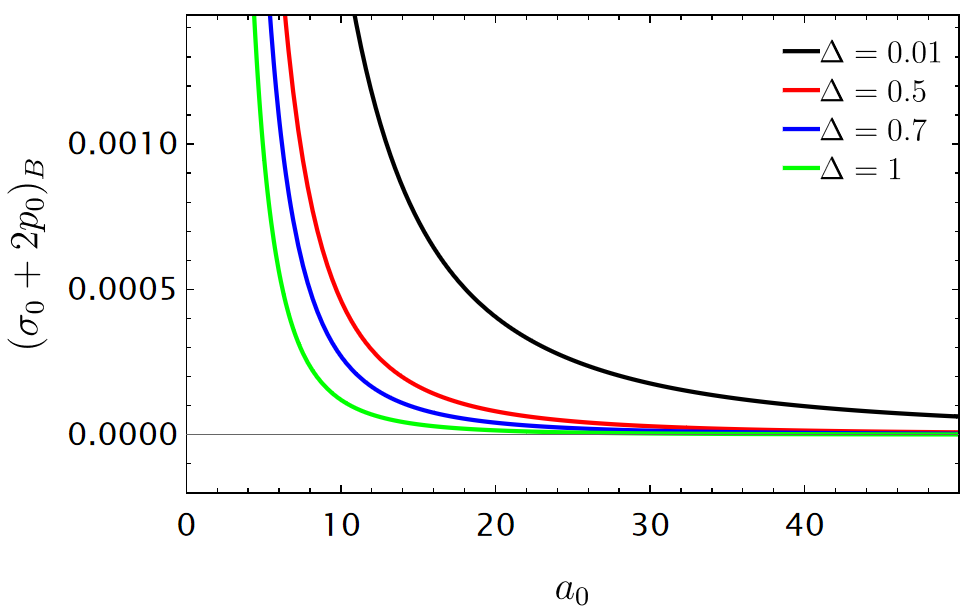}
    \caption{Static surface energy conditions for the Barrow thin-shell wormhole as functions of the throat radius $a_0$. The upper-left panel shows the surface energy density $\sigma_{B0}$, the upper-right panel displays the intrinsic shell NEC combination $\sigma_{B0}+p_{B0}$, and the lower panel shows the remaining trace SEC combination $\sigma_{B0}+2p_{B0}$. The curves correspond to $\Delta=0.01$, $0.5$, $0.7$, and $1$, with $M=1$, and are displayed only in their respective admissible domains, $a_0>r_h^{(B)}(\Delta)$.}
    \label{fig:ecbarrow}
\end{figure}

Figure~\ref{fig:ecbarrow} shows that the Barrow deformation changes both the horizon position and the near-horizon distribution of surface stresses. For every value of $\Delta$, the density starts from zero at the corresponding horizon, becomes more negative at a finite distance from the throat, and then approaches the common asymptotic behavior $\sigma_{B0}\simeq-1/(2\pi a_0)$. The NEC curves are initially positive because of the pressure divergence near the horizon, but cross into a negative regime as the throat is moved outward. The remaining trace SEC combination remains positive and decays rapidly with $a_0$, with larger values of $\Delta$ concentrating the effect of the entropy deformation more strongly near the throat.

For the barotropic shell model, substituting the Barrow lapse function into Eqs.~\eqref{eq:potential_barotropic_entropy} and \eqref{eq:w_static_entropy} gives
\begin{equation}
w_{B0}
=
-\frac{1}{2}
-
\frac{(\Delta+1)\mathcal A_B}
{4a_0^{\Delta+1}
\left(1-\dfrac{\mathcal A_B}{a_0^{\Delta+1}}\right)},
\label{eq:barrow_w0}
\end{equation}
and
\begin{equation}
V_{{\rm eff},B}(a)
=
1-\frac{\mathcal A_B}{a^{\Delta+1}}
-
\left(
1-\frac{\mathcal A_B}{a_0^{\Delta+1}}
\right)
\left(\frac{a_0}{a}\right)^{2+4w}.
\label{eq:barrow_potential}
\end{equation}
The equilibrium value $w=w_{B0}$ ensures that $V_{{\rm eff},B}(a_0)=V_{{\rm eff},B}'(a_0)=0$. The corresponding potential curvature is
\begin{equation}
V_{{\rm eff},B}''(a_0)
=
-\frac{(\Delta+1)^2\mathcal A_B}
{a_0^{\Delta+3}
\left(1-\dfrac{\mathcal A_B}{a_0^{\Delta+1}}\right)}.
\label{eq:barrow_barotropic_stability}
\end{equation}
Since $a_0>r_h^{(B)}$ implies $1-\mathcal A_B/a_0^{\Delta+1}>0$, one has $V_{{\rm eff},B}''(a_0)<0$ throughout the physical domain. Therefore, the Barrow thin-shell wormhole supported by a constant barotropic shell is linearly unstable under radial perturbations.

\begin{figure}[htp!]
    \centering
    \includegraphics[width=0.49\linewidth]{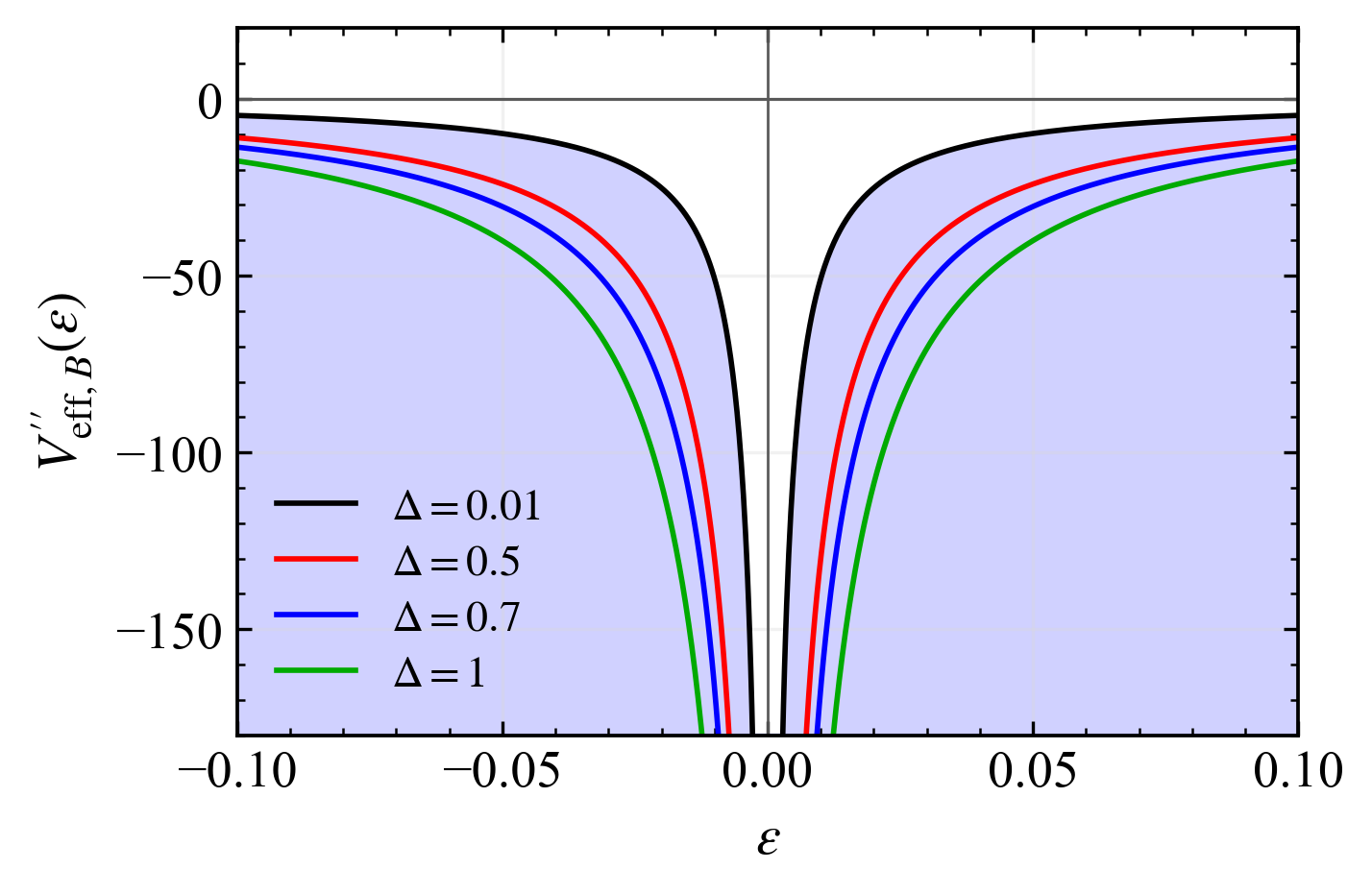}
    \includegraphics[width=0.49\linewidth]{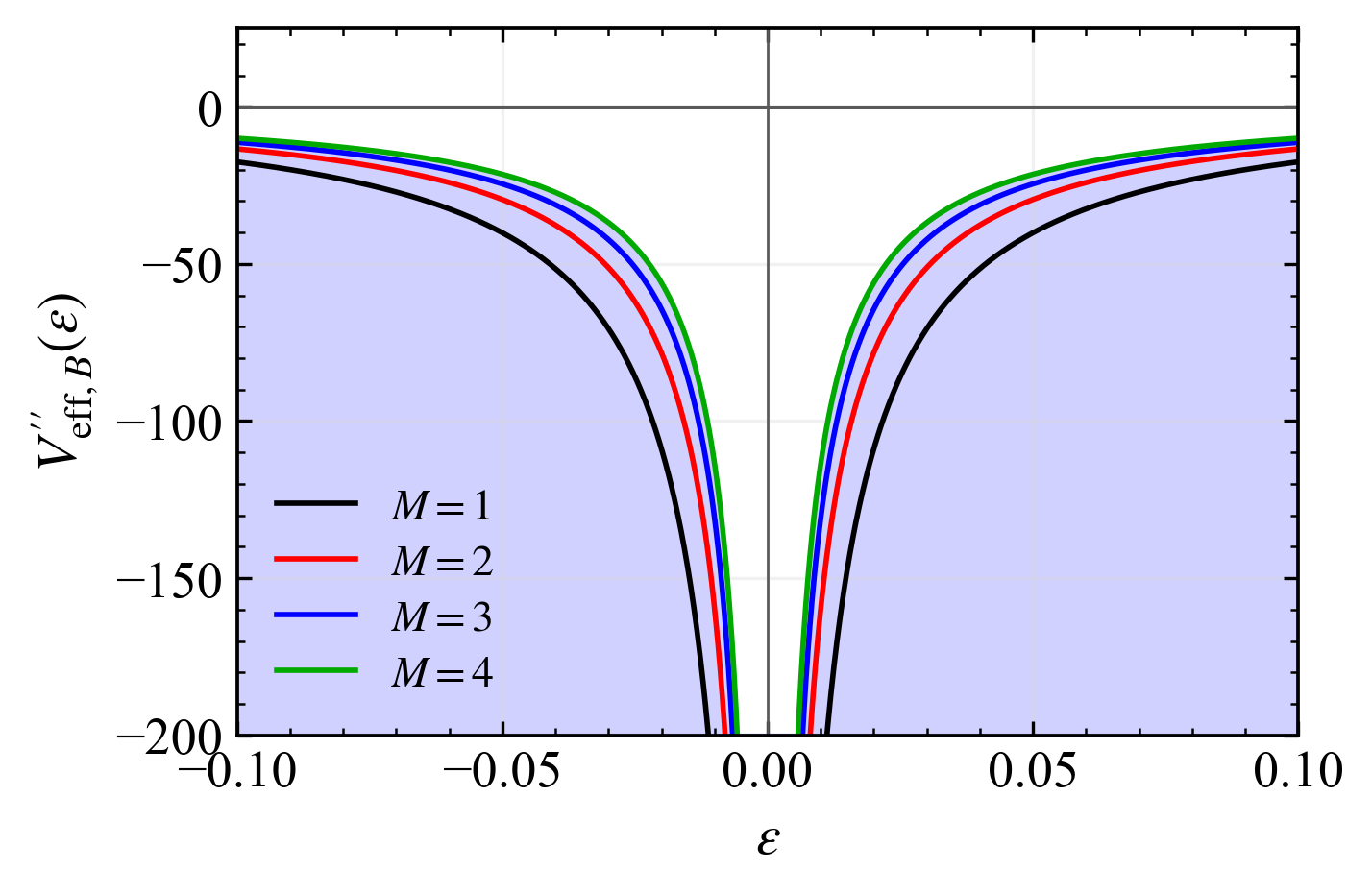}
    \caption{Effective-potential curvature $V_{{\rm eff},B}''(\mathcal E)$ for the barotropic Barrow thin-shell wormhole, with $a_0=r_h^{(B)}+|\mathcal E|$. In the left panel, $M=1$ and the curves correspond to $\Delta=0.01$, $0.5$, $0.7$, and $1$. In the right panel, $\Delta=1$ and the curves correspond to $M=1$, $2$, $3$, and $4$. In both cases, $V_{{\rm eff},B}''(\mathcal E)<0$ throughout the physical domain, indicating linear instability.}
    \label{fig:stbbarrow}
\end{figure}

Figure~\ref{fig:stbbarrow} confirms the analytic result in Eq.~\eqref{eq:barrow_barotropic_stability}. The curvature diverges to negative values as the shell approaches the corresponding Barrow horizon and tends to zero from below when the throat is displaced outward. Varying either $\Delta$ or $M$ changes the magnitude and radial scale of the instability, but neither parameter produces a stable interval within the constant-barotropic model. Thus, the Barrow modification changes the quantitative strength of the Schwarzschild-like instability without altering its sign.

For the variable Chaplygin shell model, the Barrow geometry gives the equilibrium parameter
\begin{equation}
\Omega_{B0}
=
-\frac{a_0^{n-2}}{16\pi^2}
\left[
2+
\frac{(\Delta-1)\mathcal A_B}{a_0^{\Delta+1}}
\right].
\label{eq:barrow_omega_chaplygin}
\end{equation}
For $n\neq4$, substituting Eq.~\eqref{eq:barrow_omega_chaplygin} into Eq.~\eqref{eq:chaplygin_potential} gives
\begin{equation}
\begin{aligned}
V_{{\rm eff},B}^{\rm (C)}(a)
={}&
1-\frac{\mathcal A_B}{a^{\Delta+1}}
-
\left(
1-\frac{\mathcal A_B}{a_0^{\Delta+1}}
\right)
\left(\frac{a_0}{a}\right)^2-
\frac{a_0^{n-2}}{4-n}
\left[
2+
\frac{(\Delta-1)\mathcal A_B}{a_0^{\Delta+1}}
\right]
\left[
a^{2-n}
-
\frac{a_0^{4-n}}{a^2}
\right].
\end{aligned}
\label{eq:barrow_potential_chaplygin}
\end{equation}
For $n=4$, the logarithmic branch must be considered separately:
\begin{equation}
\begin{aligned}
V_{{\rm eff},B}^{\rm (C)}(a)
={}&
1-\frac{\mathcal A_B}{a^{\Delta+1}}
-
\left(
1-\frac{\mathcal A_B}{a_0^{\Delta+1}}
\right)
\left(\frac{a_0}{a}\right)^2-
\frac{a_0^2}{a^2}
\left[
2+
\frac{(\Delta-1)\mathcal A_B}{a_0^{\Delta+1}}
\right]
\ln\left(\frac{a}{a_0}\right).
\end{aligned}
\label{eq:barrow_potential_chaplygin_n4}
\end{equation}
The parameter in Eq.~\eqref{eq:barrow_omega_chaplygin} ensures that
\begin{equation}
V_{{\rm eff},B}^{\rm (C)}(a_0)
=
V_{{\rm eff},B}^{{'}\,\rm (C)}(a_0)
=
0.
\label{eq:barrow_chaplygin_equilibrium}
\end{equation}
The potential curvature is
\begin{equation}
V_{{\rm eff},B}^{{''}\,\rm (C)}(a_0)
=
\frac{1}{a_0^2}
\left[
2(n-2)
+
(\Delta-1)(n-\Delta-3)
\frac{\mathcal A_B}{a_0^{\Delta+1}}
\right].
\label{eq:barrow_chaplygin_stability}
\end{equation}
Therefore, linear radial stability requires
\begin{equation}
2(n-2)
+
(\Delta-1)(n-\Delta-3)
\frac{\mathcal A_B}{a_0^{\Delta+1}}
>0.
\label{eq:barrow_chaplygin_stability_condition}
\end{equation}
In particular, for $n=4$,
\begin{equation}
V_{{\rm eff},B}^{{''}\,\rm (C)}(a_0)
=
\frac{1}{a_0^2}
\left[
4-
(1-\Delta)^2
\frac{\mathcal A_B}{a_0^{\Delta+1}}
\right]
>0.
\label{eq:barrow_chaplygin_stability_n4}
\end{equation}
Hence, the variable Chaplygin shell with $n=4$ is linearly stable throughout the admissible Barrow branch.

\begin{figure}[htp!]
    \centering
    \includegraphics[width=1\linewidth]{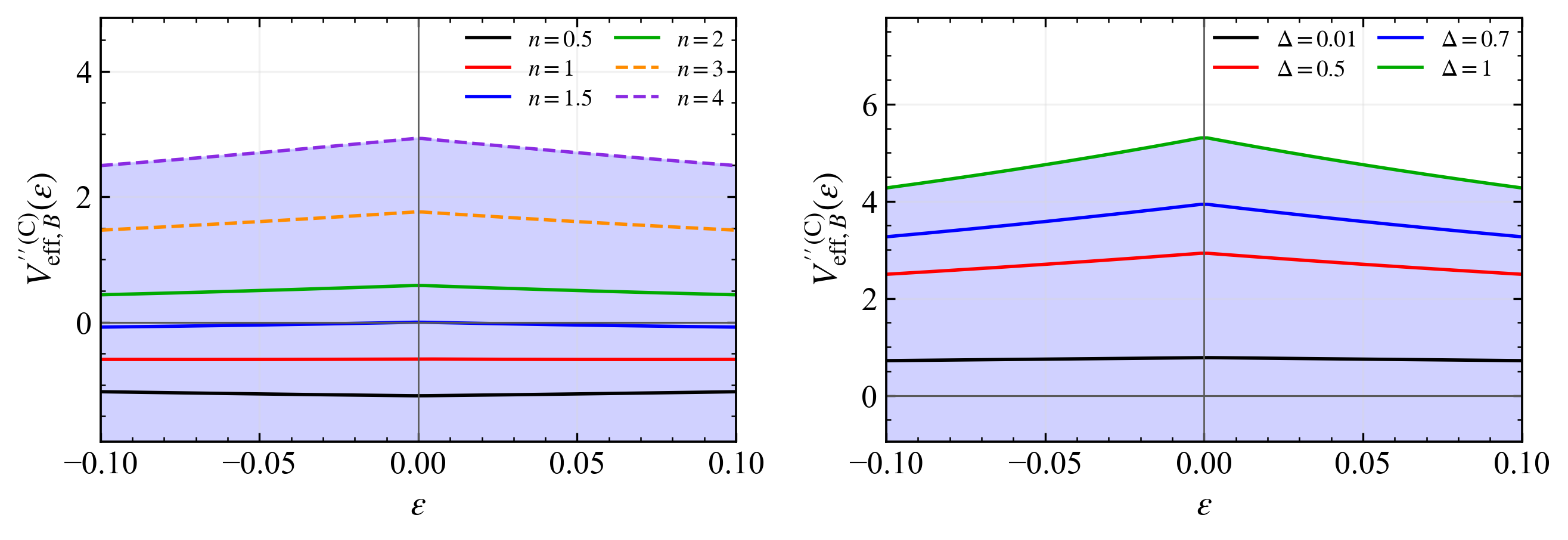}
    \caption{Effective-potential curvature $V_{{\rm eff},B}^{{''}\,\rm (C)}(\mathcal E)$ for the variable Chaplygin Barrow thin-shell wormhole, with $a_0=r_h^{(B)}+|\mathcal E|$. In the left panel, $M=1$ and $\Delta=0.5$, and the curves correspond to $n=0.5$, $1$, $1.5$, $2$, $3$, and $4$. In the right panel, $M=1$ and $n=4$, and the curves correspond to $\Delta=0.01$, $0.5$, $0.7$, and $1$. The Chaplygin parameter $\Omega$ is fixed by the static equilibrium condition in each case. Positive regions of the curvature correspond to linear radial stability.}
    \label{fig:stbchbarrow}
\end{figure}

Figure~\ref{fig:stbchbarrow} shows that the variable Chaplygin shell changes the stability pattern found for the barotropic Barrow branch. For fixed $\Delta=0.5$, the threshold value of $n$ is $\Delta+1=1.5$ in the horizon limit and approaches $n=2$ as the throat moves outward. Consequently, the curves with $n=0.5$ and $n=1$ remain unstable, while the $n=1.5$ branch is marginal at the horizon and becomes unstable immediately outside it. The $n=2$ curve is stable at finite radius and approaches marginality asymptotically, whereas all branches with $n>2$ remain stable throughout the physical domain. More generally, when $\Delta+1<n<2$, a stable near-horizon interval can occur before the curvature eventually changes sign at larger throat radii. The right panel confirms that the stable $n=4$ sector is robust under variation of the Barrow parameter: changing $\Delta$ modifies the magnitude and near-horizon scale of the curvature, but does not generate an instability.

\subsection{Tsallis--Cirto entropy}
\label{sec:tsallis_entropy}

The Tsallis--Cirto entropy is given by
\begin{equation}
\mathcal S_T(r)=\left(\pi r^2\right)^\delta,
\label{eq:tsallis_entropy}
\end{equation}
where the Bekenstein--Hawking limit is recovered for $\delta=1$. Substituting Eq.~\eqref{eq:tsallis_entropy} into Eq.~\eqref{eq:entropic_metric}, we obtain
\begin{equation}
F_T(r)=1-\frac{2M\pi^{1-\delta}}{\delta r^{2\delta-1}}.
\label{eq:tsallis_entropy_lapse}
\end{equation}
For the sector $\delta>1/2$ considered here, the horizon is located at
\begin{equation}
r_h^{(T)}=\left(\frac{2M\pi^{1-\delta}}{\delta}\right)^{\frac{1}{2\delta-1}},
\label{eq:tsallis_horizon}
\end{equation}
and the static throat must satisfy $a_0>r_h^{(T)}$. The static surface density and pressure are
\begin{equation}
\sig_{T0}=-\frac{1}{2\pi a_0}\sqrt{1-\frac{2M\pi^{1-\delta}}{\delta a_0^{2\delta-1}}},
\label{eq:tsallis_static_density}
\end{equation}
and
\begin{equation}
p_{T0}=\frac{M\pi^{1-\delta}(2\delta-1)}{4\pi\delta a_0^{2\delta}\sqrt{1-\dfrac{2M\pi^{1-\delta}}{\delta a_0^{2\delta-1}}}}+\frac{1}{4\pi a_0}\sqrt{1-\frac{2M\pi^{1-\delta}}{\delta a_0^{2\delta-1}}}.
\label{eq:tsallis_static_pressure}
\end{equation}
Thus, the shell density is negative throughout the admissible exterior region, while the tangential pressure is positive. As the throat approaches $r_h^{(T)}$, the density tends to zero from below and the pressure diverges. At large radii, the Tsallis correction decays as $r^{-(2\delta-1)}$, and the shell approaches the usual asymptotic thin-shell regime.

The static intrinsic NEC combination and the remaining trace SEC combination are, respectively,
\begin{equation}
\left(\sig_0+p_0\right)_T=
\frac{\dfrac{2M\pi^{1-\delta}(2\delta+1)}{\delta a_0^{2\delta-1}}-2}
{8\pi a_0\sqrt{1-\dfrac{2M\pi^{1-\delta}}{\delta a_0^{2\delta-1}}}}.
\label{eq:tsallis_nec_explicit}
\end{equation}
and
\begin{equation}
\left(\sig_0+2p_0\right)_T=
\frac{M\pi^{1-\delta}(2\delta-1)}
{2\pi\delta a_0^{2\delta}\sqrt{1-\dfrac{2M\pi^{1-\delta}}{\delta a_0^{2\delta-1}}}}.
\label{eq:tsallis_sec_explicit}
\end{equation}
The WEC is violated because $\sig_{T0}<0$ for every allowed static radius. The intrinsic shell NEC changes sign at a radius determined by the nonextensive parameter $\delta$, whereas the remaining trace SEC combination remains positive in the exterior domain.

\begin{figure}[htp!]
    \centering
    \includegraphics[width=0.49\linewidth]{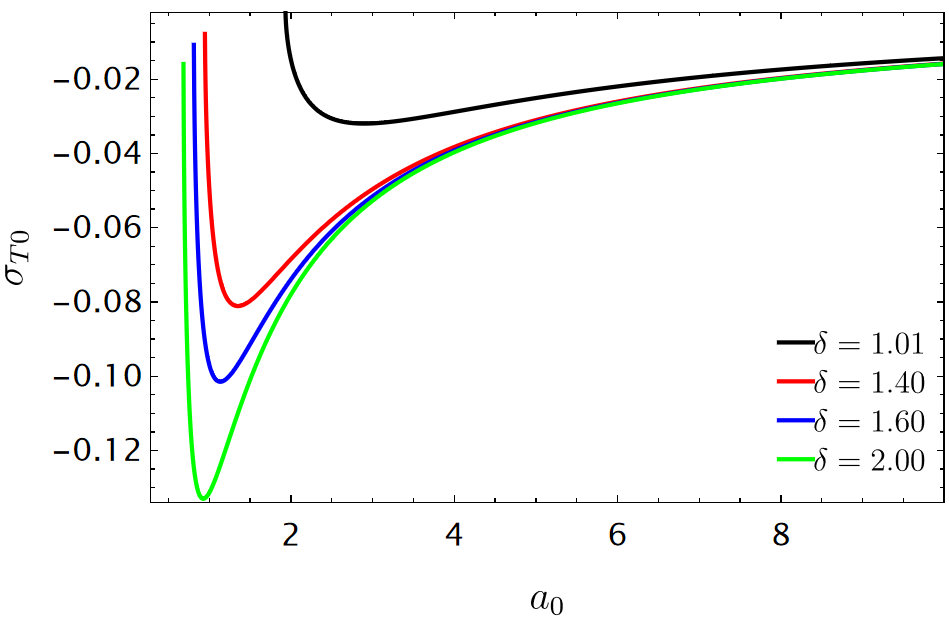}
    \includegraphics[width=0.49\linewidth]{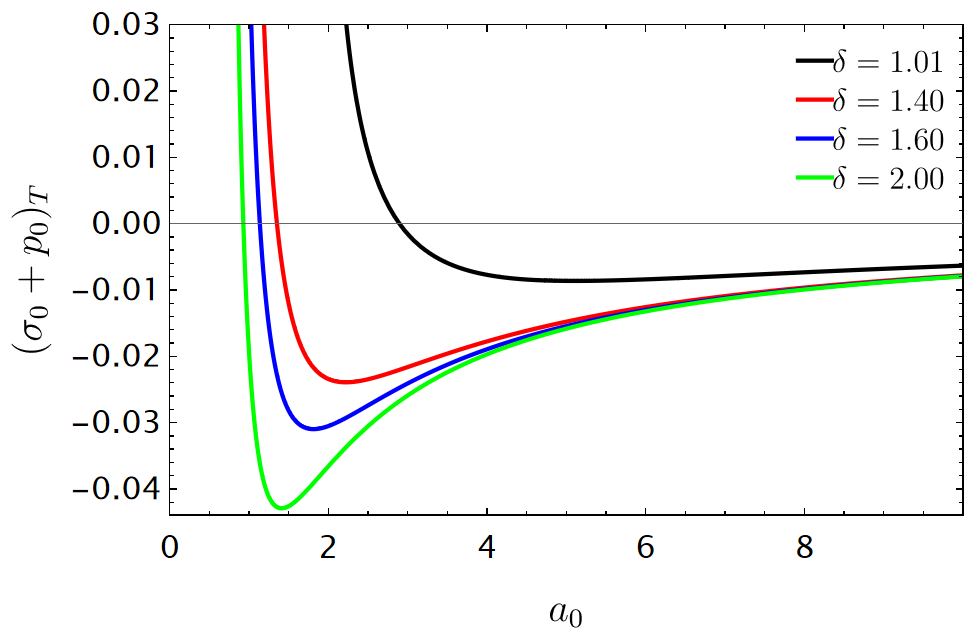}
    \includegraphics[width=0.5\linewidth]{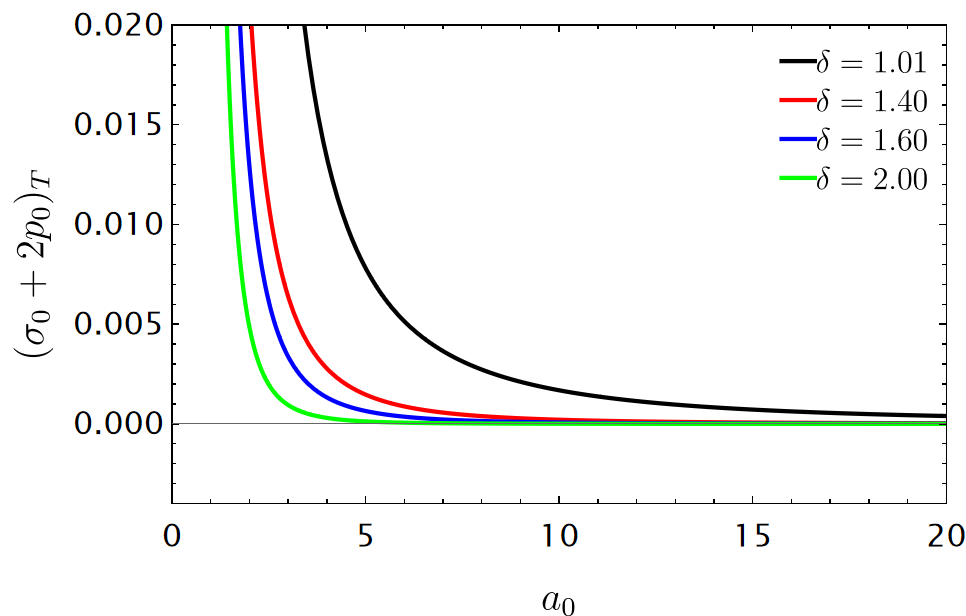}
    \caption{Static surface energy conditions for the Tsallis--Cirto thin-shell wormhole as functions of the throat radius $a_0$. The upper-left panel shows the surface energy density $\sigma_{T0}$, the upper-right panel displays the intrinsic shell NEC combination $\sigma_{T0}+p_{T0}$, and the lower panel shows the remaining trace SEC combination $\sigma_{T0}+2p_{T0}$. The curves correspond to $\delta=1.01$, $1.40$, $1.60$, and $2.00$, with $M=1$, and are shown only in their respective admissible domains, $a_0>r_h^{(T)}(\delta)$.}
    \label{fig:ectssalis}
\end{figure}

Figure~\ref{fig:ectssalis} shows that increasing $\delta$ modifies the near-horizon scale of the shell observables while preserving their common asymptotic behavior. The negative density develops a deeper minimum for larger values of $\delta$, whereas the intrinsic NEC combination remains positive sufficiently close to the horizon and becomes negative as the throat is displaced outward. Over the range shown, the corresponding zero of $\sig_{T0}+p_{T0}$ moves toward smaller radii as $\delta$ increases. The remaining trace SEC combination is positive for all curves and rapidly approaches zero away from the horizon, indicating that the Tsallis deformation mainly affects the near-throat structure of the surface stresses.

For the barotropic shell model, substitution of the Tsallis--Cirto lapse into Eqs.~\eqref{eq:potential_barotropic_entropy} and \eqref{eq:w_static_entropy} gives
\begin{equation}
w_{T0}=-\frac{1}{2}-\frac{M\pi^{1-\delta}(2\delta-1)}{2\delta a_0^{2\delta-1}\left[1-\dfrac{2M\pi^{1-\delta}}{\delta a_0^{2\delta-1}}\right]},
\label{eq:tsallis_w0}
\end{equation}
and
\begin{equation}
V_{{\rm eff},T}(a)=1-\frac{2M\pi^{1-\delta}}{\delta a^{2\delta-1}}-\left(1-\frac{2M\pi^{1-\delta}}{\delta a_0^{2\delta-1}}\right)\left(\frac{a_0}{a}\right)^{2+4w}.
\label{eq:tsallis_potential}
\end{equation}
The equilibrium value $w=w_{T0}$ ensures that $V_{{\rm eff},T}(a_0)=V_{{\rm eff},T}'(a_0)=0$. The potential curvature is
\begin{equation}
V_{{\rm eff},T}''(a_0)=-\frac{2M\pi^{1-\delta}(2\delta-1)^2}{\delta a_0^{2\delta+1}\left[1-\dfrac{2M\pi^{1-\delta}}{\delta a_0^{2\delta-1}}\right]}.
\label{eq:tsallis_barotropic_stability}
\end{equation}
Since $a_0>r_h^{(T)}$, the denominator is positive and therefore $V_{{\rm eff},T}''(a_0)<0$ throughout the physical domain. Hence, the Tsallis--Cirto thin-shell wormhole supported by a constant barotropic shell is linearly unstable under radial perturbations.

\begin{figure}[htp!]
    \centering
    \includegraphics[width=0.49\linewidth]{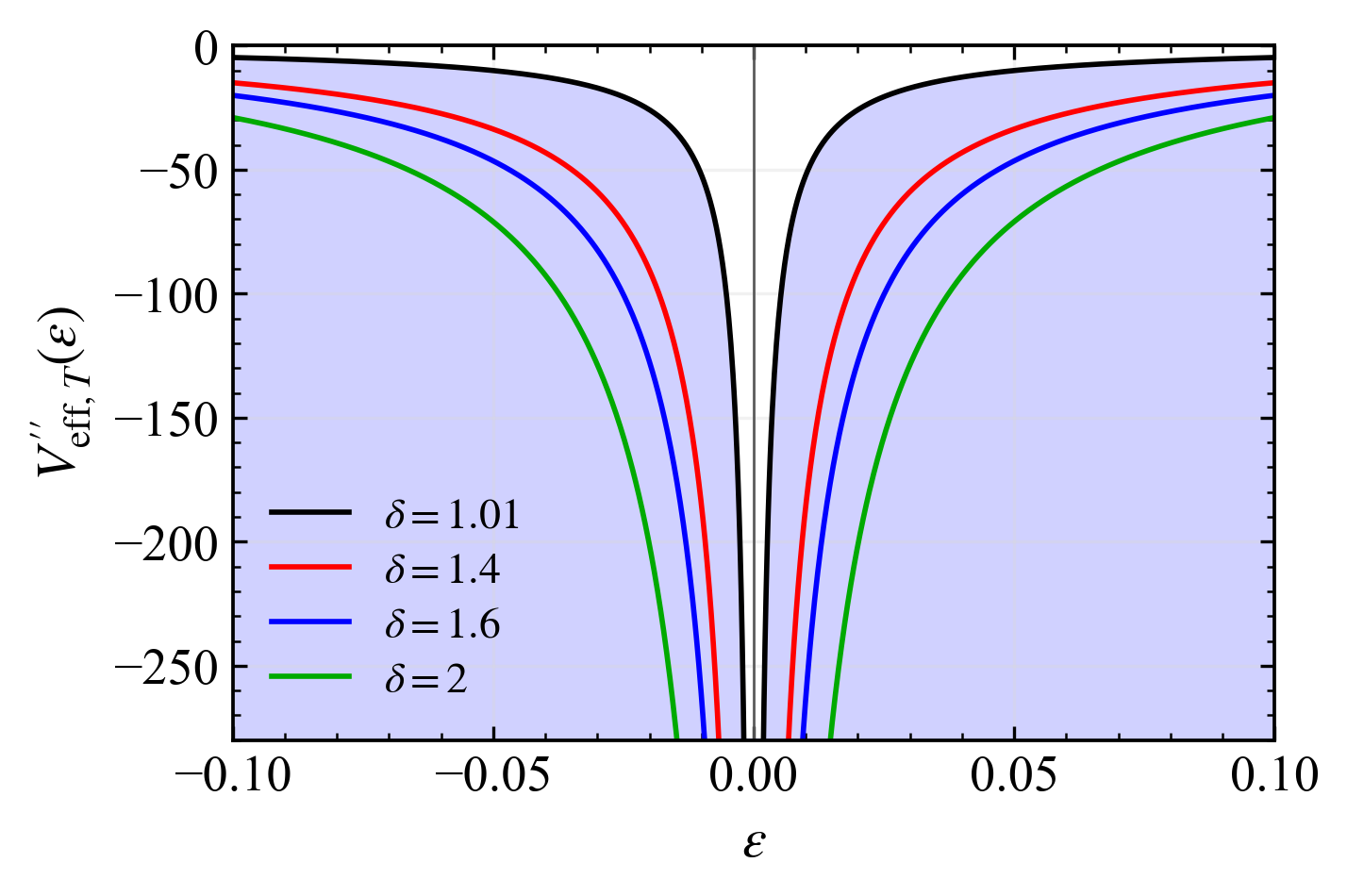}
    \includegraphics[width=0.49\linewidth]{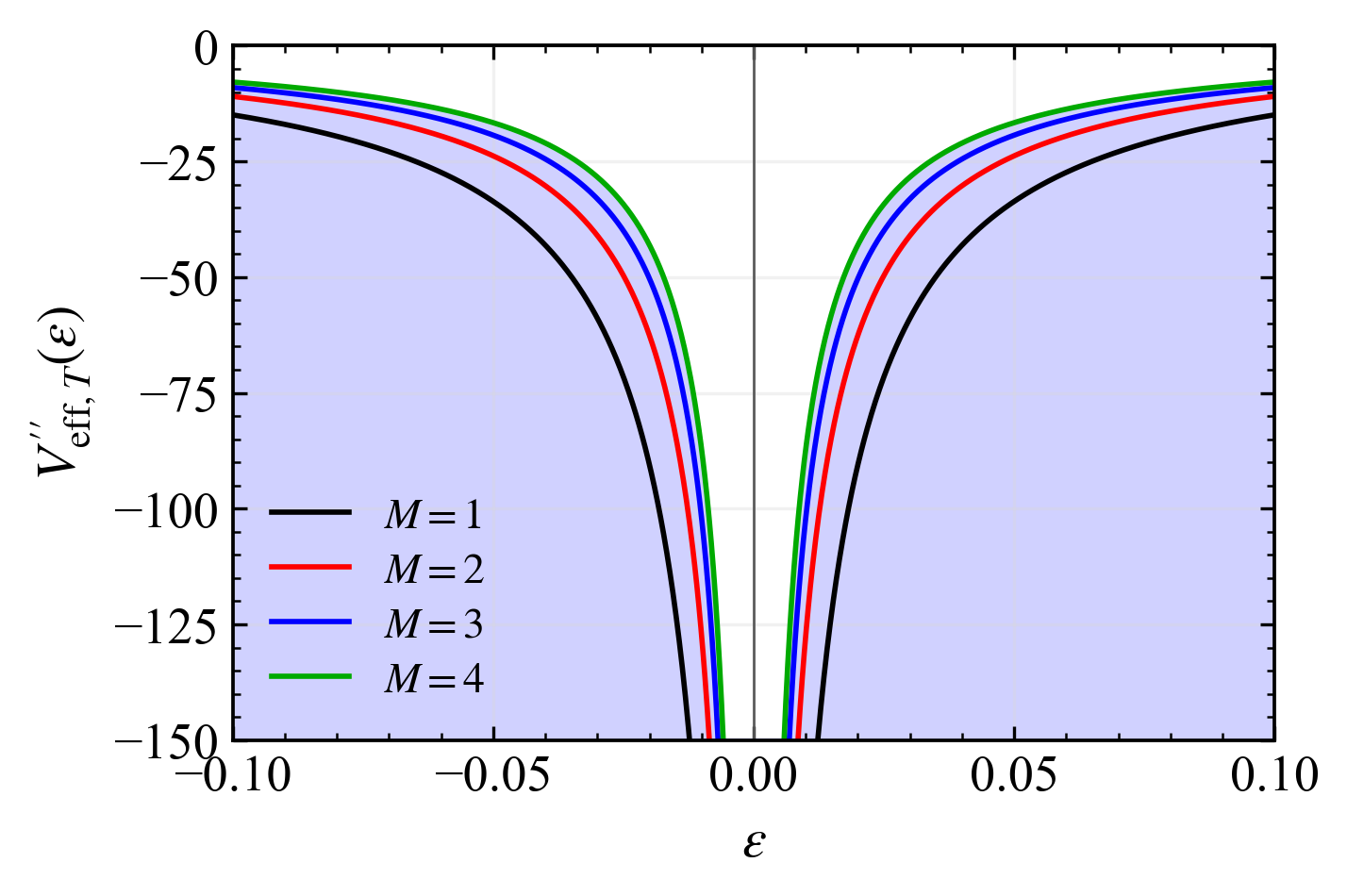}
    \caption{Effective-potential curvature $V_{{\rm eff},T}''(\mathcal E)$ for the barotropic Tsallis--Cirto thin-shell wormhole, with $a_0=r_h^{(T)}+|\mathcal E|$. In the left panel, $M=1$ and the curves correspond to $\delta=1.01$, $1.40$, $1.60$, and $2.00$. In the right panel, $\delta=1.40$ and the curves correspond to $M=1$, $2$, $3$, and $4$. In both cases, $V_{{\rm eff},T}''(\mathcal E)<0$ throughout the physical domain, indicating linear instability.}
    \label{fig:stbtsallis}
\end{figure}

Figure~\ref{fig:stbtsallis} confirms that the Tsallis--Cirto deformation does not remove the barotropic instability. The curvature becomes increasingly negative as the throat approaches the corresponding horizon and tends to zero from below for larger values of $|\mathcal E|$. The changes induced by $\delta$ and $M$ are quantitative: they alter the depth and scale of the instability, but no stable branch is generated. This behavior parallels the Schwarzschild and Barrow barotropic sectors, in which the entropy-induced modification of the lapse is insufficient to compensate for the response of a constant barotropic shell.

For the variable Chaplygin shell model, the Tsallis--Cirto geometry gives the equilibrium parameter
\begin{equation}
\Omega_{T0}=-\frac{a_0^{n-2}}{16\pi^2}\left[2+\frac{2M\pi^{1-\delta}(2\delta-3)}{\delta a_0^{2\delta-1}}\right].
\label{eq:tsallis_omega_chaplygin}
\end{equation}
For $n\neq4$, the effective potential becomes
\begin{equation}
V_{{\rm eff},T}^{\rm (C)}(a)=1-\frac{2M\pi^{1-\delta}}{\delta a^{2\delta-1}}-\left(1-\frac{2M\pi^{1-\delta}}{\delta a_0^{2\delta-1}}\right)\left(\frac{a_0}{a}\right)^2-\frac{a_0^{n-2}}{4-n}\left[2+\frac{2M\pi^{1-\delta}(2\delta-3)}{\delta a_0^{2\delta-1}}\right]\left[a^{2-n}-\frac{a_0^{4-n}}{a^2}\right].
\label{eq:tsallis_potential_chaplygin}
\end{equation}
For the special case $n=4$, the logarithmic branch is
\begin{equation}
V_{{\rm eff},T}^{\rm (C)}(a)=1-\frac{2M\pi^{1-\delta}}{\delta a^{2\delta-1}}-\left(1-\frac{2M\pi^{1-\delta}}{\delta a_0^{2\delta-1}}\right)\left(\frac{a_0}{a}\right)^2-\frac{a_0^2}{a^2}\left[2+\frac{2M\pi^{1-\delta}(2\delta-3)}{\delta a_0^{2\delta-1}}\right]\ln\left(\frac{a}{a_0}\right).
\label{eq:tsallis_potential_chaplygin_n4}
\end{equation}
The value in Eq.~\eqref{eq:tsallis_omega_chaplygin} ensures the two equilibrium conditions at $a=a_0$. The corresponding potential curvature is
\begin{equation}
V_{{\rm eff},T}^{{''}\,\rm (C)}(a_0)=\frac{1}{a_0^2}\left[2(n-2)+(2\delta-3)(n-2\delta-1)\frac{2M\pi^{1-\delta}}{\delta a_0^{2\delta-1}}\right].
\label{eq:tsallis_chaplygin_stability}
\end{equation}
The variable Chaplygin configuration is linearly stable whenever
\begin{equation}
2(n-2)+(2\delta-3)(n-2\delta-1)\frac{2M\pi^{1-\delta}}{\delta a_0^{2\delta-1}}>0.
\label{eq:tsallis_chaplygin_stability_condition}
\end{equation}
In particular, for $n=4$,
\begin{equation}
V_{{\rm eff},T}^{{''}\,\rm (C)}(a_0)=\frac{1}{a_0^2}\left[4-(2\delta-3)^2\frac{2M\pi^{1-\delta}}{\delta a_0^{2\delta-1}}\right]>0,
\label{eq:tsallis_chaplygin_stability_n4}
\end{equation}
for the interval $1\leq\delta\leq2$ considered in the numerical analysis. Thus, the $n=4$ Chaplygin branch is stable throughout the admissible Tsallis--Cirto exterior.

\begin{figure}[htp!]
    \centering
    \includegraphics[width=1\linewidth]{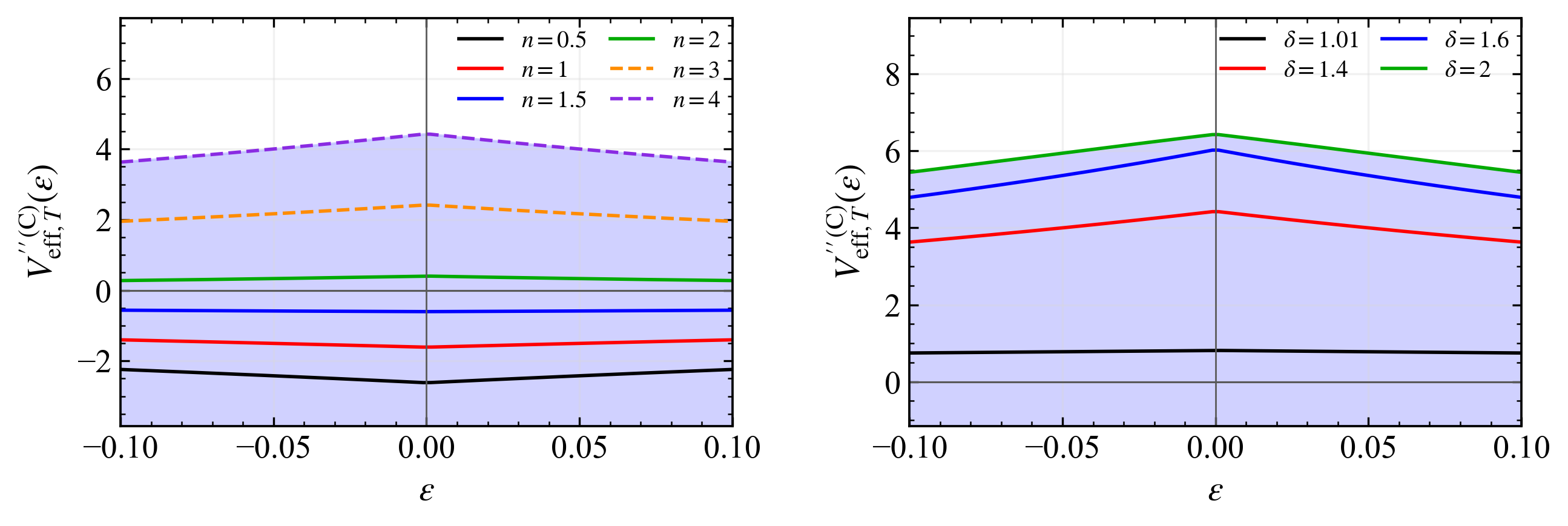}
    \caption{Effective-potential curvature $V_{{\rm eff},T}^{{''}\,\rm (C)}(\mathcal E)$ for the variable Chaplygin Tsallis--Cirto thin-shell wormhole, with $a_0=r_h^{(T)}+|\mathcal E|$. In the left panel, $M=1$ and $\delta=1.4$, and the curves correspond to $n=0.5$, $1$, $1.5$, $2$, $3$, and $4$. In the right panel, $M=1$ and $n=4$, and the curves correspond to $\delta=1.01$, $1.4$, $1.6$, and $2$. The Chaplygin parameter $\Omega$ is fixed by the static equilibrium condition in each case. Positive regions of the curvature correspond to linear radial stability.}
    \label{fig:stbchtsallis}
\end{figure}

Figure~\ref{fig:stbchtsallis} displays the qualitative change generated by the variable Chaplygin model. In contrast with the barotropic Tsallis--Cirto branch, whose curvature is negative throughout the exterior domain, the Chaplygin shell admits stable configurations and has a finite curvature in the horizon limit. For fixed $\delta=1.4$, the near-horizon stability threshold is $n=2\delta-1=1.8$, whereas the large-radius threshold is $n=2$. Consequently, the curves with $n=0.5$, $1$, and $1.5$ remain unstable, while the $n=2$, $3$, and $4$ branches are stable in the interval displayed. More generally, the window $2\delta-1<n<2$ corresponds to a stable near-horizon region followed by a transition to instability at sufficiently large throat radius. This differs from the Schwarzschild case, for which the near-horizon threshold is $n=1$, and from the Barrow example with $\Delta=0.5$, for which it is $n=1.5$. The Tsallis deformation with $\delta=1.4$ therefore shifts the onset of near-horizon stability to larger values of $n$. In the right panel, the choice $n=4$ lies safely above both thresholds for all displayed values of $\delta$, explaining why all curves remain positive. Increasing $\delta$ enhances the positive curvature in the plotted near-horizon region, indicating a stronger local restoring response, while preserving the stability of this Chaplygin branch.

\subsection{R\'enyi entropy}
\label{sec:renyi_entropy}

The R\'enyi entropy and its corresponding lapse function are given by
\begin{equation}
\mathcal S_R(r)=\frac{1}{\lambda_R}\ln\left(1+\lambda_R\pi r^2\right),
\label{eq:renyi_entropy_lapse}
\end{equation}
and
\begin{equation}
    F_R(r)=1-\frac{2M}{r}\left(1+\lambda_R\pi r^2\right).
\end{equation}
For $0<\lambda_R<1/(16\pi M^2)$, the lapse has two positive Killing horizons,
\begin{equation}
r_{h,\pm}^{(R)}=\frac{1\pm\sqrt{1-16\pi\lambda_RM^2}}{4\pi\lambda_RM},
\label{eq:renyi_horizons}
\end{equation}
with $r_{h,-}^{(R)}<r_{h,+}^{(R)}$ and $r_{h,-}^{(R)}\to2M$ as $\lambda_R\to0$. The lapse is positive only between these two roots, and therefore the shell must lie in the local static patch
\begin{equation}
r_{h,-}^{(R)}<a_0<r_{h,+}^{(R)}.
\label{eq:renyi_throat_domain}
\end{equation}
The outer zero bounds the static region because the Killing vector $\partial_t$ becomes null there and spacelike beyond it. Nevertheless, it should not be interpreted as an ordinary de Sitter cosmological horizon, since the R\'enyi lapse contains a term linear in $r$, rather than a contribution proportional to $-\Lambda r^2$. Thus, for $\lambda_R>0$, the geometry is neither asymptotically flat nor asymptotically de Sitter. This bounded positive-lapse sector is particularly relevant for the present construction, since it permits a thin-shell junction even when the corresponding smooth Morris--Thorne density reconstruction is not viable.

The static surface density and pressure are
\begin{equation}
\sig_{R0}=-\frac{1}{2\pi a_0}\sqrt{1-\frac{2M}{a_0}-2\pi\lambda_RMa_0},
\label{eq:renyi_static_density}
\end{equation}
and
\begin{equation}
p_{R0}=\frac{\dfrac{2M}{a_0^2}-2\pi\lambda_RM}{8\pi\sqrt{1-\dfrac{2M}{a_0}-2\pi\lambda_RMa_0}}+\frac{1}{4\pi a_0}\sqrt{1-\frac{2M}{a_0}-2\pi\lambda_RMa_0}.
\label{eq:renyi_static_pressure}
\end{equation}
As in the previous cases, the surface density is negative throughout the admissible domain. In contrast, the tangential pressure diverges positively near the inner horizon and negatively near the outer horizon, reflecting the opposite signs of $F_R'(r)$ at the two boundaries of the static interval.

The intrinsic NEC combination and the remaining trace SEC combination are
\begin{equation}
\left(\sig_0+p_0\right)_R=\frac{\dfrac{6M}{a_0}+2\pi\lambda_RMa_0-2}{8\pi a_0\sqrt{1-\dfrac{2M}{a_0}-2\pi\lambda_RMa_0}},
\label{eq:renyi_nec_explicit}
\end{equation}
and
\begin{equation}
\left(\sig_0+2p_0\right)_R=\frac{\dfrac{2M}{a_0^2}-2\pi\lambda_RM}{4\pi\sqrt{1-\dfrac{2M}{a_0}-2\pi\lambda_RMa_0}}.
\label{eq:renyi_sec_explicit}
\end{equation}
Since $\sig_{R0}<0$, the WEC is violated throughout the shell. The NEC combination is positive close to the inner horizon, changes sign within the static region, and becomes negative toward the outer horizon. The remaining trace SEC combination changes sign at $a_0=(\pi\lambda_R)^{-1/2}$, which coincides with the maximum of the R\'enyi lapse. Therefore, unlike the Schwarzschild, Barrow, and Tsallis--Cirto cases, the R\'enyi shell possesses an outer portion of the static interval in which both the intrinsic NEC combination and the remaining trace SEC combination are negative.

\begin{figure}[htp!]
    \centering
    \includegraphics[width=0.49\linewidth]{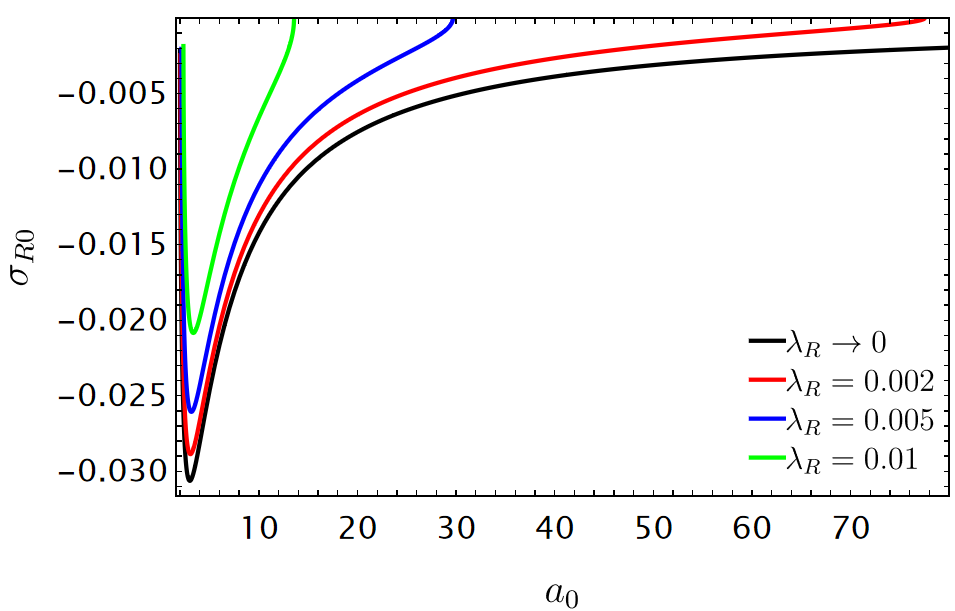}
    \includegraphics[width=0.49\linewidth]{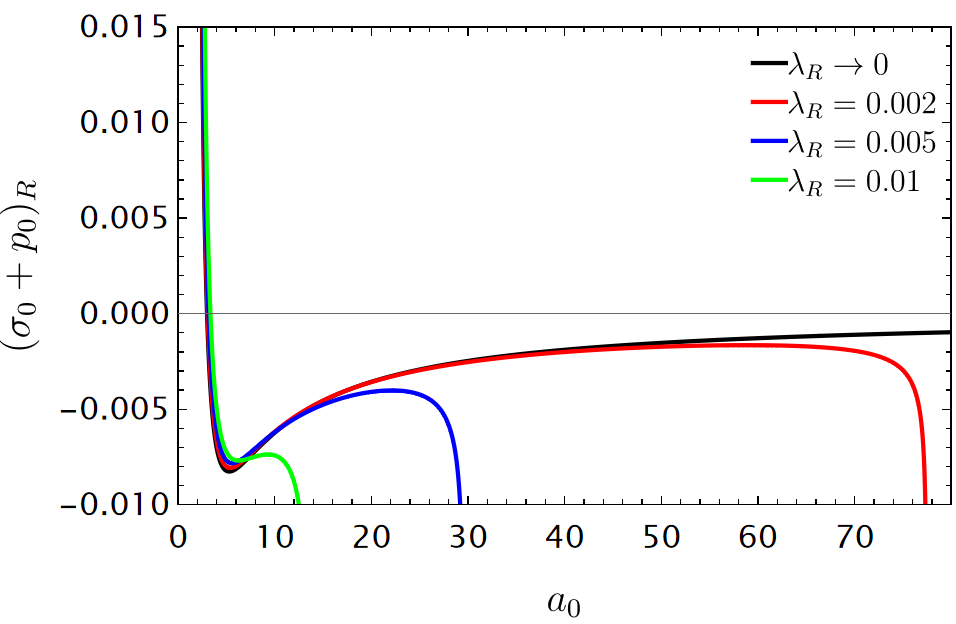}
    \includegraphics[width=0.5\linewidth]{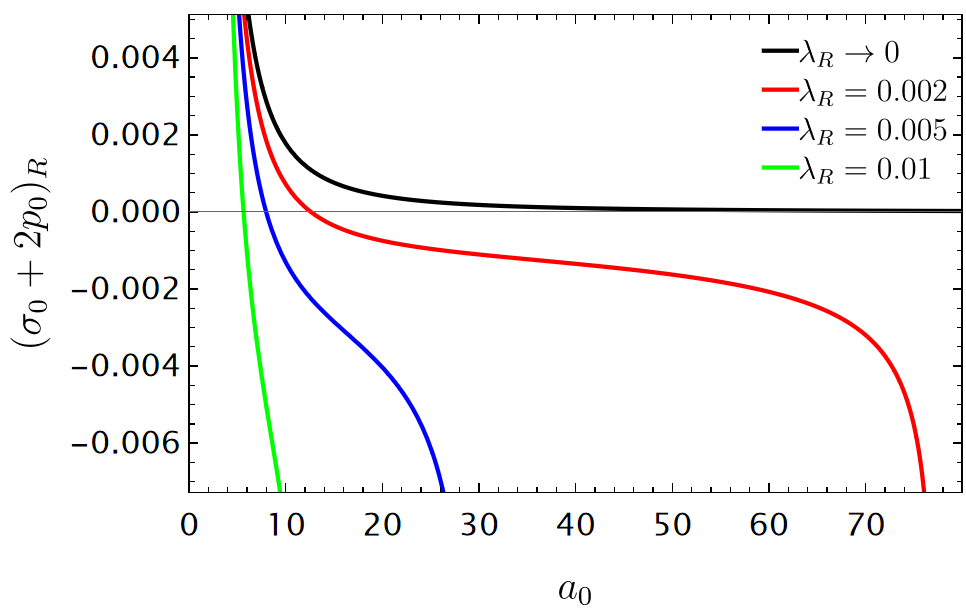}
    \caption{Static surface energy conditions for the R\'enyi thin-shell wormhole as functions of the throat radius $a_0$. The upper-left panel shows the surface energy density $\sigma_{R0}$, the upper-right panel displays the intrinsic shell NEC combination $\sigma_{R0}+p_{R0}$, and the lower panel shows the remaining trace SEC combination $\sigma_{R0}+2p_{R0}$. The curves correspond to $\lambda_R\to0$, $0.002$, $0.005$, and $0.01$, with $M=1$. For $\lambda_R>0$, the curves are displayed only in the physical static interval between the inner and outer horizons, where $F_R(a_0)>0$.}
    \label{fig:ecrenyi}
\end{figure}

Figure~\ref{fig:ecrenyi} illustrates how the R\'enyi parameter transforms the unbounded Schwarzschild exterior into a finite static region. For $\lambda_R>0$, the density vanishes at both horizons and develops a negative minimum in the interior of the allowed interval. The NEC combination begins positive near the inner horizon, crosses to negative values at an intermediate radius, and diverges negatively as the outer horizon is approached. The remaining trace SEC combination follows a related but distinct pattern: it is positive in the inner part of the static domain, vanishes where the lapse reaches its maximum, and becomes negative toward the outer boundary. Increasing $\lambda_R$ brings the outer horizon inward and compresses these transitions into a smaller radial interval.

For the barotropic shell model, the equilibrium parameter and effective potential become
\begin{equation}
w_{R0}=-\frac{1}{2}-\frac{\dfrac{2M}{a_0}-2\pi\lambda_RMa_0}{4\left(1-\dfrac{2M}{a_0}-2\pi\lambda_RMa_0\right)},
\label{eq:renyi_w0}
\end{equation}
and
\begin{equation}
V_{{\rm eff},R}(a)=1-\frac{2M}{a}\left(1+\lambda_R\pi a^2\right)-\left[1-\frac{2M}{a_0}\left(1+\lambda_R\pi a_0^2\right)\right]\left(\frac{a_0}{a}\right)^{2+4w}.
\label{eq:renyi_potential}
\end{equation}
After imposing $w=w_{R0}$, the static potential curvature is
\begin{equation}
V_{{\rm eff},R}''(a_0)=-\frac{2M}{a_0^3}-\frac{2\pi\lambda_RM}{a_0}-\frac{4M^2\left(\dfrac{1}{a_0^2}-\pi\lambda_R\right)^2}{1-\dfrac{2M}{a_0}-2\pi\lambda_RMa_0}.
\label{eq:renyi_barotropic_stability}
\end{equation}
All terms in Eq.~\eqref{eq:renyi_barotropic_stability} are non-positive in the physical static interval, and the first two are strictly negative. Hence, the barotropic R\'enyi thin-shell wormhole is linearly unstable for every admissible static radius.

\begin{figure}[htp!]
    \centering
    \includegraphics[width=0.49\linewidth]{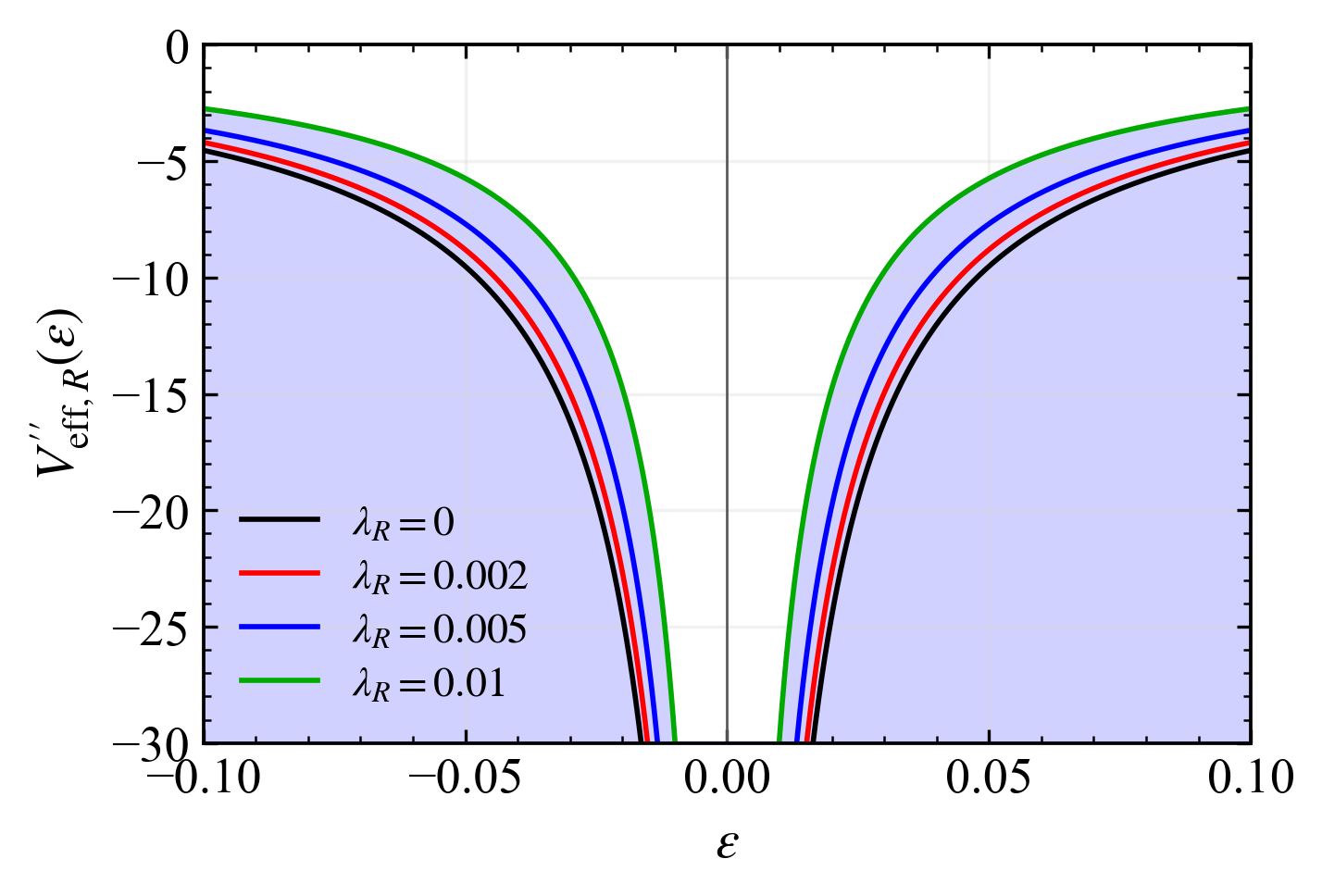}
    \includegraphics[width=0.49\linewidth]{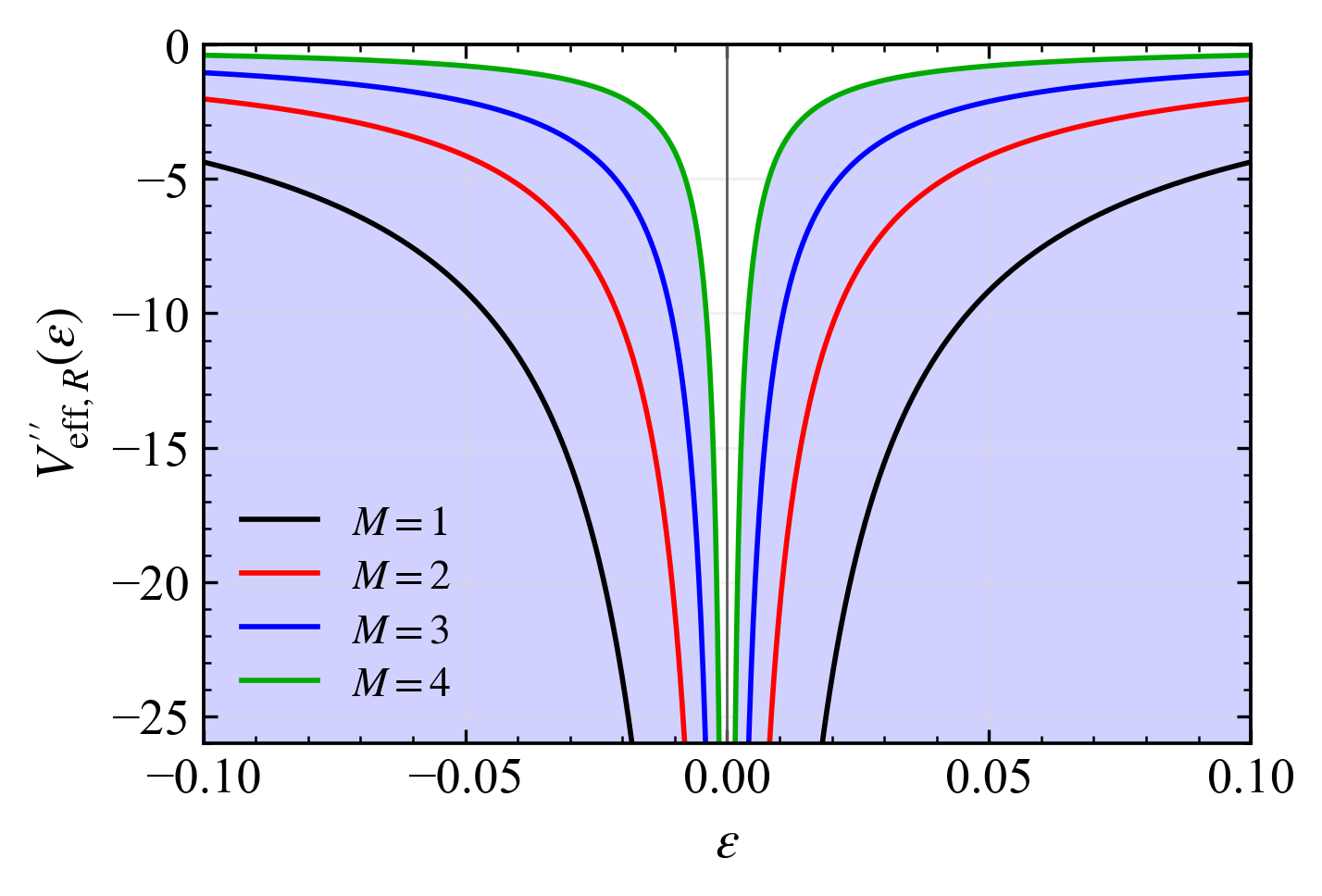}
    \caption{Effective-potential curvature $V_{{\rm eff},R}''(\mathcal E)$ for the barotropic R\'enyi thin-shell wormhole, with $a_0=r_{h,-}^{(R)}+|\mathcal E|$. In the left panel, $M=1$ and the curves correspond to $\lambda_R =0$, $0.002$, $0.005$, and $0.01$. In the right panel, $\lambda_R=0.001$ and the curves correspond to $M=1$, $2$, $3$, and $4$. For $\lambda_R>0$, the curves and shaded regions are displayed only within the physical static interval $r_{h,-}^{(R)}<a_0<r_{h,+}^{(R)}$, where $F_R(a_0)>0$. In both cases, $V_{{\rm eff},R}''(\mathcal E)<0$ throughout the admissible domain, indicating linear instability.}
    \label{fig:stbrenyi}
\end{figure}

Figure~\ref{fig:stbrenyi} confirms that the finite R\'enyi static region does not alter the qualitative conclusion obtained for the barotropic Schwarzschild, Barrow, and Tsallis--Cirto branches. The curvature becomes strongly negative as the shell approaches the inner horizon, while its magnitude decreases away from that boundary. The effect of $\lambda_R$ is to reshape the local scale of the instability and, through the outer horizon, to restrict the available radial domain. Likewise, increasing $M$ changes the magnitude of the curves but does not create a stable barotropic sector. The two mirrored sides of each profile arise solely from the parametrization $a_0=r_{h,-}^{(R)}+|\mathcal E|$ and do not represent distinct wormhole branches.

For the variable Chaplygin shell model, the static value of the equation-of-state parameter is
\begin{equation}
\Omega_{R0}=-\frac{a_0^{n-2}}{8\pi^2}\left(1-\frac{M}{a_0}-3\pi\lambda_RMa_0\right).
\label{eq:renyi_omega_chaplygin}
\end{equation}
For $n\neq4$, the corresponding effective potential is
\begin{equation}
V_{{\rm eff},R}^{\rm (C)}(a)=1-\frac{2M}{a}\left(1+\lambda_R\pi a^2\right)-\left[1-\frac{2M}{a_0}\left(1+\lambda_R\pi a_0^2\right)\right]\left(\frac{a_0}{a}\right)^2-\frac{2a_0^{n-2}}{4-n}\left(1-\frac{M}{a_0}-3\pi\lambda_RMa_0\right)\left[a^{2-n}-\frac{a_0^{4-n}}{a^2}\right].
\label{eq:renyi_potential_chaplygin}
\end{equation}
For $n=4$, the logarithmic branch takes the form
\begin{equation}
V_{{\rm eff},R}^{\rm (C)}(a)=1-\frac{2M}{a}\left(1+\lambda_R\pi a^2\right)-\left[1-\frac{2M}{a_0}\left(1+\lambda_R\pi a_0^2\right)\right]\left(\frac{a_0}{a}\right)^2-\frac{2a_0^2}{a^2}\left(1-\frac{M}{a_0}-3\pi\lambda_RMa_0\right)\ln\left(\frac{a}{a_0}\right).
\label{eq:renyi_potential_chaplygin_n4}
\end{equation}
The parameter in Eq.~\eqref{eq:renyi_omega_chaplygin} guarantees that $V_{{\rm eff},R}^{\rm (C)}(a_0)=V_{{\rm eff},R}^{{'}\,\rm (C)}(a_0)=0$. The potential curvature is
\begin{equation}
V_{{\rm eff},R}^{{''}\,\rm (C)}(a_0)=\frac{2}{a_0^3}\left[(n-2)a_0+(3-n)M-3\pi\lambda_RM(n-1)a_0^2\right].
\label{eq:renyi_chaplygin_stability}
\end{equation}
Thus, the variable Chaplygin R\'enyi shell is stable whenever the right-hand side of Eq.~\eqref{eq:renyi_chaplygin_stability} is positive. In particular, the $n=4$ branch satisfies
\begin{equation}
V_{{\rm eff},R}^{{''}\,\rm (C)}(a_0)=\frac{2}{a_0^3}\left(2a_0-M-9\pi\lambda_RMa_0^2\right).
\label{eq:renyi_chaplygin_stability_n4}
\end{equation}

\begin{figure}[htp!]
    \centering
    \includegraphics[width=1.0\linewidth]{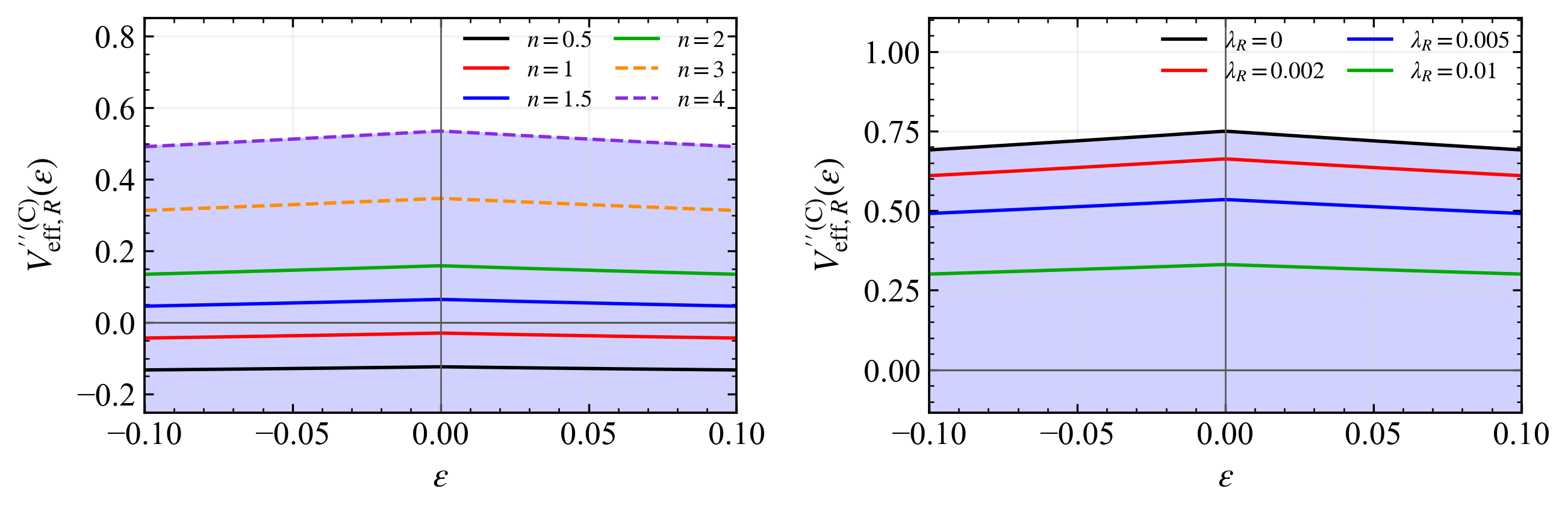}
    \caption{Effective-potential curvature $V_{{\rm eff},R}^{{''}\,\rm (C)}(\mathcal E)$ for the variable Chaplygin R\'enyi thin-shell wormhole, with $a_0=r_{h,-}^{(R)}+|\mathcal E|$. In the left panel, $M=1$ and $\lambda_R=0.005$, and the curves correspond to $n=0.5$, $1$, $1.5$, $2$, $3$, and $4$. In the right panel, $M=1$ and $n=4$, and the curves correspond to $\lambda_R=0$, $0.002$, $0.005$, and $0.01$. The Chaplygin parameter $\Omega$ is fixed by the static equilibrium condition in each case. The curves are restricted to the admissible positive-lapse region, and positive curvature indicates linear radial stability.}
    \label{fig:stbchrenyi}
\end{figure}

Figure~\ref{fig:stbchrenyi} shows that the variable Chaplygin equation of state can produce a locally stable R\'enyi thin-shell branch, in sharp contrast with the barotropic result. In the left panel, with $M=1$ and $\lambda_R=0.005$, the $n=0.5$ and $n=1$ configurations remain unstable near the inner horizon, whereas the curves with $n\geq1.5$ are positive in the displayed region. The threshold is shifted from the Schwarzschild value because the R\'enyi deformation changes both the inner horizon and the coefficient multiplying $n$ in Eq.~\eqref{eq:renyi_chaplygin_stability}. At the inner horizon, the local transition occurs at $n=r_{h,-}^{(R)}/[4M-r_{h,-}^{(R)}]$, which is approximately $1.07$ for the parameters of the left panel. This explains why $n=1$ is still unstable while the $n=1.5$ branch is already stable close to the horizon.

The comparison with the other Chaplygin sectors is instructive. For Schwarzschild, the $n=4$ branch is stable throughout the entire exterior region, while the Barrow and Tsallis--Cirto examples also retain stable $n=4$ branches over the radial domains considered. The R\'enyi geometry differs because $\lambda_R>0$ introduces an outer horizon and the negative term proportional to $\lambda_R a_0^2$ in Eq.~\eqref{eq:renyi_chaplygin_stability}. Consequently, positivity near the inner horizon does not automatically imply stability across the whole finite static interval. In particular, the $n=4$ curvature changes sign when $2a_0-M-9\pi\lambda_RMa_0^2=0$, whenever this radius lies between the two R\'enyi horizons. The right panel therefore establishes the local near-inner-horizon stability of the displayed $n=4$ branches, while also showing that increasing $\lambda_R$ weakens the restoring curvature. The symmetry with respect to $\mathcal E=0$ follows from the use of $|\mathcal E|$ in the throat parametrization, rather than from the presence of two independent static solutions.

\subsection{Kaniadakis entropy}
\label{sec:kaniadakis_entropy}

The Kaniadakis entropy is given by
\begin{equation}
\mathcal S_K(r)=\frac{1}{\kappa}\sinh\left(\kappa\pi r^2\right).
\label{eq:kaniadakis_entropy}
\end{equation}
Substituting Eq.~\eqref{eq:kaniadakis_entropy} into Eq.~\eqref{eq:entropic_metric}, we obtain the Kaniadakis lapse function,
\begin{equation}
F_K(r)=1-\frac{2M}{r}\sech\left(\kappa\pi r^2\right).
\label{eq:kaniadakis_entropy_lapse}
\end{equation}
The $\kappa\to0$ limit recovers the Bekenstein--Hawking entropy and the Schwarzschild lapse function. For $\kappa\geq0$, the horizon is determined implicitly by
\begin{equation}
r_h^{(K)}\cosh\left(\kappa\pi[r_h^{(K)}]^2\right)=2M.
\label{eq:kaniadakis_horizon}
\end{equation}
The left-hand side of Eq.~\eqref{eq:kaniadakis_horizon} increases monotonically for positive $r$, and hence the Kaniadakis geometry possesses a unique positive Killing horizon. For $\kappa>0$, one has $r_h^{(K)}<2M$, while $r_h^{(K)}\to2M$ as $\kappa\to0$. Therefore, the admissible static exterior region is simply $a_0>r_h^{(K)}$.

Writing $x_0=\kappa\pi a_0^2$, the static surface density and pressure are
\begin{equation}
\sig_{K0}=-\frac{1}{2\pi a_0}\sqrt{1-\frac{2M}{a_0}\sech x_0},
\label{eq:kaniadakis_static_density}
\end{equation}
and
\begin{equation}
p_{K0}=\frac{M\sech x_0\left(1+2x_0\tanh x_0\right)}{4\pi a_0^2\sqrt{1-\dfrac{2M}{a_0}\sech x_0}}+\frac{1}{4\pi a_0}\sqrt{1-\frac{2M}{a_0}\sech x_0}.
\label{eq:kaniadakis_static_pressure}
\end{equation}
The surface density is negative throughout the admissible exterior region, whereas the tangential pressure is positive. As $a_0\to r_h^{(K)}$, the density tends to zero from below while the pressure diverges. At large radii, the hyperbolic correction is exponentially suppressed and the shell approaches the usual asymptotic thin-shell behavior.

The static intrinsic NEC combination and the remaining trace SEC combination are
\begin{equation}
\left(\sig_0+p_0\right)_K=\frac{\dfrac{2M\sech x_0}{a_0}\left(3+2x_0\tanh x_0\right)-2}{8\pi a_0\sqrt{1-\dfrac{2M}{a_0}\sech x_0}},
\label{eq:kaniadakis_nec_explicit}
\end{equation}
and
\begin{equation}
\left(\sig_0+2p_0\right)_K=\frac{M\sech x_0\left(1+2x_0\tanh x_0\right)}{2\pi a_0^2\sqrt{1-\dfrac{2M}{a_0}\sech x_0}}.
\label{eq:kaniadakis_sec_explicit}
\end{equation}
Since $\sig_{K0}<0$, the WEC is violated throughout the physical domain. The remaining trace SEC combination remains positive because $F_K'(a_0)>0$, whereas the intrinsic NEC is satisfied only sufficiently close to the horizon and becomes negative as the throat is displaced outward.

\begin{figure}[htp!]
    \centering
    \includegraphics[width=0.49\linewidth]{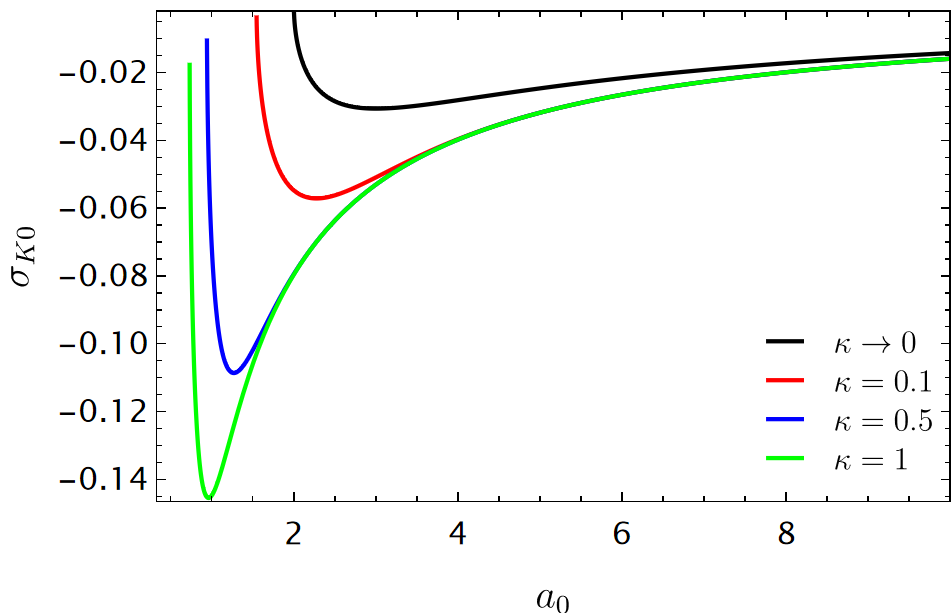}
    \includegraphics[width=0.49\linewidth]{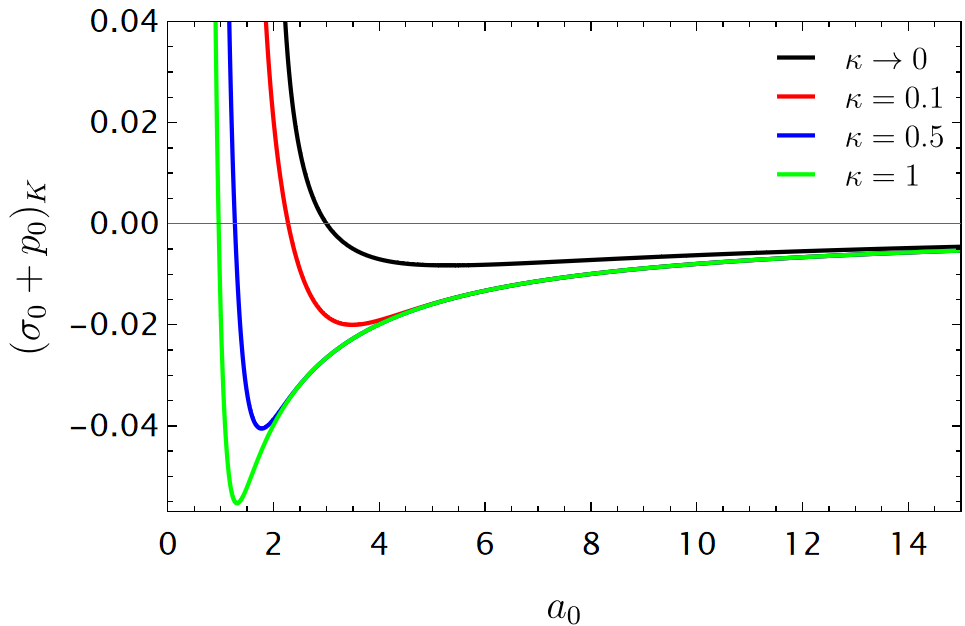}
    \includegraphics[width=0.5\linewidth]{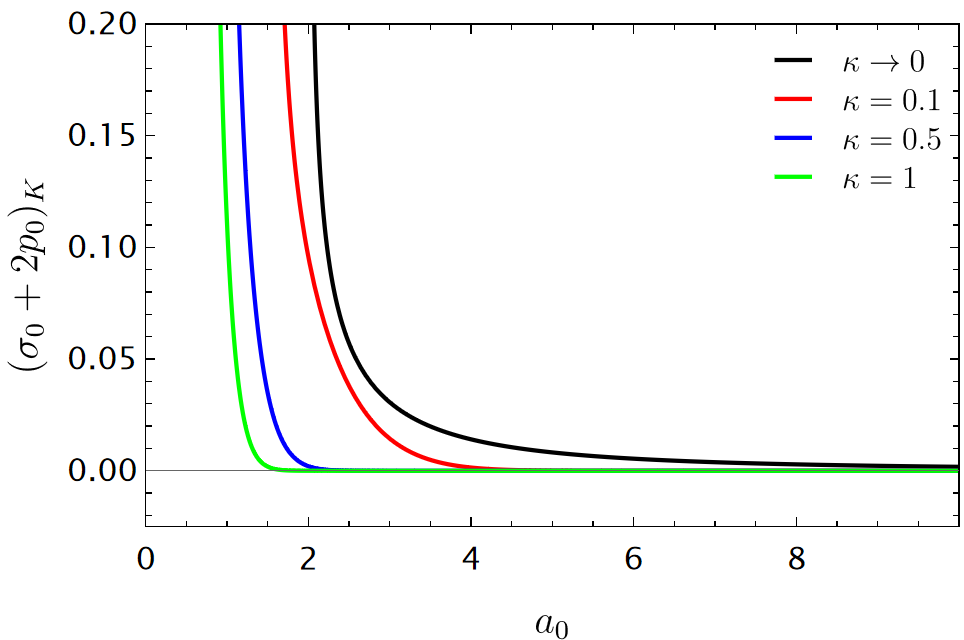}
    \caption{Static surface energy conditions for the Kaniadakis thin-shell wormhole as functions of the throat radius $a_0$. The upper-left panel shows the surface energy density $\sigma_{K0}$, the upper-right panel displays the intrinsic shell NEC combination $\sigma_{K0}+p_{K0}$, and the lower panel shows the remaining trace SEC combination $\sigma_{K0}+2p_{K0}$. The curves correspond to $\kappa\to0$, $0.1$, $0.5$, and $1$, with $M=1$, and are shown only in their respective admissible domains, $a_0>r_h^{(K)}$.}
    \label{fig:eckaniadakis}
\end{figure}

Figure~\ref{fig:eckaniadakis} shows that increasing $\kappa$ shifts the horizon inward and concentrates the surface-stress structure closer to the throat. The density reaches a deeper negative minimum as $\kappa$ increases, but all curves converge at sufficiently large radius because the hyperbolic correction becomes exponentially small. The intrinsic NEC is positive near the horizon due to the divergent pressure contribution and changes sign at a radius that moves inward as $\kappa$ grows. In contrast, the remaining trace SEC combination remains positive for every displayed curve and decays rapidly away from the throat.

For the barotropic shell model, the equilibrium parameter and effective potential are
\begin{equation}
w_{K0}=-\frac{1}{2}-\frac{M\sech x_0\left(1+2x_0\tanh x_0\right)}{2a_0\left(1-\dfrac{2M}{a_0}\sech x_0\right)},
\label{eq:kaniadakis_w0}
\end{equation}
and
\begin{equation}
V_{{\rm eff},K}(a)=1-\frac{2M}{a}\sech\left(\kappa\pi a^2\right)-\left[1-\frac{2M}{a_0}\sech x_0\right]\left(\frac{a_0}{a}\right)^{2+4w}.
\label{eq:kaniadakis_potential}
\end{equation}
After imposing $w=w_{K0}$, the potential curvature becomes
\begin{equation}
V_{{\rm eff},K}''(a_0)=\frac{2M\sech x_0}{a_0^3}\left[-1+4x_0^2\left(1-2\tanh^2x_0\right)-\frac{\dfrac{2M\sech x_0}{a_0}\left(1+2x_0\tanh x_0\right)^2}{1-\dfrac{2M}{a_0}\sech x_0}\right].
\label{eq:kaniadakis_barotropic_stability}
\end{equation}
For the Kaniadakis branch considered here, $V_{{\rm eff},K}''(a_0)<0$ throughout the admissible domain. Hence, the localized hyperbolic deformation does not produce a stable barotropic thin-shell configuration.

\begin{figure}[htp!]
    \centering
    \includegraphics[width=0.49\linewidth]{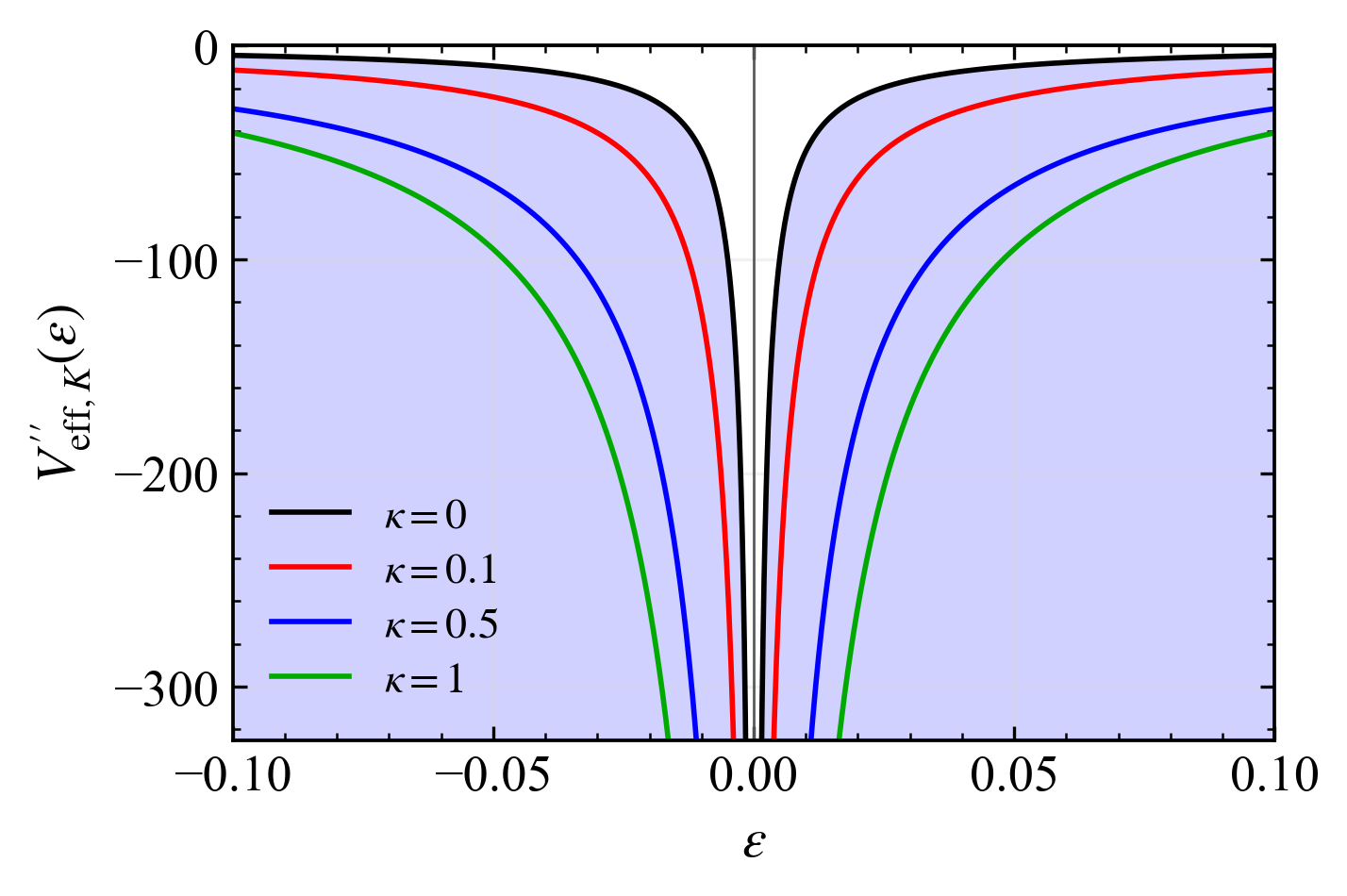}
    \includegraphics[width=0.49\linewidth]{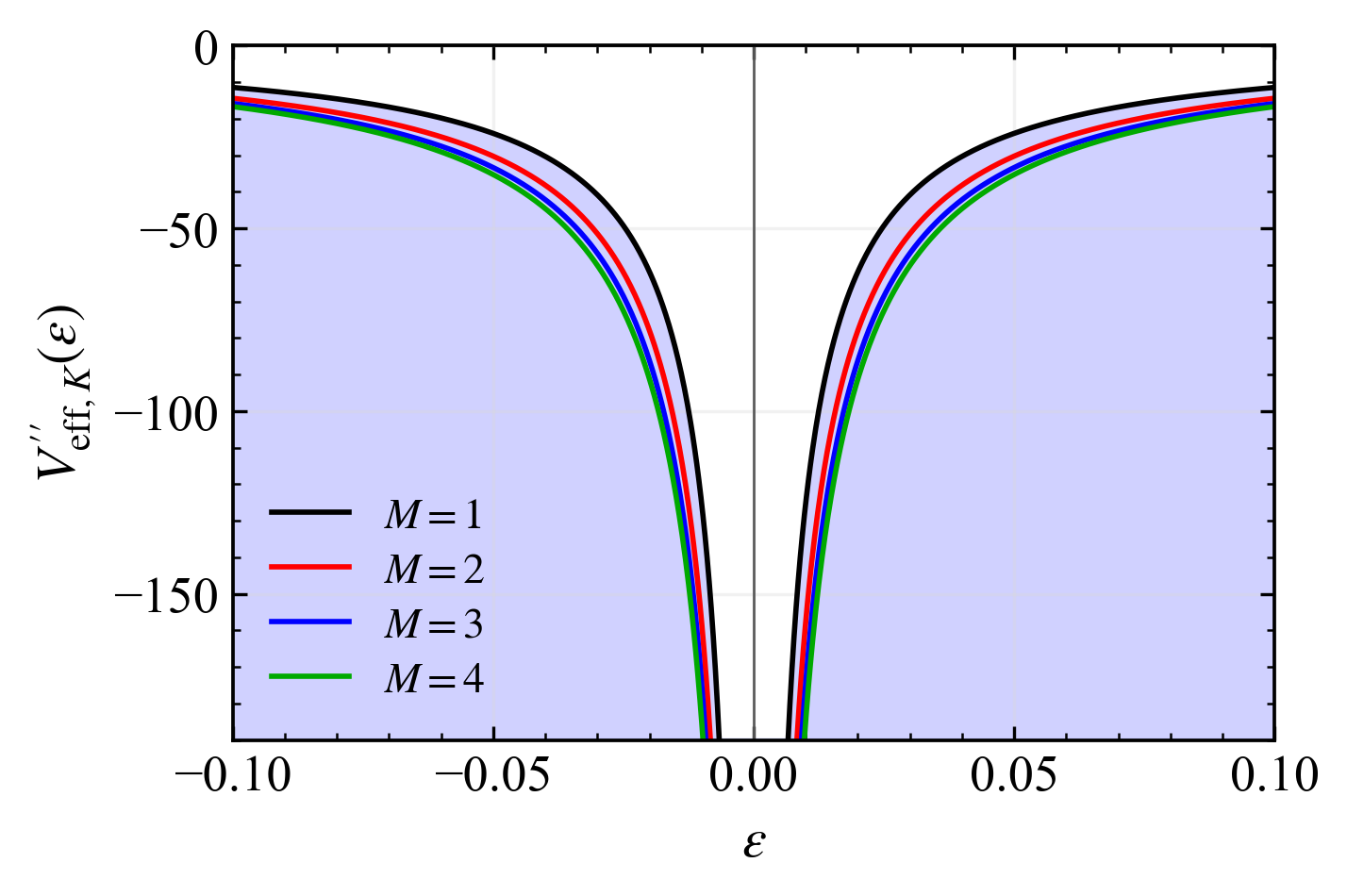}
    \caption{Effective-potential curvature $V_{{\rm eff},K}''(\mathcal E)$ for the barotropic Kaniadakis thin-shell wormhole, with $a_0=r_h^{(K)}+|\mathcal E|$. In the left panel, $M=1$ and the curves correspond to $\kappa\to0$, $0.1$, $0.5$, and $1$. In the right panel, $\kappa=0.1$ and the curves correspond to $M=1$, $2$, $3$, and $4$. In both cases, $V_{{\rm eff},K}''(\mathcal E)<0$ throughout the physical domain $a_0>r_h^{(K)}$, indicating linear instability.}
    \label{fig:stbkaniadakis}
\end{figure}

Figure~\ref{fig:stbkaniadakis} confirms that the barotropic instability remains present despite the strong local deformation of the Kaniadakis lapse. The curvature is most negative near the horizon and approaches zero from below as the shell moves outward. Increasing $\kappa$ enhances the magnitude of the near-horizon instability, whereas increasing the mass at fixed $\kappa$ mainly rescales the profile. This behavior parallels the Schwarzschild, Barrow, Tsallis--Cirto, and R\'enyi barotropic branches: changing the seed geometry modifies the strength of the instability but does not alter its sign.

For the variable Chaplygin shell model, define
\begin{equation}
\mathcal C_{K0}=1-\frac{M}{a_0}\sech x_0+2\kappa\pi M a_0\sech x_0\tanh x_0.
\label{eq:kaniadakis_C}
\end{equation}
The static equation-of-state parameter is then
\begin{equation}
\Omega_{K0}=-\frac{a_0^{n-2}}{8\pi^2}\mathcal C_{K0}.
\label{eq:kaniadakis_omega_chaplygin}
\end{equation}
For $n\neq4$, the corresponding effective potential is
\begin{equation}
V_{{\rm eff},K}^{\rm (C)}(a)=1-\frac{2M}{a}\sech\left(\kappa\pi a^2\right)-\left[1-\frac{2M}{a_0}\sech x_0\right]\left(\frac{a_0}{a}\right)^2-\frac{2a_0^{n-2}\mathcal C_{K0}}{4-n}\left[a^{2-n}-\frac{a_0^{4-n}}{a^2}\right].
\label{eq:kaniadakis_potential_chaplygin}
\end{equation}
For the special case $n=4$, one obtains
\begin{equation}
V_{{\rm eff},K}^{\rm (C)}(a)=1-\frac{2M}{a}\sech\left(\kappa\pi a^2\right)-\left[1-\frac{2M}{a_0}\sech x_0\right]\left(\frac{a_0}{a}\right)^2-\frac{2a_0^2\mathcal C_{K0}}{a^2}\ln\left(\frac{a}{a_0}\right).
\label{eq:kaniadakis_potential_chaplygin_n4}
\end{equation}
The value in Eq.~\eqref{eq:kaniadakis_omega_chaplygin} guarantees that $V_{{\rm eff},K}^{\rm (C)}(a_0)=V_{{\rm eff},K}^{{'}\,\rm (C)}(a_0)=0$. The potential curvature is
\begin{equation}
V_{{\rm eff},K}^{{''}\,\rm (C)}(a_0)=\frac{1}{a_0^2}\left[2(n-2)+\frac{2M\sech x_0}{a_0}\left(3-n+2nx_0\tanh x_0+4x_0^2\left(1-2\tanh^2x_0\right)\right)\right].
\label{eq:kaniadakis_chaplygin_stability}
\end{equation}
The variable Chaplygin shell is linearly stable whenever $V_{{\rm eff},K}^{{''}\,\rm (C)}(a_0)>0$. In particular, for $n=4$,
\begin{equation}
V_{{\rm eff},K}^{{''}\,\rm (C)}(a_0)=\frac{1}{a_0^2}\left[4+\frac{2M\sech x_0}{a_0}\left(-1+8x_0\tanh x_0+4x_0^2\left(1-2\tanh^2x_0\right)\right)\right].
\label{eq:kaniadakis_chaplygin_stability_n4}
\end{equation}

\begin{figure}[htp!]
    \centering
    \includegraphics[width=1.0\linewidth]{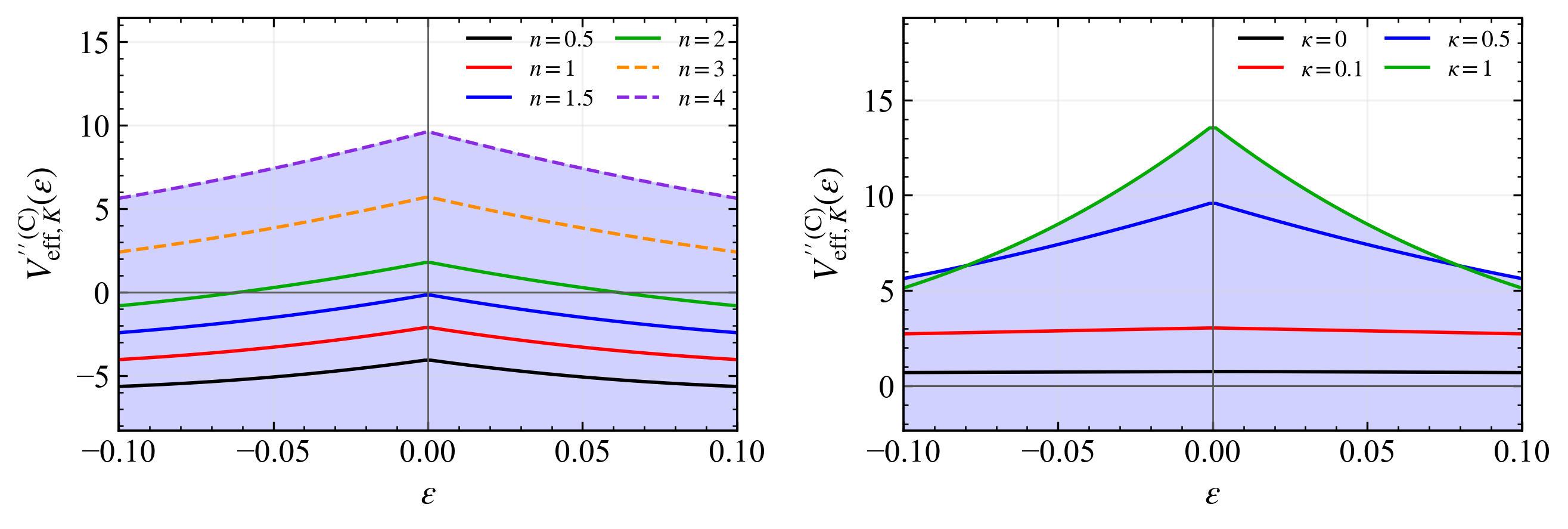}
    \caption{Effective-potential curvature $V_{{\rm eff},K}^{{''}\,\rm (C)}(\mathcal E)$ for the variable Chaplygin Kaniadakis thin-shell wormhole, with $a_0=r_h^{(K)}+|\mathcal E|$. In the left panel, $M=1$ and $\kappa=0.5$, and the curves correspond to $n=0.5$, $1$, $1.5$, $2$, $3$, and $4$. In the right panel, $M=1$ and $n=4$, and the curves correspond to $\kappa=0$, $0.1$, $0.5$, and $1$. The Chaplygin parameter $\Omega$ is fixed by the static equilibrium condition in each case. Positive regions of the curvature correspond to linear radial stability.}
    \label{fig:stbchkaniadakis}
\end{figure}

Figure~\ref{fig:stbchkaniadakis} reveals a substantially different outcome from the barotropic analysis. For fixed $M=1$ and $\kappa=0.5$, the branches with $n=0.5$ and $n=1$ are unstable, while the $n=1.5$ curve remains close to the transition between the two regimes. The $n=2$ branch exhibits a finite near-horizon stable interval, but its curvature decreases and becomes negative as the throat is displaced outward. On the other hand, the $n=3$ and $n=4$ branches remain positive over the plotted range.

This behavior reflects the localized nature of the Kaniadakis deformation. In the Schwarzschild case, the Chaplygin threshold changes smoothly from $n=1$ near the horizon to $n=2$ at large radius. The Barrow and Tsallis--Cirto sectors shift the near-horizon threshold through their respective power-law deformations, but the stable high-$n$ branches remain comparatively simple. Here, the hyperbolic functions introduce a more pronounced radial dependence: they can generate a stable window for $n=2$ close to the throat without preserving it farther away. Thus, the Kaniadakis correction affects not only the onset of stability but also the radial extent of the stable domain.

The right panel shows that the $n=4$ branch is robustly stable for all displayed values of $\kappa$. The Schwarzschild limit $\kappa=0$ already yields positive curvature, while nonzero $\kappa$ substantially increases the near-horizon restoring curvature. The curves decrease away from $\mathcal E=0$ because the hyperbolic correction becomes progressively suppressed with radius. As in the preceding stability diagrams, the apparent symmetry under $\mathcal E\to-\mathcal E$ results from the parametrization $a_0=r_h^{(K)}+|\mathcal E|$ and does not represent two independent static branches.

\subsection{Logarithmically corrected entropy}
\label{sec:log_entropy}

The logarithmically corrected entropy is given by
\begin{equation}
\mathcal S_{\log}(r)=\pi r^2+\lambda_{\log}\ln\left(\pi r^2\right).
\label{eq:log_entropy}
\end{equation}
Substituting Eq.~\eqref{eq:log_entropy} into Eq.~\eqref{eq:entropic_metric}, we obtain the logarithmic lapse function,
\begin{equation}
F_{\log}(r)=1-\frac{2\pi M r}{\lambda_{\log}+\pi r^2}.
\label{eq:log_entropy_lapse}
\end{equation}
For $0<\lambda_{\log}<\pi M^2$, the lapse has two positive Killing horizons,
\begin{equation}
r_{h,\pm}^{(\log)}=M\pm\sqrt{M^2-\frac{\lambda_{\log}}{\pi}}.
\label{eq:log_horizons}
\end{equation}
The lapse is positive in the intervals $0<r<r_{h,-}^{(\log)}$ and $r>r_{h,+}^{(\log)}$, whereas it is negative between the horizons. We construct the wormhole from two copies of the outer static branch, and therefore the throat must satisfy
\begin{equation}
a_0>r_{h,+}^{(\log)}.
\label{eq:log_throat_domain}
\end{equation}
The Schwarzschild limit is recovered when $\lambda_{\log}\to0$, for which $r_{h,-}^{(\log)}\to0$ and $r_{h,+}^{(\log)}\to2M$. At $\lambda_{\log}=\pi M^2$, the horizons coalesce at $r_h=M$, while $\lambda_{\log}>\pi M^2$ gives no real horizons. We consequently restrict the present analysis to $0\leq\lambda_{\log}<\pi M^2$. Negative values of $\lambda_{\log}$ are not considered, since the lapse develops a finite-radius pole at $r_{\rm s}=\sqrt{-\lambda_{\log}/\pi}$.

The static surface density and pressure are
\begin{equation}
\sig_{\log0}=-\frac{1}{2\pi a_0}\sqrt{1-\frac{2\pi M a_0}{\lambda_{\log}+\pi a_0^2}},
\label{eq:log_static_density}
\end{equation}
and
\begin{equation}
p_{\log0}=\frac{M\left(\pi a_0^2-\lambda_{\log}\right)}{4\left(\lambda_{\log}+\pi a_0^2\right)^2\sqrt{1-\dfrac{2\pi M a_0}{\lambda_{\log}+\pi a_0^2}}}+\frac{1}{4\pi a_0}\sqrt{1-\frac{2\pi M a_0}{\lambda_{\log}+\pi a_0^2}}.
\label{eq:log_static_pressure}
\end{equation}
Thus, the shell density is negative on the entire outer static branch, while the tangential pressure is positive. Near $r_{h,+}^{(\log)}$, the density vanishes from below and the pressure diverges; at larger radii, the logarithmic correction becomes subdominant and the standard asymptotic thin-shell behavior is recovered.

The static intrinsic NEC combination and the remaining trace SEC combination are
\begin{equation}
\left(\sig_0+p_0\right)_{\log}=\frac{\dfrac{2\pi M a_0\left(3\pi a_0^2+\lambda_{\log}\right)}{\left(\lambda_{\log}+\pi a_0^2\right)^2}-2}{8\pi a_0\sqrt{1-\dfrac{2\pi M a_0}{\lambda_{\log}+\pi a_0^2}}},
\label{eq:log_nec_explicit}
\end{equation}
and
\begin{equation}
\left(\sig_0+2p_0\right)_{\log}=\frac{M\left(\pi a_0^2-\lambda_{\log}\right)}{2\left(\lambda_{\log}+\pi a_0^2\right)^2\sqrt{1-\dfrac{2\pi M a_0}{\lambda_{\log}+\pi a_0^2}}}.
\label{eq:log_sec_explicit}
\end{equation}
The WEC is violated because $\sig_{\log0}<0$. On the outer branch, $\pi a_0^2>\lambda_{\log}$, so the remaining trace SEC combination remains positive. The intrinsic NEC is satisfied sufficiently close to the outer horizon and becomes negative as the throat is moved outward.

\begin{figure}[htp!]
    \centering
    \includegraphics[width=0.49\linewidth]{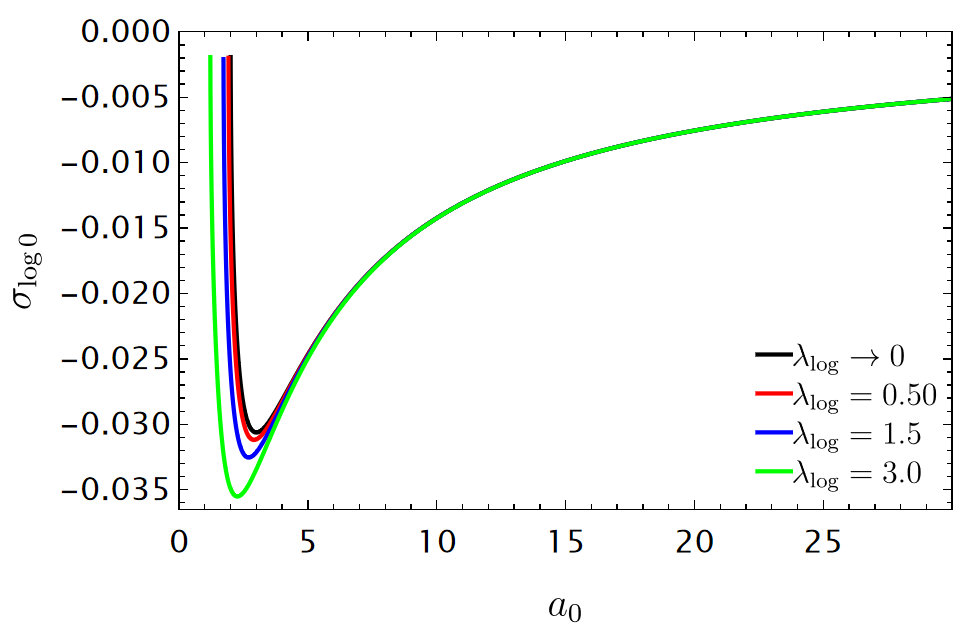}
    \includegraphics[width=0.49\linewidth]{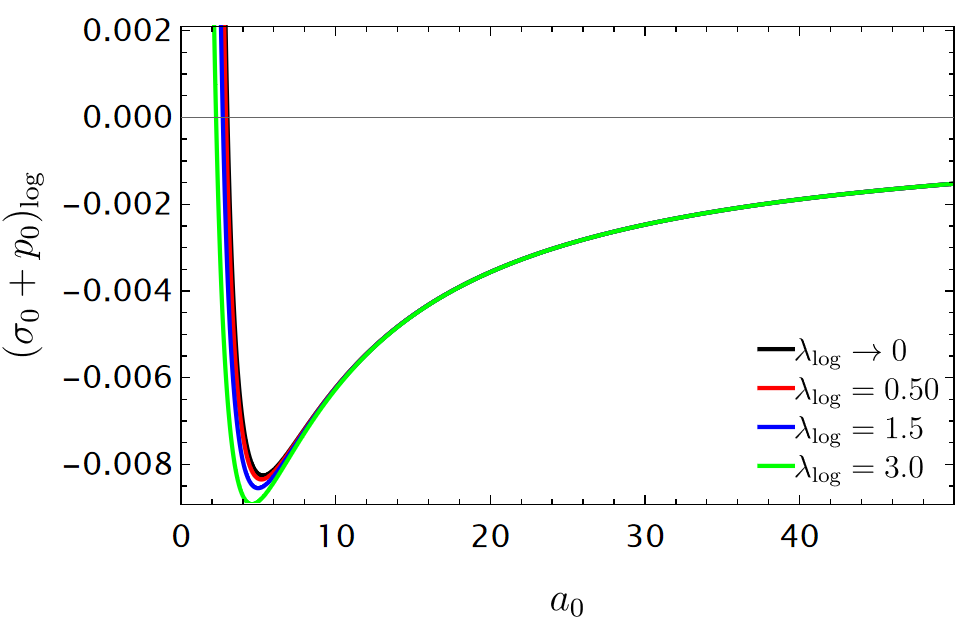}
    \includegraphics[width=0.5\linewidth]{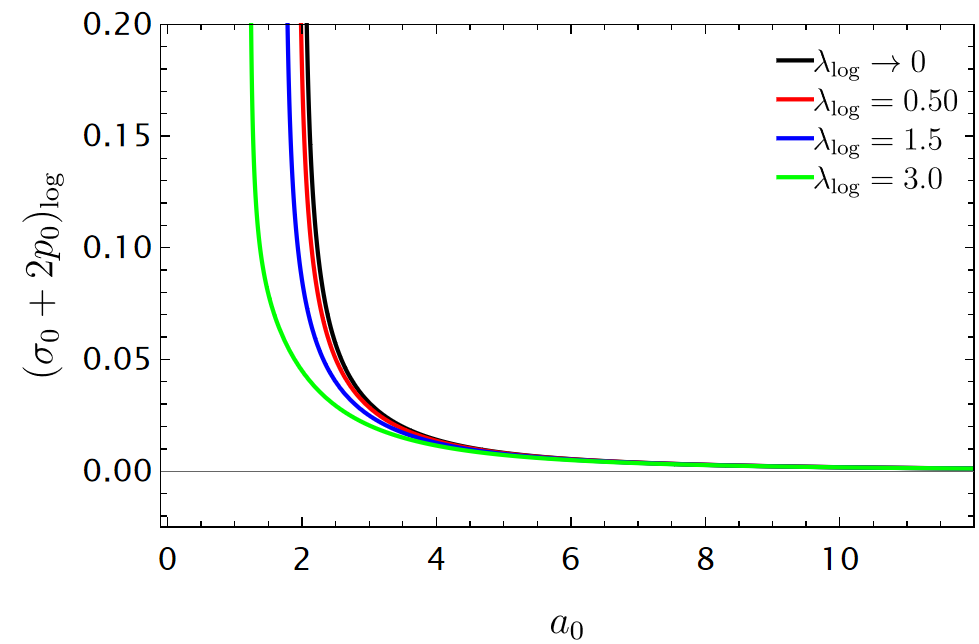}
    \caption{Static surface energy conditions for the logarithmic thin-shell wormhole as functions of the throat radius $a_0$. The upper-left panel shows the surface energy density $\sigma_{\log0}$, the upper-right panel displays the intrinsic shell NEC combination $\sigma_{\log0}+p_{\log0}$, and the lower panel shows the remaining trace SEC combination $\sigma_{\log0}+2p_{\log0}$. The curves correspond to $\lambda_{\log}\to0$, $0.50$, $1.5$, and $3.0$, with $M=1$, and are displayed on the outer static branch, $a_0>r_{h,+}^{(\log)}$.}
    \label{fig:eclog}
\end{figure}

Figure~\ref{fig:eclog} shows that the logarithmic correction primarily changes the near-horizon location and depth of the surface-stress profiles. The negative density reaches its minimum progressively closer to the throat as $\lambda_{\log}$ increases, while all curves approach the Schwarzschild-like behavior at large radius. The intrinsic NEC is positive close to the outer horizon because the pressure term dominates there, but its zero is displaced by the correction scale before the combination settles into a negative regime. The remaining trace SEC combination remains positive throughout the outer branch and falls rapidly to zero, consistently with the positive sign of $F'_{\log}$ in this region.

For the barotropic shell model, the equilibrium parameter and effective potential are
\begin{equation}
w_{\log0}=-\frac{1}{2}-\frac{\pi M a_0\left(\pi a_0^2-\lambda_{\log}\right)}{2\left(\lambda_{\log}+\pi a_0^2\right)^2\left[1-\dfrac{2\pi M a_0}{\lambda_{\log}+\pi a_0^2}\right]},
\label{eq:log_w0}
\end{equation}
and
\begin{equation}
V_{{\rm eff},\log}(a)=1-\frac{2\pi M a}{\lambda_{\log}+\pi a^2}-\left(1-\frac{2\pi M a_0}{\lambda_{\log}+\pi a_0^2}\right)\left(\frac{a_0}{a}\right)^{2+4w}.
\label{eq:log_potential}
\end{equation}
The equilibrium value $w=w_{\log0}$ ensures that $V_{{\rm eff},\log}(a_0)=V_{{\rm eff},\log}'(a_0)=0$. The resulting potential curvature is
\begin{equation}
V_{{\rm eff},\log}''(a_0)=-\frac{4\pi^2 M a_0\left(\pi a_0^2-3\lambda_{\log}\right)}{\left(\lambda_{\log}+\pi a_0^2\right)^3}+\frac{2\pi M\left(\pi a_0^2-\lambda_{\log}\right)}{a_0\left(\lambda_{\log}+\pi a_0^2\right)^2}-\frac{4\pi^2M^2\left(\pi a_0^2-\lambda_{\log}\right)^2}{\left(\lambda_{\log}+\pi a_0^2\right)^4\left[1-\dfrac{2\pi M a_0}{\lambda_{\log}+\pi a_0^2}\right]}.
\label{eq:log_barotropic_stability}
\end{equation}
For the outer static branch and the parameter range considered here, $V_{{\rm eff},\log}''(a_0)<0$ throughout the admissible domain. Hence, the logarithmic deformation changes the scale of the instability but does not stabilize a constant-barotropic shell.

\begin{figure}[htp!]
    \centering
    \includegraphics[width=0.49\linewidth]{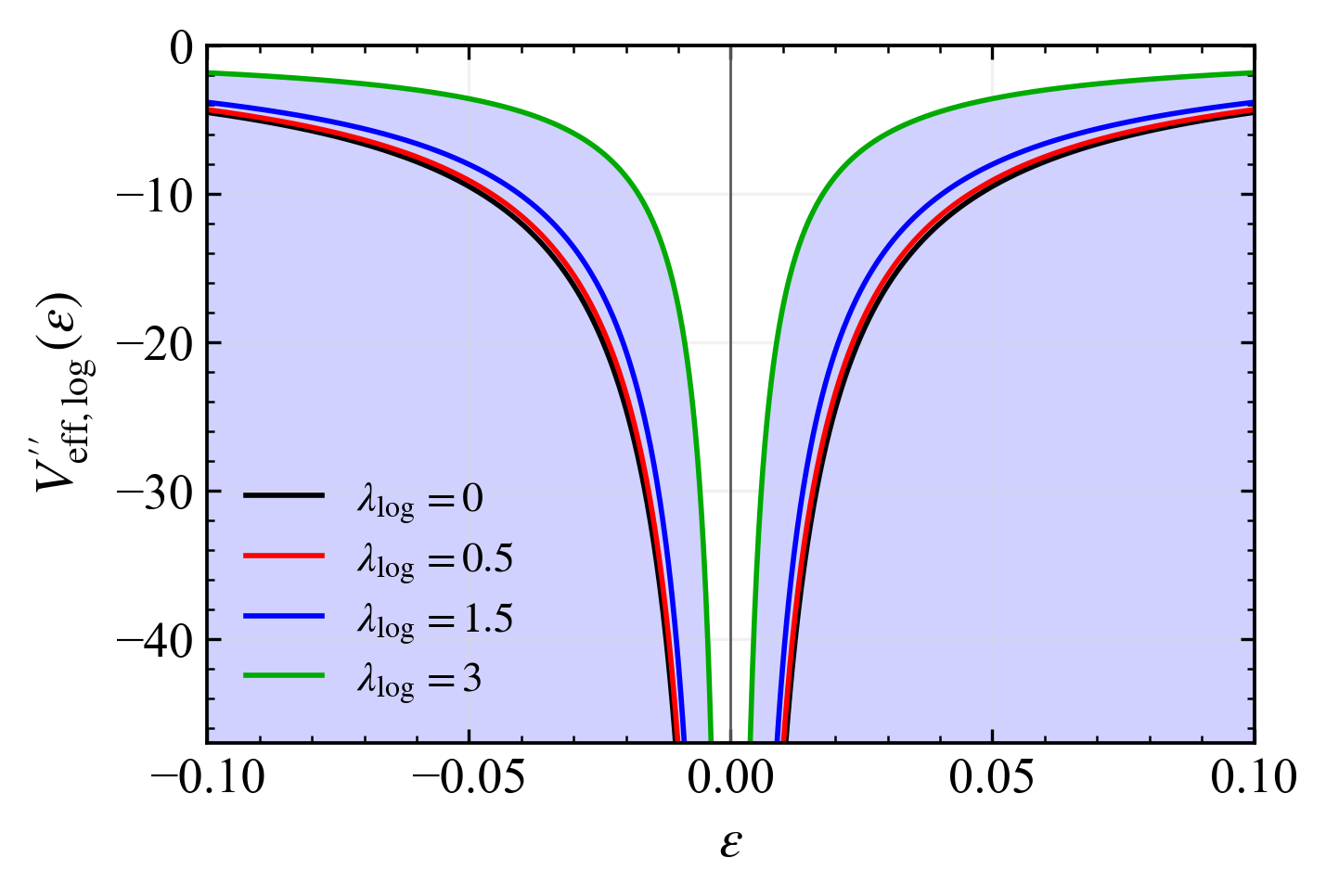}
    \includegraphics[width=0.49\linewidth]{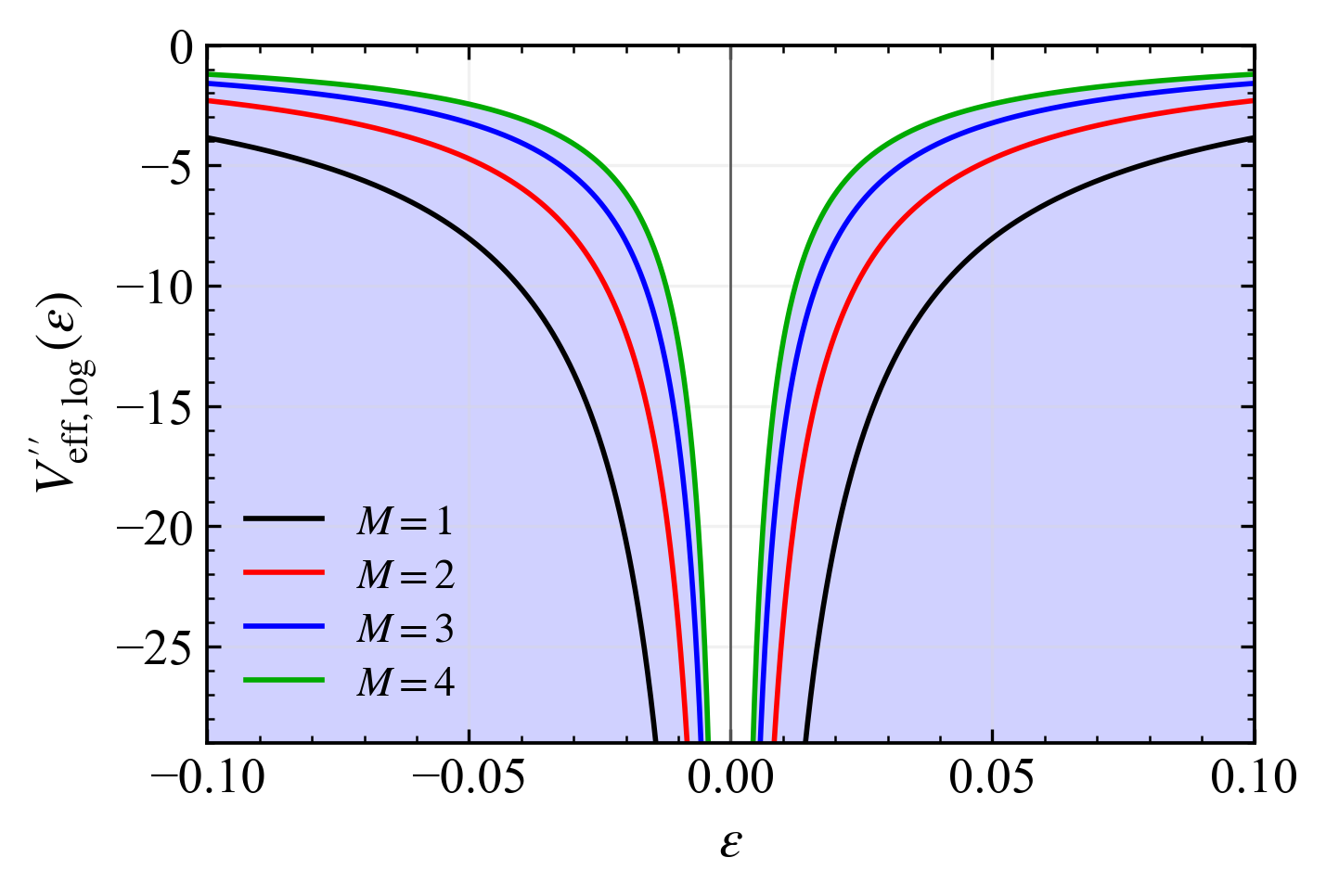}
    \caption{Effective-potential curvature $V_{{\rm eff},\log}''(\mathcal E)$ for the barotropic logarithmic thin-shell wormhole, with $a_0=r_{h,+}^{(\log)}+|\mathcal E|$. In the left panel, $M=1$ and the curves correspond to $\lambda_{\log}\to0$, $0.50$, $1.5$, and $3.0$. In the right panel, $\lambda_{\log}=1.5$ and the curves correspond to $M=1$, $2$, $3$, and $4$. The curves and shaded regions are displayed only on the outer static branch, $a_0>r_{h,+}^{(\log)}$, where $F_{\log}(a_0)>0$. In both cases, $V_{{\rm eff},\log}''(\mathcal E)<0$ throughout the admissible domain, indicating linear instability.}
    \label{fig:stblog}
\end{figure}

Figure~\ref{fig:stblog} shows that the logarithmic lapse retains the same barotropic stability character found in the preceding models. The potential curvature becomes sharply negative when the shell is brought close to the outer horizon and relaxes toward zero from below as the throat is moved outward. Varying $\lambda_{\log}$ changes the horizon position and the depth of the near-horizon instability, while changing $M$ chiefly rescales the curves. Thus, similarly to the Schwarzschild, Barrow, Tsallis--Cirto, R\'enyi, and Kaniadakis cases, no stable interval emerges for a constant barotropic equation of state.

For the variable Chaplygin shell model, the static parameter is
\begin{equation}
\Omega_{\log0}=-\frac{a_0^{n-2}}{8\pi^2}\left[1-\frac{\pi M a_0\left(\pi a_0^2+3\lambda_{\log}\right)}{\left(\lambda_{\log}+\pi a_0^2\right)^2}\right].
\label{eq:log_omega_chaplygin}
\end{equation}
For $n\neq4$, the effective potential becomes
\begin{equation}
V_{{\rm eff},\log}^{\rm (C)}(a)=1-\frac{2\pi M a}{\lambda_{\log}+\pi a^2}-\left(1-\frac{2\pi M a_0}{\lambda_{\log}+\pi a_0^2}\right)\left(\frac{a_0}{a}\right)^2-\frac{2a_0^{n-2}}{4-n}\left[1-\frac{\pi M a_0\left(\pi a_0^2+3\lambda_{\log}\right)}{\left(\lambda_{\log}+\pi a_0^2\right)^2}\right]\left[a^{2-n}-\frac{a_0^{4-n}}{a^2}\right].
\label{eq:log_potential_chaplygin}
\end{equation}
For $n=4$, the logarithmic branch takes the form
\begin{equation}
V_{{\rm eff},\log}^{\rm (C)}(a)=1-\frac{2\pi M a}{\lambda_{\log}+\pi a^2}-\left(1-\frac{2\pi M a_0}{\lambda_{\log}+\pi a_0^2}\right)\left(\frac{a_0}{a}\right)^2-\frac{2a_0^2}{a^2}\left[1-\frac{\pi M a_0\left(\pi a_0^2+3\lambda_{\log}\right)}{\left(\lambda_{\log}+\pi a_0^2\right)^2}\right]\ln\left(\frac{a}{a_0}\right).
\label{eq:log_potential_chaplygin_n4}
\end{equation}
The value in Eq.~\eqref{eq:log_omega_chaplygin} ensures that $V_{{\rm eff},\log}^{\rm (C)}(a_0)=V_{{\rm eff},\log}^{{'}\,\rm (C)}(a_0)=0$. The corresponding curvature is
\begin{equation}
V_{{\rm eff},\log}^{{''}\,\rm (C)}(a_0)=-\frac{4\pi^2 M a_0\left(\pi a_0^2-3\lambda_{\log}\right)}{\left(\lambda_{\log}+\pi a_0^2\right)^3}+\frac{2\pi M(n+1)\left(\pi a_0^2-\lambda_{\log}\right)}{a_0\left(\lambda_{\log}+\pi a_0^2\right)^2}+\frac{2(n-2)}{a_0^2}\left(1-\frac{2\pi M a_0}{\lambda_{\log}+\pi a_0^2}\right).
\label{eq:log_chaplygin_stability}
\end{equation}
The variable Chaplygin shell is linearly stable whenever $V_{{\rm eff},\log}^{{''}\,\rm (C)}(a_0)>0$. In particular, for $n=4$,
\begin{equation}
V_{{\rm eff},\log}^{{''}\,\rm (C)}(a_0)=-\frac{4\pi^2 M a_0\left(\pi a_0^2-3\lambda_{\log}\right)}{\left(\lambda_{\log}+\pi a_0^2\right)^3}+\frac{10\pi M\left(\pi a_0^2-\lambda_{\log}\right)}{a_0\left(\lambda_{\log}+\pi a_0^2\right)^2}+\frac{4}{a_0^2}\left(1-\frac{2\pi M a_0}{\lambda_{\log}+\pi a_0^2}\right).
\label{eq:log_chaplygin_stability_n4}
\end{equation}

\begin{figure}[htp!]
    \centering
    \includegraphics[width=1.0\linewidth]{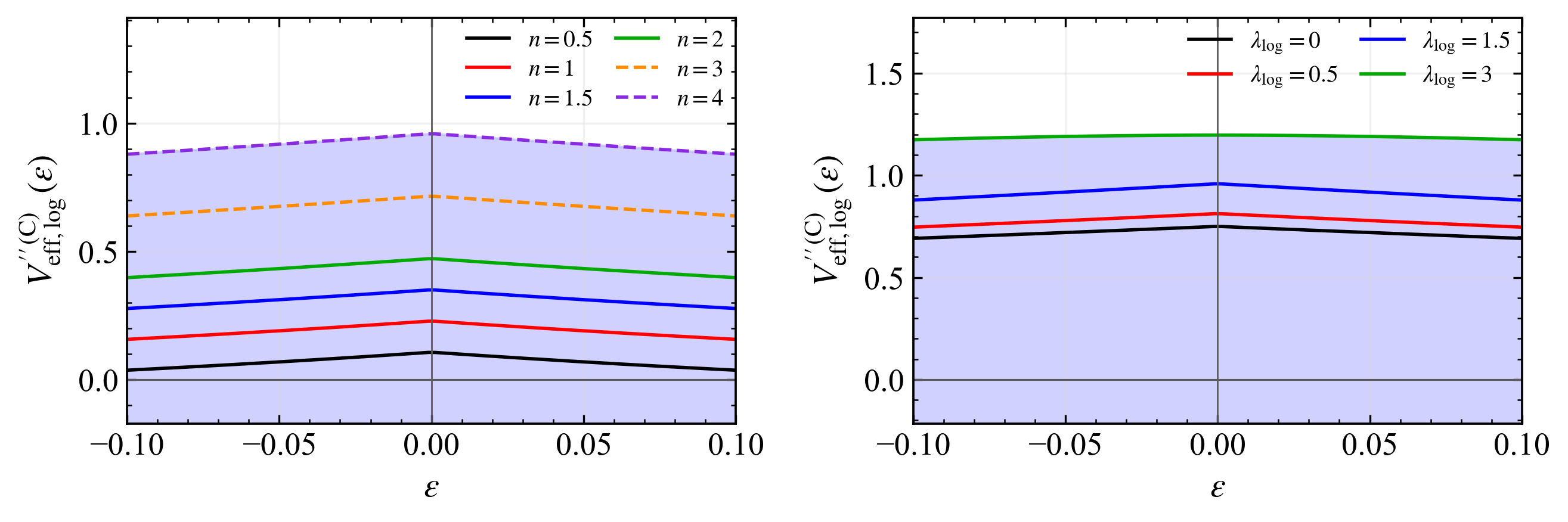}
    \caption{Effective-potential curvature $V_{{\rm eff},\log}^{{''}\,\rm (C)}(\mathcal E)$ for the variable Chaplygin logarithmic thin-shell wormhole, with $a_0=r_{h,+}^{(\log)}+|\mathcal E|$. In the left panel, $M=1$ and $\lambda_{\log}=1.5$, and the curves correspond to $n=0.5$, $1$, $1.5$, $2$, $3$, and $4$. In the right panel, $M=1$ and $n=4$, and the curves correspond to $\lambda_{\log}=0$, $0.5$, $1.5$, and $3$. The Chaplygin parameter $\Omega$ is fixed by the static equilibrium condition in each case. Positive regions of the curvature correspond to linear radial stability.}
    \label{fig:stbchlog}
\end{figure}

Figure~\ref{fig:stbchlog} highlights a distinctive feature of the logarithmic deformation. For $M=1$ and $\lambda_{\log}=1.5$, all the values $n=0.5$, $1$, $1.5$, $2$, $3$, and $4$ shown in the left panel have positive curvature in the near-horizon interval displayed. This is unlike the Schwarzschild, Barrow, Tsallis--Cirto, and Kaniadakis examples, where the low-$n$ branches remain unstable near the selected horizon for the corresponding parameter choices. The logarithmic term lowers the local stability threshold: at the outer horizon of the left panel, Eq.~\eqref{eq:log_chaplygin_stability} gives a critical value $n\simeq0.06$, so even the $n=0.5$ branch is locally stable.

This near-horizon stabilization does not remove the large-radius distinction between the Chaplygin branches. Since Eq.~\eqref{eq:log_chaplygin_stability} approaches $2(n-2)/a_0^2$ as $a_0$ increases, the curves with $n<2$ eventually cross into an unstable regime, although those transitions lie outside the narrow interval shown in the left panel. The $n=2$ branch tends toward marginality at large radius, whereas the $n=3$ and $n=4$ branches remain stable. The behavior therefore differs from the R\'enyi case, where the outer horizon truncates the static region, and from the Kaniadakis case, where the localized hyperbolic correction can restrict the stable window even for intermediate values of $n$.

The right panel shows that the stable $n=4$ branch persists from the Schwarzschild limit $\lambda_{\log}=0$ to the largest displayed logarithmic correction. Increasing $\lambda_{\log}$ raises the local curvature near the horizon, indicating a stronger restoring response, while all curves recover the same qualitative decay away from the throat.

\subsection{Loop-quantum-gravity-inspired entropy}
\label{sec:lqg_entropy}

We consider the normalized loop-quantum-gravity-inspired entropy
\begin{equation}
\mathcal S_L(r)=\frac{\exp\!\left[(1-q)\alpha_L\pi r^2\right]-1}{1-q},
\qquad
F_L(r)=1-\frac{4\pi M}{\mathcal S'_L(r)}=1-\frac{2M}{\alpha_L r}\exp\!\left[(q-1)\alpha_L\pi r^2\right],
\label{eq:lqg_entropy_lapse}
\end{equation}
where $q$ is the nonextensive parameter and $\alpha_L$ specifies the normalization inherited from the underlying area spectrum. Its radial derivative is
\begin{equation}
F'_L(r)=\frac{2M}{\alpha_L}\exp\!\left[(q-1)\alpha_L\pi r^2\right]\left[\frac{1}{r^2}-2\pi\alpha_L(q-1)\right].
\label{eq:lqg_lapse_prime}
\end{equation}

We assume $\alpha_L>0$, which guarantees $\mathcal S'_L(r)>0$ for $r>0$. Since $\alpha_L$ introduces a simultaneous rescaling of the Schwarzschild mass scale, we set $\alpha_L=1$ in the qualitative discussion below. With this normalization, the limit $q\to1$ yields
\begin{equation}
F_L(r)\longrightarrow1-\frac{2M}{r},
\end{equation}
thereby recovering the Schwarzschild geometry.

For $q\neq1$, the horizon radii are conveniently written as
\begin{equation}
r_{h,k}^{(L)}=\left[-\frac{W_k\!\left(-\dfrac{8\pi(q-1)M^2}{\alpha_L}\right)}{2\pi\alpha_L(q-1)}\right]^{1/2},
\label{eq:lqg_horizons}
\end{equation}
where $W_k(z)$ is the Lambert function, defined by $W_k(z)\exp[W_k(z)]=z$, and $k$ labels its branches. For $q=1$, the single horizon is located at
\begin{equation}
r_h^{(L)}=\frac{2M}{\alpha_L}.
\end{equation}

The LQG-inspired geometry naturally separates into two qualitatively distinct sectors. For $q<1$, only the principal branch $W_0$ contributes to Eq.~\eqref{eq:lqg_horizons}, yielding a unique positive horizon. Moreover, $F'_L(r)>0$ for all $r>0$, so that the lapse increases monotonically from $-\infty$ near the origin to unity at spatial infinity. The corresponding thin-shell construction is therefore defined on the exterior static region
\begin{equation}
a_0>r_h^{(L)}.
\end{equation}
Writing $\kappa=(1-q)\alpha_L$, the lapse becomes
\begin{equation}
F_L(r)=1-\frac{2M}{\alpha_L r}\exp\!\left(-\kappa\pi r^2\right).
\end{equation}
Hence, the correction to the Schwarzschild term is exponentially suppressed at large radius, as in the Kaniadakis geometry. Although the two lapse functions are not identical, they share the same single-horizon Schwarzschild-like causal structure and the same type of static exterior region.

Conversely, for $q>1$, the lapse is negative both near the origin and at large radius. A static region can then occur only between two horizons. Two distinct positive roots exist provided that
\begin{equation}
1<q<1+\frac{\alpha_L}{8\pi e M^2},
\label{eq:lqg_q_bound}
\end{equation}
with the branches $W_0$ and $W_{-1}$ determining the inner and outer horizons, respectively. At the limiting value in Eq.~\eqref{eq:lqg_q_bound}, the two roots merge into a degenerate horizon. For $q>1$ sufficiently close to unity, the lapse admits the expansion
\begin{equation}
F_L(r)=1-\frac{2M}{\alpha_L r}-2\pi M(q-1)r+\mathcal O\!\left[(q-1)^2r^3\right].
\label{eq:lqg_renyi_expansion}
\end{equation}
For $\alpha_L=1$, this reproduces the R\'enyi lapse at first order under the local identification $\lambda_R\simeq q-1$. Accordingly, the $q>1$ branch displays the same two-horizon structure found in the positive-parameter R\'enyi sector, with the admissible static interval given by
\begin{equation}
r_{h,-}^{(L)}<a_0<r_{h,+}^{(L)}.
\end{equation}

The LQG-inspired entropy does not introduce a qualitatively new horizon structure within the cases considered here: depending on the sign of $q$, its positive-lapse domain follows the single-horizon or two-horizon pattern already illustrated by the Kaniadakis and R\'enyi sectors, respectively. Its $q<1$ sector is Kaniadakis-like, with a single horizon and an asymptotically flat static exterior, whereas its $q>1$ sector is R\'enyi-like, with a finite static region bounded by two horizons. Our numerical evaluation of the surface energy conditions and of $V_{\rm eff}''(a_0)$ confirms that the corresponding qualitative behavior follows that already obtained for these two representative cases. We therefore do not present separate energy-condition and stability figures for the LQG-inspired entropy.

\subsection{Exponentially corrected entropy}
\label{sec:exp_entropy}

The exponentially corrected entropy is given by
\begin{equation}
\mathcal S_{\exp}(r)=\pi r^2+\eta\exp\left(-\pi r^2\right).
\label{eq:exp_entropy}
\end{equation}
Substituting Eq.~\eqref{eq:exp_entropy} into Eq.~\eqref{eq:entropic_metric}, we obtain the exponentially corrected lapse function,
\begin{equation}
F_{\exp}(r)=1-\frac{2M}{r\left[1-\eta\exp\left(-\pi r^2\right)\right]}.
\label{eq:exp_entropy_lapse}
\end{equation}
Exponentially suppressed entropy contributions can be regarded as phenomenological quantum-inspired or semiclassical corrections that are relevant only when the horizon scale approaches the microscopic scale implicit in the deformation parameter \cite{ChatterjeeGhosh2020}. We restrict the analysis to $0\leq\eta\leq1$, for which $1-\eta\exp(-\pi r^2)>0$ at every finite $r>0$, and no additional finite-radius pole appears in the bulk geometry. The derivative of the lapse is
\begin{equation}
F'_{\exp}(r)=
\frac{
2M\left[
1+\left(2\pi r^2-1\right)
\eta\exp\left(-\pi r^2\right)
\right]
}
{
r^2\left[
1-\eta\exp\left(-\pi r^2\right)
\right]^2
}.
\label{eq:exp_lapse_prime}
\end{equation}
For the parameter range considered here, $F'_{\exp}(r)>0$ for $r>0$, so the lapse is monotonic and possesses a unique positive Killing horizon. Its radius is determined implicitly by
\begin{equation}
r_h^{(\exp)}
\left[
1-\eta\exp\left(-\pi[r_h^{(\exp)}]^2\right)
\right]
=
2M.
\label{eq:exp_horizon}
\end{equation}
The thin-shell construction is therefore performed on the exterior static region,
\begin{equation}
a_0>r_h^{(\exp)}.
\label{eq:exp_throat_domain}
\end{equation}
The Schwarzschild geometry is recovered for $\eta=0$, with $r_h^{(\exp)}=2M$. For $\eta>0$, the horizon is displaced outward, whereas the correction rapidly vanishes for large radii. In particular, when $\eta\exp(-4\pi M^2)\ll1$, one has $r_h^{(\exp)}\simeq2M+2M\eta\exp(-4\pi M^2)$. Consequently, the deformation is negligible for macroscopic masses and becomes physically relevant in the small-mass regime, where the horizon can probe the scale at which the exponential entropy correction is appreciable. This interpretation concerns the effective bulk geometry; the shell dynamics itself remains described classically through the Darmois--Israel formalism.

The static surface density and pressure are
\begin{equation}
\sig_{\exp0}
=
-\frac{1}{2\pi a_0}
\sqrt{
1-\frac{2M}
{a_0\left[
1-\eta\exp\left(-\pi a_0^2\right)
\right]}
}.
\label{eq:exp_static_density}
\end{equation}
and
\begin{equation}
\begin{aligned}
p_{\exp0}
={}&
\frac{
M\left[
1+\left(2\pi a_0^2-1\right)
\eta\exp\left(-\pi a_0^2\right)
\right]
}
{
4\pi a_0^2
\left[
1-\eta\exp\left(-\pi a_0^2\right)
\right]^2
\sqrt{
1-\dfrac{2M}
{a_0\left[
1-\eta\exp\left(-\pi a_0^2\right)
\right]}
}
}
+
\frac{1}{4\pi a_0}
\sqrt{
1-\frac{2M}
{a_0\left[
1-\eta\exp\left(-\pi a_0^2\right)
\right]}
}.
\end{aligned}
\label{eq:exp_static_pressure}
\end{equation}
The density is negative throughout the admissible exterior region, whereas the tangential pressure is positive. When the throat approaches the horizon, $\sig_{\exp0}$ tends to zero from below and $p_{\exp0}$ diverges. The exponential deformation modifies these quantities only in a narrow region around the horizon, since every correction proportional to $\eta\exp(-\pi a_0^2)$ is rapidly suppressed as the shell is moved outward.

The static intrinsic NEC combination and the remaining trace SEC combination are
\begin{equation}
\begin{aligned}
\left(\sig_0+p_0\right)_{\exp}
={}&
\frac{
\dfrac{
2M\left[
1+\left(2\pi a_0^2-1\right)
\eta\exp\left(-\pi a_0^2\right)
\right]
}
{
a_0\left[
1-\eta\exp\left(-\pi a_0^2\right)
\right]^2
}
-
2\left[
1-\dfrac{2M}
{a_0\left[
1-\eta\exp\left(-\pi a_0^2\right)
\right]}
\right]
}
{
8\pi a_0
\sqrt{
1-\dfrac{2M}
{a_0\left[
1-\eta\exp\left(-\pi a_0^2\right)
\right]}
}
}.
\end{aligned}
\label{eq:exp_nec_explicit}
\end{equation}
and
\begin{equation}
\begin{aligned}
\left(\sig_0+2p_0\right)_{\exp}
={}&
\frac{
M\left[
1+\left(2\pi a_0^2-1\right)
\eta\exp\left(-\pi a_0^2\right)
\right]
}
{
2\pi a_0^2
\left[
1-\eta\exp\left(-\pi a_0^2\right)
\right]^2
\sqrt{
1-\dfrac{2M}
{a_0\left[
1-\eta\exp\left(-\pi a_0^2\right)
\right]}
}
}.
\end{aligned}
\label{eq:exp_sec_explicit}
\end{equation}
Since $\sig_{\exp0}<0$, the WEC is violated throughout the static branch. The remaining trace SEC combination is positive because the lapse is monotonic, while the intrinsic NEC is satisfied only near the horizon, where the pressure contribution dominates, and becomes negative at larger throat radii.

\begin{figure}[htp!]
    \centering
    \includegraphics[width=0.49\linewidth]{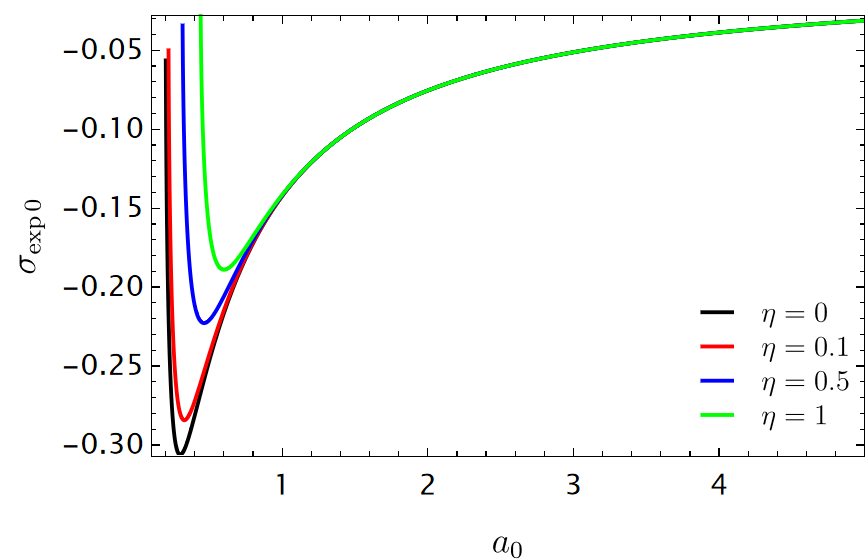}
    \includegraphics[width=0.49\linewidth]{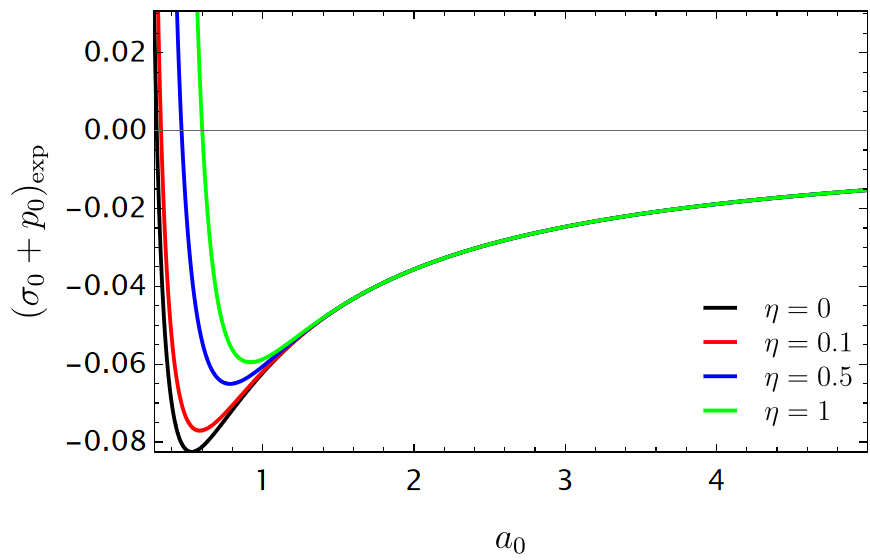}
    \includegraphics[width=0.5\linewidth]{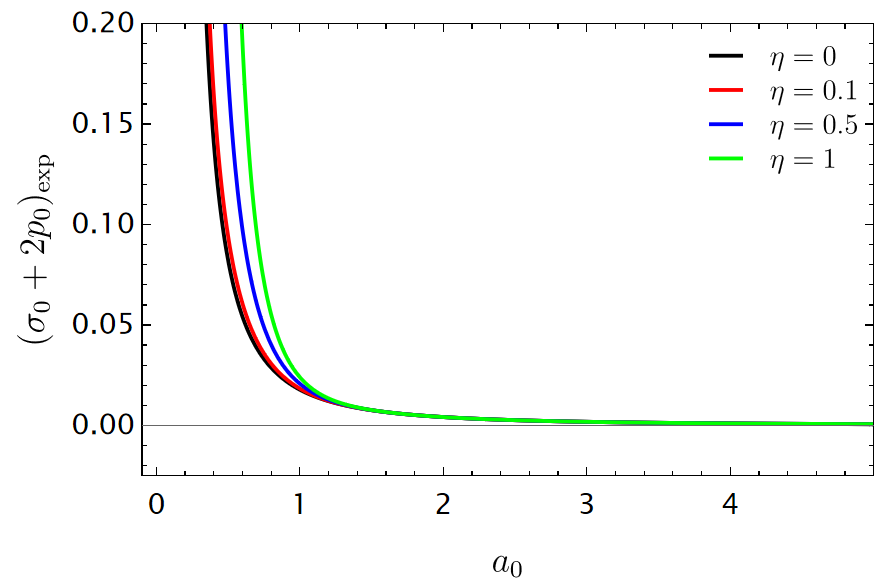}
    \caption{Static surface energy conditions for the exponentially corrected thin-shell wormhole as functions of the throat radius $a_0$. The upper-left panel shows the surface energy density $\sigma_{\exp0}$, the upper-right panel displays the intrinsic shell NEC combination $\sigma_{\exp0}+p_{\exp0}$, and the lower panel shows the remaining trace SEC combination $\sigma_{\exp0}+2p_{\exp0}$. The curves correspond to $\eta=0$, $0.1$, $0.5$, and $1$, with $M=0.1$, and are shown only in their respective admissible domains, $a_0>r_h^{(\exp)}$.}
    \label{fig:ecexp}
\end{figure}

Figure~\ref{fig:ecexp} shows how the entropy deformation redistributes the shell stresses in the small-mass regime. When $\eta$ increases, the horizon moves outward and the negative density minimum becomes less pronounced, indicating that the exponential correction softens the amount of negative surface energy required at the most demanding part of the exterior branch. The same trend appears in the intrinsic NEC combination, whose negative trough becomes shallower as the correction is strengthened. This effect should not be interpreted as a removal of exotic matter, since the density remains negative, but rather as a local renormalization of the stress scale near a small horizon. The rapid convergence of the curves at larger $a_0$ reflects the exponential suppression of the correction and the recovery of the Schwarzschild-like regime.

For the barotropic shell model, the equilibrium parameter and effective potential are
\begin{equation}
\begin{aligned}
w_{\exp0}
={}&
-\frac{1}{2}
-
\frac{
M\left[
1+\left(2\pi a_0^2-1\right)
\eta\exp\left(-\pi a_0^2\right)
\right]
}
{
2a_0
\left[
1-\eta\exp\left(-\pi a_0^2\right)
\right]^2
\left[
1-\dfrac{2M}
{a_0\left[
1-\eta\exp\left(-\pi a_0^2\right)
\right]}
\right]
}.
\end{aligned}
\label{eq:exp_w0}
\end{equation}
and
\begin{equation}
\begin{aligned}
V_{{\rm eff},\exp}(a)
={}&
1-\frac{2M}
{a\left[
1-\eta\exp\left(-\pi a^2\right)
\right]}
-
\left[
1-\frac{2M}
{a_0\left[
1-\eta\exp\left(-\pi a_0^2\right)
\right]}
\right]
\left(\frac{a_0}{a}\right)^{2+4w}.
\end{aligned}
\label{eq:exp_potential}
\end{equation}
The equilibrium value $w=w_{\exp0}$ ensures that $V_{{\rm eff},\exp}(a_0)=V_{{\rm eff},\exp}'(a_0)=0$. The corresponding potential curvature is
\begin{equation}
\begin{aligned}
V_{{\rm eff},\exp}''(a_0)
={}&
\frac{
2M
}{
a_0^3
\left[
1-\eta\exp\left(-\pi a_0^2\right)
\right]^3
}
\Bigg\{
2\pi a_0^2
\eta\exp\left(-\pi a_0^2\right)
\left[
1-\eta\exp\left(-\pi a_0^2\right)
\right]
\left(3-2\pi a_0^2\right)
\\
&-
\left[
1-\eta\exp\left(-\pi a_0^2\right)
\right]
\left[
1+\left(2\pi a_0^2-1\right)
\eta\exp\left(-\pi a_0^2\right)
\right]
\\
&-
4\pi a_0^2
\eta\exp\left(-\pi a_0^2\right)
\left[
1+\left(2\pi a_0^2-1\right)
\eta\exp\left(-\pi a_0^2\right)
\right]
\Bigg\}
\\
&-
\frac{
4M^2
\left[
1+\left(2\pi a_0^2-1\right)
\eta\exp\left(-\pi a_0^2\right)
\right]^2
}
{
a_0^4
\left[
1-\eta\exp\left(-\pi a_0^2\right)
\right]^4
\left[
1-\dfrac{2M}
{a_0\left[
1-\eta\exp\left(-\pi a_0^2\right)
\right]}
\right]
}.
\end{aligned}
\label{eq:exp_barotropic_stability}
\end{equation}
For the parameter range considered here, $V_{{\rm eff},\exp}''(a_0)<0$ throughout the admissible domain. Thus, the exponential modification of the seed geometry does not by itself stabilize a shell with a constant barotropic response.

\begin{figure}[htp!]
    \centering
    \includegraphics[width=0.49\linewidth]{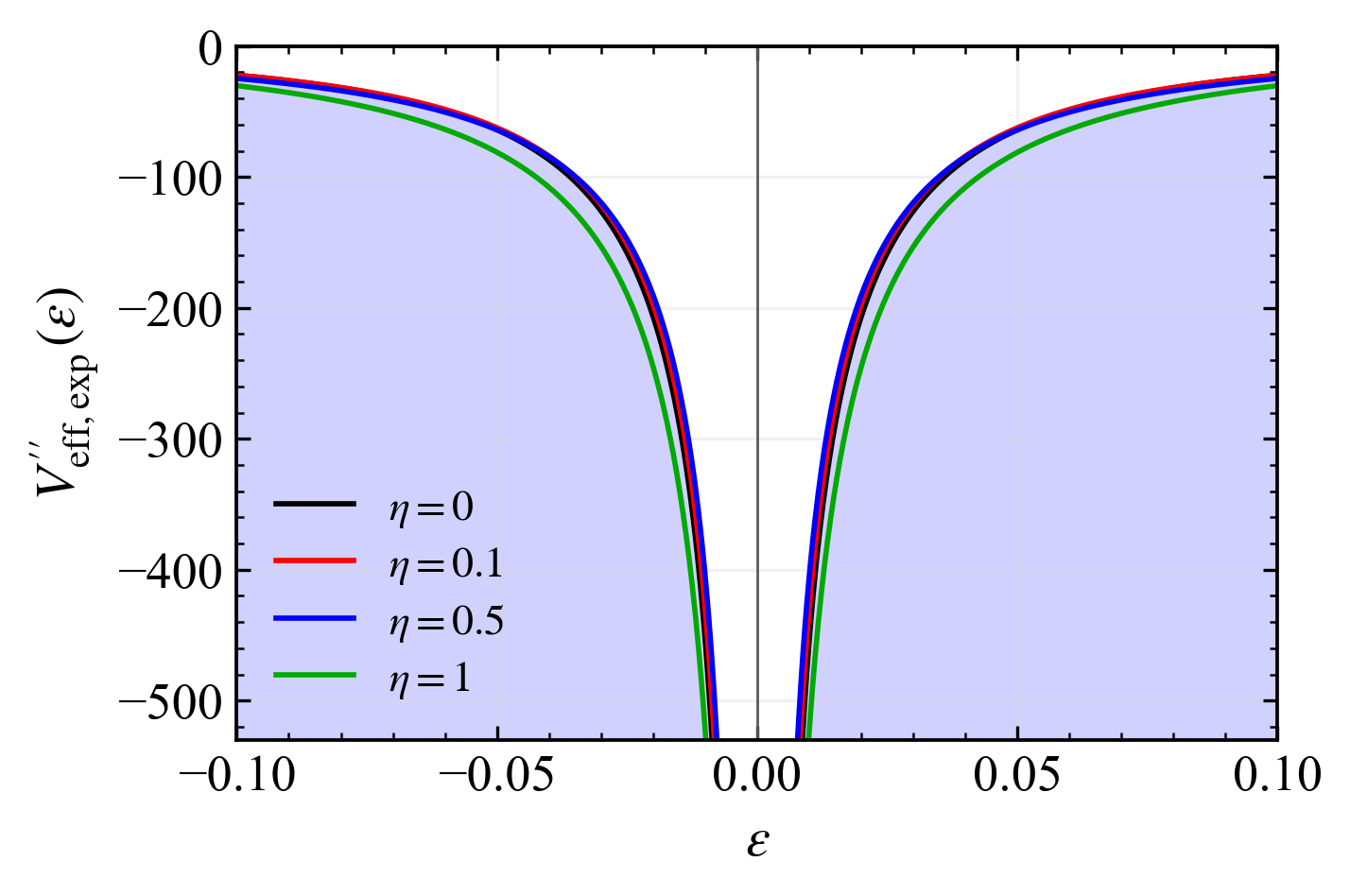}
    \includegraphics[width=0.49\linewidth]{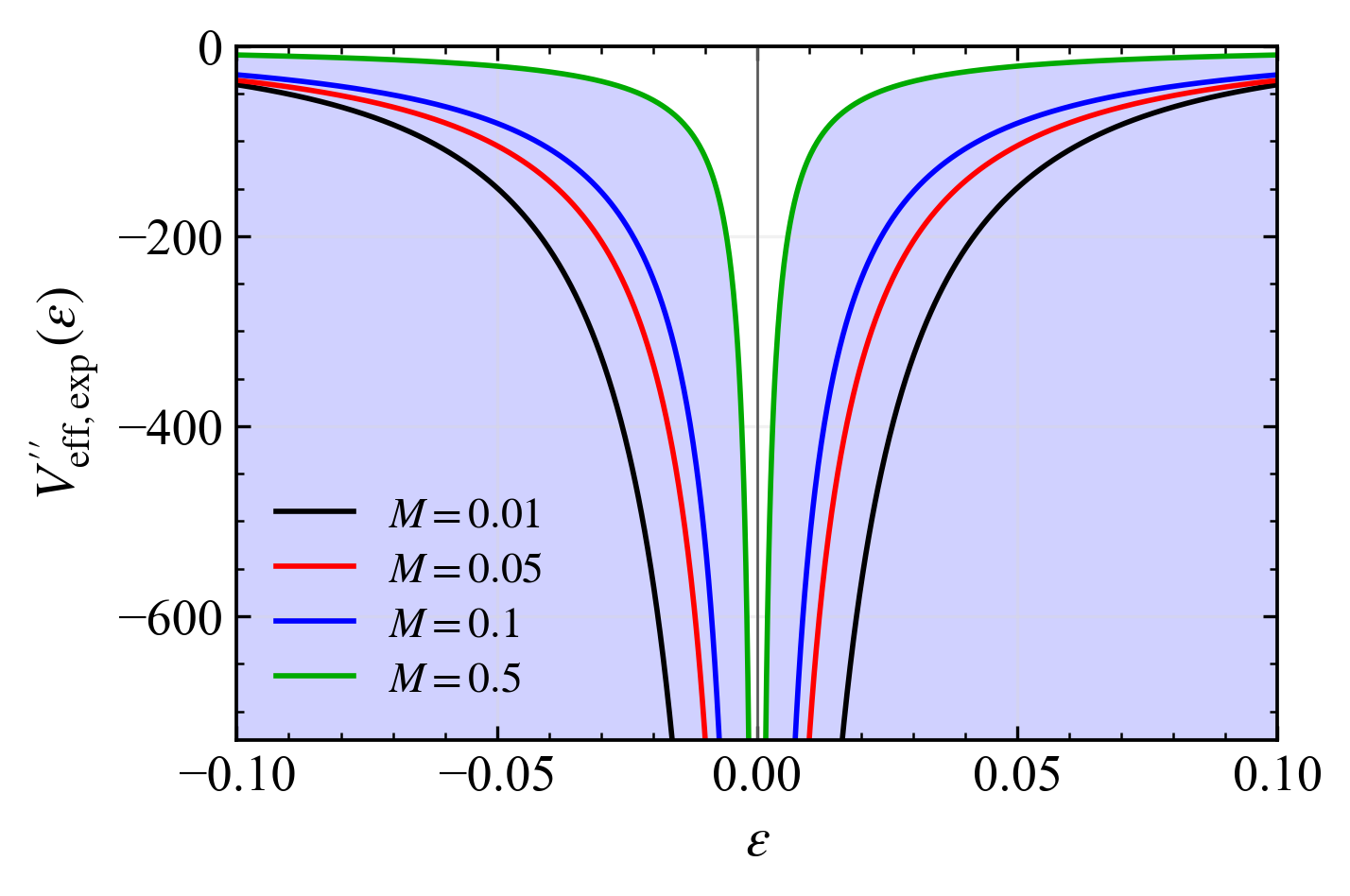}
    \caption{Effective-potential curvature $V_{{\rm eff},\exp}''(\mathcal E)$ for the barotropic exponentially corrected thin-shell wormhole, with $a_0=r_h^{(\exp)}+|\mathcal E|$. In the left panel, $M=0.1$ and the curves correspond to $\eta=0$, $0.1$, $0.5$, and $1$. In the right panel, $\eta=1$ and the curves correspond to $M=0.01$, $0.05$, $0.1$, and $0.5$. The curves and shaded regions are displayed only in the exterior static domain, $a_0>r_h^{(\exp)}$, where $F_{\exp}(a_0)>0$. In both cases, $V_{{\rm eff},\exp}''(\mathcal E)<0$ throughout the admissible domain, indicating linear instability.}
    \label{fig:stbexp}
\end{figure}

Figure~\ref{fig:stbexp} separates two physical effects. The left panel shows that changing $\eta$ modifies the local geometry close to the horizon, but every barotropic branch remains unstable because the potential curvature never becomes positive. The right panel exposes the scale dependence more directly. For small masses, the horizon lies in the region where $\eta\exp(-\pi r^2)$ is still appreciable, and the deformation visibly reshapes the instability curve. As the mass grows, the horizon probes larger radii, where the exponential factor is strongly suppressed and the geometry becomes practically indistinguishable from its classical counterpart. Therefore, the figure supports a semiclassical interpretation in which the correction is relevant for small black holes but has no macroscopic effect on the barotropic stability conclusion.

For the variable Chaplygin shell model, the static equation-of-state parameter is
\begin{equation}
\begin{aligned}
\Omega_{\exp0}
={}&
-\frac{a_0^{n-2}}{8\pi^2}
\Bigg[
1-
\frac{
M\left\{
1-\left(2\pi a_0^2+1\right)
\eta\exp\left(-\pi a_0^2\right)
\right\}
}
{
a_0
\left[
1-\eta\exp\left(-\pi a_0^2\right)
\right]^2
}
\Bigg].
\end{aligned}
\label{eq:exp_omega_chaplygin}
\end{equation}
For $n\neq4$, the effective potential becomes
\begin{equation}
\begin{aligned}
V_{{\rm eff},\exp}^{\rm (C)}(a)
={}&
1-\frac{2M}
{a\left[
1-\eta\exp\left(-\pi a^2\right)
\right]}
\\
&-
\left[
1-\frac{2M}
{a_0\left[
1-\eta\exp\left(-\pi a_0^2\right)
\right]}
\right]
\left(\frac{a_0}{a}\right)^2
\\
&-
\frac{2a_0^{n-2}}{4-n}
\Bigg[
1-
\frac{
M\left\{
1-\left(2\pi a_0^2+1\right)
\eta\exp\left(-\pi a_0^2\right)
\right\}
}
{
a_0
\left[
1-\eta\exp\left(-\pi a_0^2\right)
\right]^2
}
\Bigg]
\left[
a^{2-n}-\frac{a_0^{4-n}}{a^2}
\right].
\end{aligned}
\label{eq:exp_potential_chaplygin}
\end{equation}
For the special case $n=4$, the potential becomes
\begin{equation}
\begin{aligned}
V_{{\rm eff},\exp}^{\rm (C)}(a)
={}&
1-\frac{2M}
{a\left[
1-\eta\exp\left(-\pi a^2\right)
\right]}
\\
&-
\left[
1-\frac{2M}
{a_0\left[
1-\eta\exp\left(-\pi a_0^2\right)
\right]}
\right]
\left(\frac{a_0}{a}\right)^2
\\
&-
\frac{2a_0^2}{a^2}
\Bigg[
1-
\frac{
M\left\{
1-\left(2\pi a_0^2+1\right)
\eta\exp\left(-\pi a_0^2\right)
\right\}
}
{
a_0
\left[
1-\eta\exp\left(-\pi a_0^2\right)
\right]^2
}
\Bigg]
\ln\left(\frac{a}{a_0}\right).
\end{aligned}
\label{eq:exp_potential_chaplygin_n4}
\end{equation}
The parameter in Eq.~\eqref{eq:exp_omega_chaplygin} ensures that $V_{{\rm eff},\exp}^{\rm (C)}(a_0)=V_{{\rm eff},\exp}^{{'}\,\rm (C)}(a_0)=0$. The potential curvature is
\begin{equation}
\begin{aligned}
V_{{\rm eff},\exp}^{{''}\,\rm (C)}(a_0)
={}&
\frac{
2M
}{
a_0^3
\left[
1-\eta\exp\left(-\pi a_0^2\right)
\right]^3
}
\Bigg\{
2\pi a_0^2
\eta\exp\left(-\pi a_0^2\right)
\left[
1-\eta\exp\left(-\pi a_0^2\right)
\right]
\left(3-2\pi a_0^2\right)
\\
&+
(n-1)
\left[
1-\eta\exp\left(-\pi a_0^2\right)
\right]
\left[
1+\left(2\pi a_0^2-1\right)
\eta\exp\left(-\pi a_0^2\right)
\right]
\\
&-
4\pi a_0^2
\eta\exp\left(-\pi a_0^2\right)
\left[
1+\left(2\pi a_0^2-1\right)
\eta\exp\left(-\pi a_0^2\right)
\right]
\Bigg\}
\\
&+
\frac{2(n-2)}{a_0^2}
\left[
1-\frac{2M}
{a_0\left[
1-\eta\exp\left(-\pi a_0^2\right)
\right]}
\right].
\end{aligned}
\label{eq:exp_chaplygin_stability}
\end{equation}
The variable Chaplygin shell is linearly stable whenever $V_{{\rm eff},\exp}^{{''}\,\rm (C)}(a_0)>0$. In particular, the $n=4$ branch satisfies
\begin{equation}
\begin{aligned}
V_{{\rm eff},\exp}^{{''}\,\rm (C)}(a_0)
={}&
\frac{
2M
}{
a_0^3
\left[
1-\eta\exp\left(-\pi a_0^2\right)
\right]^3
}
\Bigg\{
2\pi a_0^2
\eta\exp\left(-\pi a_0^2\right)
\left[
1-\eta\exp\left(-\pi a_0^2\right)
\right]
\left(3-2\pi a_0^2\right)
\\
&+
3
\left[
1-\eta\exp\left(-\pi a_0^2\right)
\right]
\left[
1+\left(2\pi a_0^2-1\right)
\eta\exp\left(-\pi a_0^2\right)
\right]
\\
&-
4\pi a_0^2
\eta\exp\left(-\pi a_0^2\right)
\left[
1+\left(2\pi a_0^2-1\right)
\eta\exp\left(-\pi a_0^2\right)
\right]
\Bigg\}
\\
&+
\frac{4}{a_0^2}
\left[
1-\frac{2M}
{a_0\left[
1-\eta\exp\left(-\pi a_0^2\right)
\right]}
\right].
\end{aligned}
\label{eq:exp_chaplygin_stability_n4}
\end{equation}

\begin{figure}[htp!]
    \centering
    \includegraphics[width=1.0\linewidth]{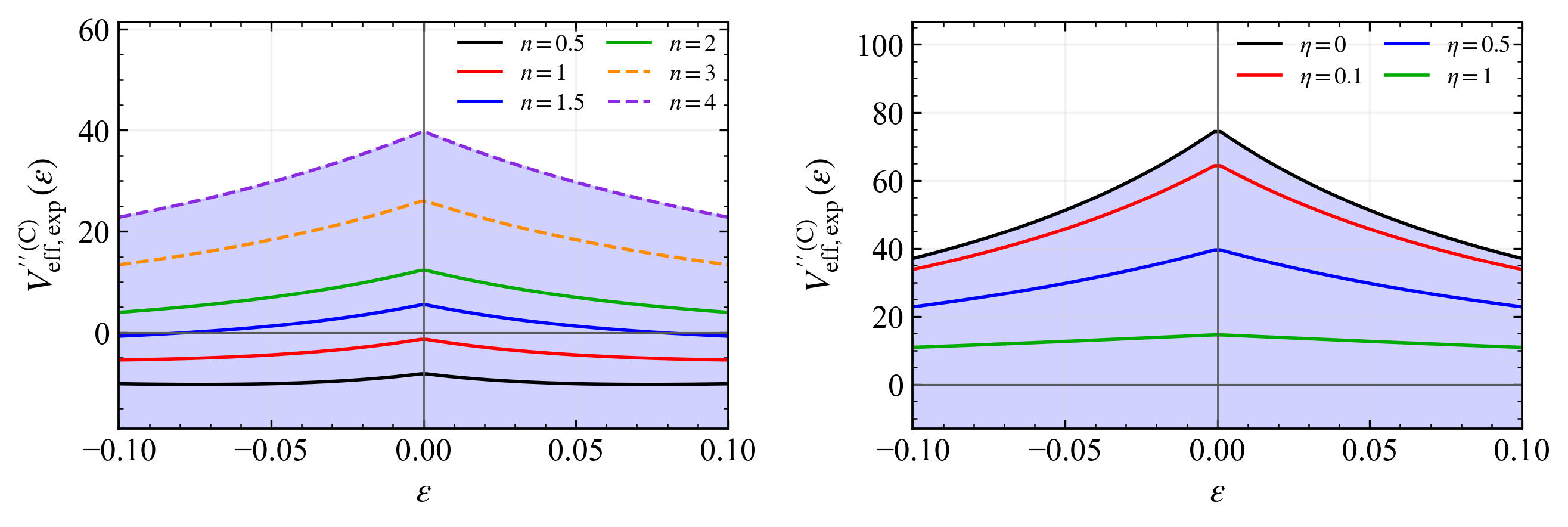}
    \caption{Effective-potential curvature $V_{{\rm eff},\exp}^{{''}\,\rm (C)}(\mathcal E)$ for the variable Chaplygin exponentially corrected thin-shell wormhole, with $a_0=r_h^{(\exp)}+|\mathcal E|$. In the left panel, $M=0.1$ and $\eta=0.5$, and the curves correspond to $n=0.5$, $1$, $1.5$, $2$, $3$, and $4$. In the right panel, $M=0.1$ and $n=4$, and the curves correspond to $\eta=0$, $0.1$, $0.5$, and $1$. The Chaplygin parameter $\Omega$ is fixed by the static equilibrium condition in each case. Positive regions of the curvature correspond to linear radial stability.}
    \label{fig:stbchexp}
\end{figure}

Figure~\ref{fig:stbchexp} shows that the exponential correction acts as a localized modification of the shell dynamics rather than as a long-range restructuring of the exterior geometry. In the left panel, where $M=0.1$ and $\eta=0.5$, the curves with $n=0.5$ and $n=1$ remain unstable, whereas the $n=1.5$, $2$, $3$, and $4$ branches have positive curvature in the near-horizon interval displayed. Thus, the variable Chaplygin response can provide a restoring contribution that is absent in the barotropic sector, but this stabilization is not generated by the bulk deformation alone: the barotropic shell remains unstable for the same geometry.

The physical role of the exponential entropy correction is particularly clear in this small-mass regime. Since the horizon radius is sufficiently small, the factor $\eta\exp(-\pi a_0^2)$ is not yet strongly suppressed, and the effective geometry retains information about the short-distance correction. The resulting change in $V_{{\rm eff},\exp}^{{''}\,\rm (C)}$ is concentrated around $\mathcal E=0$, namely, around throat configurations closest to the horizon. This contrasts with entropy deformations that retain power-law or logarithmic contributions over a broad exterior range. Here, once the shell is moved away from the horizon, the exponential factor rapidly fades and the dynamics progressively returns to the ordinary large-radius Chaplygin behavior. The correction therefore modifies the local stiffness of the shell rather than producing a persistent change throughout the exterior region.

The right panel makes this distinction more evident. For the already stable $n=4$ branch, increasing $\eta$ decreases the positive curvature, although it does not change its sign in the interval shown. Hence, the exponential contribution does not act as an additional universal stabilizing mechanism: it weakens the local restoring force while preserving the stable Chaplygin branch. The fact that the effect is visible for $M=0.1$ is physically significant. For a macroscopic mass, the horizon would occur at larger radius and the factor $\exp(-4\pi M^2)$ would be exceedingly small, causing the curves to approach the classical result. In this sense, the figure supports a semiclassical interpretation in which the entropy correction is relevant for small black holes or microscopic horizon scales, while its influence becomes negligible in the macroscopic regime.

\section{Conclusions}
\label{sec:conclusions}

In this work, we developed a unified framework for constructing symmetric thin-shell wormholes from entropy-induced black-hole geometries. Starting from a generic entropy function $\mathcal S(r)$, the associated lapse function was used as the seed for a cut-and-paste construction, allowing the junction conditions, surface stresses, energy-condition combinations, conservation law, and radial effective potential to be expressed directly in terms of the entropy derivatives. In this way, modified black-hole entropies were translated into definite geometric and dynamical properties of the wormhole throat.

A first general conclusion is that the symmetric construction retains an unavoidable exotic component on the shell. Whenever the throat lies in a positive-lapse region, the static surface energy density is negative, independently of the entropy prescription adopted for the seed geometry. Entropy deformations can nevertheless modify the radial distribution of the surface stresses, the position of the admissible static domain, and the region in which the intrinsic null energy condition on the shell is satisfied. Therefore, the relevant effect of the entropy correction is not the elimination of exotic matter, but the way in which it redistributes and localizes the exotic surface contribution.

The analysis also makes clear that the horizon structure inherited from the entropy-induced geometry is physically important for the thin-shell construction. Some entropy prescriptions preserve a Schwarzschild-like exterior with a single horizon and an unbounded positive-lapse region, whereas others generate a finite static interval bounded by two horizons. This distinction affects both the admissible location of the throat and the range over which the radial dynamics can be meaningfully investigated. In this sense, the thin-shell framework is especially useful because it can exploit positive-lapse sectors even when the same entropy deformation would not naturally support a smooth Morris--Thorne wormhole geometry.

For the radial stability analysis, the constant barotropic model provides a common reference behavior. Within the parameter domains considered here, all examined constant-barotropic branches are linearly radially unstable, with the geometric deformations modifying the scale and near-horizon intensity of the instability without generating a stable branch. This result concerns local stability under spherically symmetric radial perturbations about the static equilibrium radius, as diagnosed by the sign of $\Veff''(a_0)$; it does not address global nonlinear stability or stability against nonspherical perturbations.

The variable Chaplygin model leads to a qualitatively different conclusion. In this case, the radial response of the shell can produce stable configurations, with the stability domains controlled jointly by the entropy-induced lapse, the throat position, and the exponent governing the radial dependence of the equation of state. For each selected equilibrium radius $a_0$, however, the equilibrium condition fixes a generally distinct value $\Omega=\Omega_0(a_0)$; thus, the stability curves represent a family of equilibrated Chaplygin shells rather than a single shell with fixed $\Omega$ over the whole plotted interval. The entropy deformation affects the threshold between stable and unstable branches, while the large-radius behavior is governed by the asymptotic structure of the shell model. This separation between bulk geometry and shell response is one of the main outcomes of the present analysis: modified entropy alone does not determine stability, but it changes the geometric environment in which a given shell equation of state operates.

The comparison among the entropy sectors further shows that not all deformations act over the same physical scale. Power-law and logarithmic modifications can affect the shell dynamics over an extended exterior region, whereas hyperbolic and exponential corrections are more localized near the horizon. In particular, the exponentially corrected model provides a natural semiclassical interpretation: its contribution is appreciable for small horizon scales, while it becomes rapidly suppressed for macroscopic masses. Thus, entropy-induced thin-shell wormholes offer a useful setting for distinguishing long-range thermodynamic deformations from localized quantum-inspired corrections at the level of junction dynamics.

Several extensions of this work deserve further investigation. A natural continuation is the study of more general shell equations of state, including nonlinear, anisotropic, dissipative, or perturbatively defined responses. It would also be relevant to analyze nonsymmetric junctions, charged or rotating entropy-induced geometries, and the effects of cosmological backgrounds. Beyond linear radial stability, nonlinear evolutions of the throat could clarify the fate of perturbed stable configurations. Finally, the relation between the entropy-induced effective matter sector in the bulk and the exotic matter confined to the shell should be explored in greater detail, particularly in connection with thermodynamic stability, quantum inequalities, and possible microscopic interpretations of the entropy corrections.

\section*{Acknowledgments}
EO would like to thank Fundação Cearense de Apoio ao Desenvolvimento Científico e Tecnológico (FUNCAP), through grant BP6-0241-00335.01.00/25.

\bibliographystyle{apsrev4-1}
\bibliography{ref}

@article{MorrisThorne1988,
   author = {Michael S. Morris and Kip S. Thorne},
   doi = {10.1119/1.15620},
   issn = {0002-9505},
   issue = {5},
   journal = {American Journal of Physics},
   month = {5},
   pages = {395-412},
   publisher = {American Association of Physics Teachers},
   title = {Wormholes in spacetime and their use for interstellar travel: A tool for teaching general relativity},
   volume = {56},
   url = {https://pubs.aip.org/ajp/article/56/5/395/1044276/Wormholes-in-spacetime-and-their-use-for},
   year = {1988}
}

@article{Visser1989,
   author = {Matt Visser},
   doi = {10.1103/PhysRevD.39.3182},
   issn = {0556-2821},
   issue = {10},
   journal = {Physical Review D},
   month = {5},
   pages = {3182-3184},
   title = {Traversable wormholes: Some simple examples},
   volume = {39},
   url = {https://link.aps.org/doi/10.1103/PhysRevD.39.3182},
   year = {1989}
}

@article{Visser1989Surgical,
   author = {Matt Visser},
   doi = {10.1016/0550-3213(89)90100-4},
   issn = {05503213},
   issue = {1},
   journal = {Nuclear Physics B},
   month = {12},
   pages = {203-212},
   title = {Traversable wormholes from surgically modified Schwarzschild spacetimes},
   volume = {328},
   url = {https://linkinghub.elsevier.com/retrieve/pii/0550321389901004},
   year = {1989}
}

@book{VisserBook,
   author = {Matt Visser},
   city = {Woodbury, New York},
   isbn = {978-1-56396-653-8},
   publisher = {AIP Press},
   title = {Lorentzian Wormholes: From Einstein to Hawking},
   year = {1995}
}

@book{Darmois1927,
  author = {Darmois, Georges},
  title = {Les \'equations de la gravitation einsteinienne},
  publisher = {Gauthier-Villars},
  address = {Paris},
  year = {1927}
}

@article{Lanczos1924,
  author = {Lanczos, Cornelius},
  title = {Fl{\"a}chenhafte Verteilung der Materie in der Einsteinschen Gravitationstheorie},
  journal = {Ann. Phys.},
  volume = {379},
  pages = {518--540},
  year = {1924},
  doi = {10.1002/andp.19243791403}
}

@article{Israel1966,
   author = {W. Israel},
   doi = {10.1007/BF02710419},
   issn = {0369-3554},
   issue = {1},
   journal = {Il Nuovo Cimento B Series 10},
   month = {7},
   pages = {1-14},
   title = {Singular hypersurfaces and thin shells in general relativity},
   volume = {44},
   url = {http://link.springer.com/10.1007/BF02710419},
   year = {1966}
}

@article{PoissonVisser1995,
   author = {Eric Poisson and Matt Visser},
   doi = {10.1103/PhysRevD.52.7318},
   issn = {0556-2821},
   issue = {12},
   journal = {Physical Review D},
   month = {12},
   pages = {7318-7321},
   title = {Thin-shell wormholes: Linearization stability},
   volume = {52},
   url = {https://link.aps.org/doi/10.1103/PhysRevD.52.7318},
   year = {1995}
}

@article{Eiroa2008Stability,
   author = {Ernesto F. Eiroa},
   doi = {10.1103/PhysRevD.78.024018},
   issn = {1550-7998},
   issue = {2},
   journal = {Physical Review D},
   month = {7},
   pages = {024018},
   title = {Stability of thin-shell wormholes with spherical symmetry},
   volume = {78},
   url = {https://link.aps.org/doi/10.1103/PhysRevD.78.024018},
   year = {2008}
}

@article{GarciaLoboVisser2012,
   author = {Nadiezhda Montelongo Garcia and Francisco S. N. Lobo and Matt Visser},
   doi = {10.1103/PhysRevD.86.044026},
   issn = {1550-7998},
   issue = {4},
   journal = {Physical Review D},
   month = {8},
   pages = {044026},
   title = {Generic spherically symmetric dynamic thin-shell traversable wormholes in standard general relativity},
   volume = {86},
   url = {https://link.aps.org/doi/10.1103/PhysRevD.86.044026},
   year = {2012}
}

@article{LoboCrawford2004,
  author = {Lobo, Francisco S. N. and Crawford, Paulo},
  title = {Linearized stability analysis of thin-shell wormholes with a cosmological constant},
  journal = {Class. Quantum Gravity},
  volume = {21},
  pages = {391--404},
  year = {2004},
  doi = {10.1088/0264-9381/21/2/004}
}

@article{EiroaRomero2004,
   author = {Ernesto F. Eiroa and Gustavo E. Romero},
   doi = {10.1023/B:GERG.0000016916.79221.24},
   issn = {0001-7701},
   issue = {4},
   journal = {General Relativity and Gravitation},
   month = {4},
   pages = {651-659},
   title = {Linearized Stability of Charged Thin-Shell Wormholes},
   volume = {36},
   url = {https://link.springer.com/10.1023/B:GERG.0000016916.79221.24},
   year = {2004}
}

@article{EiroaSimeone2005,
  author = {Eiroa, Ernesto F. and Simeone, Claudio},
  title = {Thin-shell wormholes in dilaton gravity},
  journal = {Phys. Rev. D},
  volume = {71},
  pages = {127501},
  year = {2005},
  doi = {10.1103/PhysRevD.71.127501}
}

@article{Varela2015,
   author = {Victor Varela},
   doi = {10.1103/PhysRevD.92.044002},
   issn = {1550-7998},
   issue = {4},
   journal = {Physical Review D},
   month = {8},
   pages = {044002},
   title = {Note on linearized stability of Schwarzschild thin-shell wormholes with variable equations of state},
   volume = {92},
   url = {https://link.aps.org/doi/10.1103/PhysRevD.92.044002},
   year = {2015}
}

@article{Eiroa2007Chaplygin,
   author = {Ernesto F. Eiroa and Claudio Simeone},
   doi = {10.1103/PhysRevD.76.024021},
   issn = {1550-7998},
   issue = {2},
   journal = {Physical Review D},
   month = {7},
   pages = {024021},
   title = {Stability of Chaplygin gas thin-shell wormholes},
   volume = {76},
   url = {https://link.aps.org/doi/10.1103/PhysRevD.76.024021},
   year = {2007}
}

@article{Eiroa2009Chaplygin,
   author = {Ernesto F. Eiroa},
   doi = {10.1103/PhysRevD.80.044033},
   issn = {1550-7998},
   issue = {4},
   journal = {Physical Review D},
   month = {8},
   pages = {044033},
   title = {Thin-shell wormholes with a generalized Chaplygin gas},
   volume = {80},
   url = {https://link.aps.org/doi/10.1103/PhysRevD.80.044033},
   year = {2009}
}

@article{ForghaniMazharimousaviHalilsoy2019,
   author = {S. Danial Forghani and S. Habib Mazharimousavi and M. Halilsoy},
   doi = {10.1142/S0218271819501426},
   issn = {0218-2718},
   issue = {11},
   journal = {International Journal of Modern Physics D},
   month = {8},
   pages = {1950142},
   title = {Thermodynamic stability of a Schwarzschild thin-shell wormhole},
   volume = {28},
   url = {https://www.worldscientific.com/doi/abs/10.1142/S0218271819501426},
   year = {2019}
}

@article{EiroaFigueroaAguirre2024,
   author = {Ernesto F. Eiroa and Griselda Figueroa-Aguirre and Miguel L. Peñafiel and Santiago Esteban Perez Bergliaffa},
   doi = {10.1140/epjc/s10052-024-13465-3},
   issn = {1434-6052},
   issue = {11},
   journal = {The European Physical Journal C},
   month = {11},
   pages = {1160},
   title = {Dynamical and thermodynamical stability of a charged thin-shell wormhole},
   volume = {84},
   url = {https://link.springer.com/10.1140/epjc/s10052-024-13465-3},
   year = {2024}
}

@article{Javed2024PolymerTSW,
   author = {Faisal Javed and Arfa Waseem and Ghulam Fatima and Bander Almutairi},
   doi = {10.1016/j.dark.2024.101605},
   issn = {22126864},
   journal = {Physics of the Dark Universe},
   month = {12},
   pages = {101605},
   title = {Stability of thin-shell wormholes via polymer black hole in loop quantum gravity},
   volume = {46},
   url = {https://linkinghub.elsevier.com/retrieve/pii/S2212686424001870},
   year = {2024}
}

@article{Bekenstein1973,
   author = {Jacob D. Bekenstein},
   doi = {10.1103/PhysRevD.7.2333},
   issn = {0556-2821},
   issue = {8},
   journal = {Physical Review D},
   month = {4},
   pages = {2333-2346},
   title = {Black Holes and Entropy},
   volume = {7},
   url = {https://link.aps.org/doi/10.1103/PhysRevD.7.2333},
   year = {1973}
}

@article{Hawking1975,
   author = {S. W. Hawking},
   doi = {10.1007/BF02345020},
   issn = {0010-3616},
   issue = {3},
   journal = {Communications In Mathematical Physics},
   month = {8},
   pages = {199-220},
   title = {Particle creation by black holes},
   volume = {43},
   url = {http://link.springer.com/10.1007/BF02345020},
   year = {1975}
}

@article{GibbonsHawking1977,
   author = {G. W. Gibbons and S. W. Hawking},
   doi = {10.1103/PhysRevD.15.2752},
   issn = {0556-2821},
   issue = {10},
   journal = {Physical Review D},
   month = {5},
   pages = {2752-2756},
   title = {Action integrals and partition functions in quantum gravity},
   volume = {15},
   url = {https://link.aps.org/doi/10.1103/PhysRevD.15.2752},
   year = {1977}
}

@article{Wald1993,
   author = {Robert M. Wald},
   doi = {10.1103/PhysRevD.48.R3427},
   issn = {0556-2821},
   issue = {8},
   journal = {Physical Review D},
   month = {10},
   pages = {R3427-R3431},
   title = {Black hole entropy is the Noether charge},
   volume = {48},
   url = {https://link.aps.org/doi/10.1103/PhysRevD.48.R3427},
   year = {1993}
}

@article{Jacobson1995,
   author = {Ted Jacobson},
   doi = {10.1103/PhysRevLett.75.1260},
   issn = {0031-9007},
   issue = {7},
   journal = {Physical Review Letters},
   month = {6},
   pages = {1260-1263},
   title = {Thermodynamics of Spacetime: The Einstein Equation of State},
   volume = {75},
   url = {http://arxiv.org/abs/gr-qc/9504004 http://dx.doi.org/10.1103/PhysRevLett.75.1260},
   year = {1995}
}

@article{Padmanabhan2010,
   author = {T. Padmanabhan},
   doi = {10.1088/0034-4885/73/4/046901},
   issn = {0034-4885},
   issue = {4},
   journal = {Reports on Progress in Physics},
   month = {1},
   pages = {046901},
   title = {Thermodynamical Aspects of Gravity: New insights},
   volume = {73},
   url = {http://arxiv.org/abs/0911.5004 http://dx.doi.org/10.1088/0034-4885/73/4/046901},
   year = {2010}
}

@article{Verlinde2011,
   author = {Erik P Verlinde},
   doi = {10.1007/JHEP04(2011)029},
   issue = {4},
   journal = {Journal of High Energy Physics},
   pages = {29},
   title = {On the Origin of Gravity and the Laws of Newton},
   volume = {2011},
   year = {2011}
}

@article{Anand2026,
   author = {Ankit Anand and Sahil Devdutt and Kimet Jusufi and Emmanuel N. Saridakis},
   doi = {10.1140/epjc/s10052-026-15768-z},
   issn = {1434-6052},
   issue = {5},
   journal = {The European Physical Journal C},
   month = {5},
   pages = {534},
   title = {Effective matter sectors from modified entropies},
   volume = {86},
   url = {https://link.springer.com/10.1140/epjc/s10052-026-15768-z},
   year = {2026}
}

@article{Barrow2020,
   author = {John D. Barrow},
   doi = {10.1016/j.physletb.2020.135643},
   issn = {03702693},
   journal = {Physics Letters B},
   month = {9},
   pages = {135643},
   publisher = {Elsevier},
   title = {The area of a rough black hole},
   volume = {808},
   url = {https://linkinghub.elsevier.com/retrieve/pii/S0370269320304469},
   year = {2020}
}

@article{Tsallis1988,
   author = {Constantino Tsallis},
   doi = {10.1007/BF01016429},
   issn = {0022-4715},
   issue = {1-2},
   journal = {Journal of Statistical Physics},
   month = {7},
   pages = {479-487},
   publisher = {Springer},
   title = {Possible generalization of Boltzmann-Gibbs statistics},
   volume = {52},
   url = {http://link.springer.com/10.1007/BF01016429},
   year = {1988}
}

@article{TsallisCirto2013,
   author = {Constantino Tsallis and Leonardo J. L. Cirto},
   doi = {10.1140/epjc/s10052-013-2487-6},
   issn = {1434-6044},
   issue = {7},
   journal = {The European Physical Journal C},
   month = {6},
   pages = {2487},
   title = {Black hole thermodynamical entropy},
   volume = {73},
   url = {http://arxiv.org/abs/1202.2154 http://dx.doi.org/10.1140/epjc/s10052-013-2487-6},
   year = {2013}
}

@inproceedings{Renyi1961,
  author = {R{\'e}nyi, Alfr{\'e}d},
  title = {On measures of entropy and information},
  booktitle = {Proceedings of the Fourth Berkeley Symposium on Mathematical Statistics and Probability},
  volume = {1},
  pages = {547--561},
  year = {1961},
  publisher = {University of California Press}
}

@article{CzinnerIguchi2016,
  author = {Czinner, Viktor G. and Iguchi, Hideo},
  title = {R{\'e}nyi entropy and the thermodynamic stability of black holes},
  journal = {Phys. Lett. B},
  volume = {752},
  pages = {306--310},
  year = {2016},
  doi = {10.1016/j.physletb.2015.11.061}
}

@article{Kaniadakis2002,
   author = {G. Kaniadakis},
   doi = {10.1103/PhysRevE.66.056125},
   issn = {1063-651X},
   issue = {5},
   journal = {Physical Review E},
   month = {11},
   pages = {056125},
   publisher = {APS},
   title = {Statistical mechanics in the context of special relativity},
   volume = {66},
   url = {https://link.aps.org/doi/10.1103/PhysRevE.66.056125},
   year = {2002}
}

@article{KaulMajumdar2000,
   author = {Romesh K. Kaul and Parthasarathi Majumdar},
   doi = {10.1103/PhysRevLett.84.5255},
   issn = {0031-9007},
   issue = {23},
   journal = {Physical Review Letters},
   month = {6},
   pages = {5255-5257},
   title = {Logarithmic correction to the Bekenstein-Hawking entropy},
   volume = {84},
   url = {http://arxiv.org/abs/gr-qc/0002040 http://dx.doi.org/10.1103/PhysRevLett.84.5255},
   year = {2000}
}

@article{DasMajumdarBhaduri2002,
   author = {Saurya Das and Parthasarathi Majumdar and Rajat K. Bhaduri},
   doi = {10.1088/0264-9381/19/9/302},
   issn = {0264-9381},
   issue = {9},
   journal = {Classical and Quantum Gravity},
   month = {3},
   pages = {2355-2367},
   title = {General Logarithmic Corrections to Black Hole Entropy},
   volume = {19},
   url = {http://arxiv.org/abs/hep-th/0111001 http://dx.doi.org/10.1088/0264-9381/19/9/302},
   year = {2002}
}

@article{ChatterjeeGhosh2020,
   author = {Ayan Chatterjee and Amit Ghosh},
   doi = {10.1103/PhysRevLett.125.041302},
   issn = {0031-9007},
   issue = {4},
   journal = {Physical Review Letters},
   month = {7},
   pages = {041302},
   title = {Exponential corrections to black hole entropy},
   volume = {125},
   url = {http://arxiv.org/abs/2007.15401 http://dx.doi.org/10.1103/PhysRevLett.125.041302},
   year = {2020}
}

@misc{ReboucasEntropyMT2026,
  author        = {Rebou{\c c}as, Jonathan A. and Lustosa, Francisco Bento and Muniz, Celio R.},
  title         = {Traversable Wormholes Supported by Entropy-Inspired Effective Matter Sectors},
  year          = {2026},
  eprint        = {2606.00178},
  archivePrefix = {arXiv},
  primaryClass  = {gr-qc},
  url           = {https://arxiv.org/abs/2606.00178}
}

@article{Ellis1973,
   author = {Homer G. Ellis},
   doi = {10.1063/1.1666161},
   issn = {0022-2488},
   issue = {1},
   journal = {Journal of Mathematical Physics},
   month = {1},
   pages = {104-118},
   title = {Ether flow through a drainhole: A particle model in general relativity},
   volume = {14},
   url = {https://pubs.aip.org/jmp/article/14/1/104/223726/Ether-flow-through-a-drainhole-A-particle-model-in},
   year = {1973}
}

@article{Bronnikov1973,
   author = {K A Bronnikov},
   journal = {Acta Physica Polonica B},
   pages = {251-266},
   title = {Scalar-Tensor Theory and Scalar Charge},
   volume = {4},
   url = {https://inspirehep.net/files/1a28c080a733a1b776867157a30efd12},
   year = {1973}
}

@article{Lobo2005,
   author = {Francisco S. N. Lobo},
   doi = {10.1103/PhysRevD.71.084011},
   issn = {1550-7998},
   issue = {8},
   journal = {Physical Review D},
   month = {4},
   pages = {084011},
   title = {Phantom energy traversable wormholes},
   volume = {71},
   url = {http://arxiv.org/abs/gr-qc/0502099 http://dx.doi.org/10.1103/PhysRevD.71.084011},
   year = {2005}
}

@article{Sushkov2005,
   author = {S. V. Sushkov},
   doi = {10.1103/PhysRevD.71.043520},
   issn = {1550-7998},
   issue = {4},
   journal = {Physical Review D},
   month = {2},
   pages = {043520},
   title = {Wormholes supported by a phantom energy},
   volume = {71},
   url = {http://arxiv.org/abs/gr-qc/0502084 http://dx.doi.org/10.1103/PhysRevD.71.043520},
   year = {2005}
}

@article{Harko2013,
   author = {Tiberiu Harko and Francisco S. N. Lobo and M. K. Mak and Sergey V. Sushkov},
   doi = {10.1103/PhysRevD.87.067504},
   issn = {1550-7998},
   issue = {6},
   journal = {Physical Review D},
   month = {3},
   pages = {067504},
   title = {Modified-gravity wormholes without exotic matter},
   volume = {87},
   url = {http://arxiv.org/abs/1301.6878 http://dx.doi.org/10.1103/PhysRevD.87.067504},
   year = {2013}
}

@article{Radhakrishnan2024,
   author = {Ramesh Radhakrishnan and Patrick Brown and Jacob Matulevich and Eric Davis and Delaram Mirfendereski and Gerald Cleaver},
   doi = {10.3390/sym16081007},
   issn = {2073-8994},
   issue = {8},
   journal = {Symmetry},
   month = {8},
   pages = {1007},
   title = {A Review of Stable, Traversable Wormholes in f(R) Gravity Theories},
   volume = {16},
   url = {https://www.mdpi.com/2073-8994/16/8/1007},
   year = {2024}
}

@article{Santos2024,
   author = {A. C. L. Santos and R. V. Maluf and C. R. Muniz},
   doi = {10.1016/j.aop.2024.169775},
   issn = {00034916},
   journal = {Annals of Physics},
   month = {8},
   pages = {169775},
   title = {Generating 4-dimensional Wormholes with Yang-Mills Casimir Sources},
   volume = {469},
   url = {http://arxiv.org/abs/2405.08774 http://dx.doi.org/10.1016/j.aop.2024.169775},
   year = {2024}
}

@article{Alencar2021,
   author = {G. Alencar and V. B. Bezerra and C. R. Muniz and H. S. Vieira},
   doi = {10.3390/universe7070238},
   issn = {2218-1997},
   issue = {7},
   journal = {Universe},
   month = {7},
   pages = {238},
   title = {Ellis–Bronnikov Wormholes in Asymptotically Safe Gravity},
   volume = {7},
   url = {https://www.mdpi.com/2218-1997/7/7/238},
   year = {2021}
}

@article{Cruz2024,
   author = {M.B. Cruz and R.M.P. Neves and Celio R. Muniz},
   doi = {10.1088/1475-7516/2024/05/016},
   issn = {1475-7516},
   issue = {05},
   journal = {Journal of Cosmology and Astroparticle Physics},
   month = {5},
   pages = {016},
   title = {Traversable wormholes from Loop Quantum Gravity},
   volume = {2024},
   url = {https://iopscience.iop.org/article/10.1088/1475-7516/2024/05/016},
   year = {2024}
}

@article{Muniz2025,
   author = {C.R. Muniz and M.B. Cruz and R.M.P. Neves and Mushayydha Farooq and M. Zubair},
   doi = {10.1088/1475-7516/2025/07/015},
   issn = {1475-7516},
   issue = {07},
   journal = {Journal of Cosmology and Astroparticle Physics},
   month = {7},
   pages = {015},
   title = {Hot Casimir wormholes in Einstein Gauss-Bonnet gravity},
   volume = {2025},
   url = {https://iopscience.iop.org/article/10.1088/1475-7516/2025/07/015},
   year = {2025}
}

@article{Crispim2026,
   author = {T. M. Crispim and G. Alencar and C. R. Muniz},
   doi = {10.1088/1361-6382/ae74ac},
   issn = {0264-9381},
   issue = {11},
   journal = {Classical and Quantum Gravity},
   month = {5},
   pages = {115013},
   title = {Field Sources for Generalized Ellis-Bronnikov Wormhole},
   volume = {43},
   url = {http://arxiv.org/abs/2410.11147 http://dx.doi.org/10.1088/1361-6382/ae74ac},
   year = {2026}
}

@article{Silva2025,
   author = {Marcos V. de S. Silva and G. Alencar and R. N. Costa Filho and R. M. P. Neves and Celio R. Muniz},
   doi = {10.1140/epjp/s13360-025-06214-2},
   issn = {2190-5444},
   issue = {4},
   journal = {The European Physical Journal Plus},
   month = {4},
   pages = {289},
   title = {Traversable Wormholes Sourced by Dark Matter in Loop Quantum Cosmology},
   volume = {140},
   url = {http://arxiv.org/abs/2411.12063 http://dx.doi.org/10.1140/epjp/s13360-025-06214-2},
   year = {2025}
}

@article{Mustafa2023,
   author = {G. Mustafa and Ibrar Hussain and Farruh Atamurotov and Wu Ming Liu},
   doi = {10.1140/epjp/s13360-023-03775-y},
   issn = {21905444},
   issue = {2},
   journal = {European Physical Journal Plus},
   month = {2},
   publisher = {Springer Science and Business Media Deutschland GmbH},
   title = {Imprints of dark matter on wormhole geometry in modified teleparallel gravity},
   volume = {138},
   year = {2023}
}

@article{Khatri2025,
   author = {Mohan Khatri and Pradyumn Kumar Sahoo},
   doi = {10.1016/j.dark.2025.102042},
   issn = {22126864},
   journal = {Physics of the Dark Universe},
   month = {9},
   publisher = {Elsevier B.V.},
   title = {Existence of wormhole in Dekel–Zhao dark matter halo},
   volume = {49},
   year = {2025}
}

@article{ReboucasPade2026,
   author = {Jonathan Alves Rebouças and Francisco Tiago Barboza Sampaio and Francisco Bento Lustosa and Leonardo Tavares de Oliveira and C. R. Muniz},
   doi = {10.1002/andp.202500636},
   issn = {0003-3804},
   issue = {3},
   journal = {Annalen der Physik},
   month = {3},
   pages = {e00636},
   title = {A New Padé Approach to Modeling Wormholes in Dekel‐Zhao Dark Matter Halos},
   volume = {538},
   url = {https://onlinelibrary.wiley.com/doi/10.1002/andp.202500636},
   year = {2026}
}

@article{NojiriOdintsovFaraoni2022GeneralizedEntropy,
  author        = {Nojiri, Shin'ichi and Odintsov, Sergei D. and Faraoni, Valerio},
  title         = {From nonextensive statistics and black hole entropy to the holographic dark universe},
  journal       = {Physical Review D},
  volume        = {105},
  number        = {4},
  pages         = {044042},
  year          = {2022},
  doi           = {10.1103/PhysRevD.105.044042},
  eprint        = {2201.02424},
  archivePrefix = {arXiv},
  primaryClass  = {gr-qc},
  url           = {https://doi.org/10.1103/PhysRevD.105.044042}
}

@article{NojiriOdintsovPaul2022GeneralizedEntropy,
  author        = {Nojiri, Shin'ichi and Odintsov, Sergei D. and Paul, Tanmoy},
  title         = {Early and late universe holographic cosmology from a new generalized entropy},
  journal       = {Physics Letters B},
  volume        = {831},
  pages         = {137189},
  year          = {2022},
  doi           = {10.1016/j.physletb.2022.137189},
  eprint        = {2205.08876},
  archivePrefix = {arXiv},
  primaryClass  = {gr-qc},
  url           = {https://doi.org/10.1016/j.physletb.2022.137189}
}

@article{NojiriOdintsovFaraoni2021AreaLaw,
  author        = {Nojiri, Shin'ichi and Odintsov, Sergei D. and Faraoni, Valerio},
  title         = {Area-law versus {R{\'e}nyi} and {Tsallis} black hole entropies},
  journal       = {Physical Review D},
  volume        = {104},
  number        = {8},
  pages         = {084030},
  year          = {2021},
  doi           = {10.1103/PhysRevD.104.084030},
  eprint        = {2109.05315},
  archivePrefix = {arXiv},
  primaryClass  = {gr-qc},
  url           = {https://doi.org/10.1103/PhysRevD.104.084030}
}

@article{ElizaldeNojiriOdintsov2025GeneralisedEntropy,
  author        = {Elizalde, Emilio and Nojiri, Shin'ichi and Odintsov, Sergei D.},
  title         = {Black Hole Thermodynamics and Generalised Non-Extensive Entropy},
  journal       = {Universe},
  volume        = {11},
  number        = {2},
  pages         = {60},
  year          = {2025},
  doi           = {10.3390/universe11020060},
  eprint        = {2502.05801},
  archivePrefix = {arXiv},
  primaryClass  = {gr-qc},
  url           = {https://doi.org/10.3390/universe11020060}
}

@article{NojiriOdintsovFolomeev2024Wormholes,
  author        = {Nojiri, Shin'ichi and Odintsov, Sergei D. and Folomeev, Vladimir},
  title         = {Wormholes inside stars and black holes},
  journal       = {Physical Review D},
  volume        = {109},
  number        = {10},
  pages         = {104007},
  year          = {2024},
  doi           = {10.1103/PhysRevD.109.104007},
  eprint        = {2401.15868},
  archivePrefix = {arXiv},
  primaryClass  = {gr-qc},
  url           = {https://doi.org/10.1103/PhysRevD.109.104007}
}

@article{LuDiGennaroOng2025VaryingG,
  author        = {L{\"u}, Hengxin and Di Gennaro, Sofia and Ong, Yen Chin},
  title         = {Generalized entropy implies varying-{G}: Horizon area dependent field equations and black hole--cosmology coupling},
  journal       = {Annals of Physics},
  volume        = {474},
  pages         = {169914},
  year          = {2025},
  doi           = {10.1016/j.aop.2024.169914},
  eprint        = {2407.00484},
  archivePrefix = {arXiv},
  primaryClass  = {gr-qc},
  url           = {https://doi.org/10.1016/j.aop.2024.169914}
}

@misc{FiglioliaJizbaLambiase2026ThermodynamicGravity,
  author        = {Figliolia, Marco and Jizba, Petr and Lambiase, Gaetano},
  title         = {Thermodynamic Gravity with Non-Extensive Horizon Entropy and Topological Calibration},
  year          = {2026},
  eprint        = {2602.20430},
  archivePrefix = {arXiv},
  primaryClass  = {gr-qc},
  url           = {https://arxiv.org/abs/2602.20430}
}

@article{DaiMinicStojkovic2020Formation,
  author        = {Dai, De-Chang and Minic, Djordje and Stojkovic, Dejan},
  title         = {How to form a wormhole},
  journal       = {The European Physical Journal C},
  volume        = {80},
  number        = {12},
  pages         = {1103},
  year          = {2020},
  doi           = {10.1140/epjc/s10052-020-08698-x},
  eprint        = {2010.03947},
  archivePrefix = {arXiv},
  primaryClass  = {gr-qc},
  url           = {https://doi.org/10.1140/epjc/s10052-020-08698-x}
}

@article{DaiMinicStojkovic2018DeSitter,
  author        = {Dai, De-Chang and Minic, Djordje and Stojkovic, Dejan},
  title         = {New wormhole solution in de Sitter space},
  journal       = {Physical Review D},
  volume        = {98},
  number        = {12},
  pages         = {124026},
  year          = {2018},
  doi           = {10.1103/PhysRevD.98.124026},
  eprint        = {1810.03432},
  archivePrefix = {arXiv},
  primaryClass  = {hep-th},
  url           = {https://doi.org/10.1103/PhysRevD.98.124026}
}

@article{DaiStojkovic2019Observing,
  author        = {Dai, De-Chang and Stojkovic, Dejan},
  title         = {Observing a wormhole},
  journal       = {Physical Review D},
  volume        = {100},
  number        = {8},
  pages         = {083513},
  year          = {2019},
  doi           = {10.1103/PhysRevD.100.083513},
  eprint        = {1910.00429},
  archivePrefix = {arXiv},
  primaryClass  = {gr-qc},
  url           = {https://doi.org/10.1103/PhysRevD.100.083513}
}

@article{SimonettiEtAl2021SensitiveSearch,
  author        = {Simonetti, John H. and Kavic, Michael J. and Minic, Djordje and Stojkovic, Dejan and Dai, De-Chang},
  title         = {Sensitive searches for wormholes},
  journal       = {Physical Review D},
  volume        = {104},
  number        = {8},
  pages         = {L081502},
  year          = {2021},
  doi           = {10.1103/PhysRevD.104.L081502},
  eprint        = {2007.12184},
  archivePrefix = {arXiv},
  primaryClass  = {gr-qc},
  url           = {https://doi.org/10.1103/PhysRevD.104.L081502}
}

@article{BambiStojkovic2021Astrophysical,
  author        = {Bambi, Cosimo and Stojkovic, Dejan},
  title         = {Astrophysical Wormholes},
  journal       = {Universe},
  volume        = {7},
  number        = {5},
  pages         = {136},
  year          = {2021},
  doi           = {10.3390/universe7050136},
  eprint        = {2105.00881},
  archivePrefix = {arXiv},
  primaryClass  = {gr-qc},
  url           = {https://doi.org/10.3390/universe7050136}
}

\end{document}